\documentclass[10pt,twocolumn,twoside]{IEEEtran}
\usepackage{epsfig,graphics,color,subfigure,wrapfig,hyperref,url}
\usepackage{amssymb,amsmath,algorithm,times}
\usepackage{slashbox}

\begin{document}
\newcommand{\st}{k}%{|T|}%
\newcommand{\sd}{u}%{|\Delta|}%
\newcommand{\sde}{e}%{|\Delta_e|}%
\newcommand{\sn}{s}%{|N|}%\check{}
\newcommand{\iter}{r}
\newcommand{\Tdp}{\check{T}_d} %{T_d'}
\newcommand{\Sp}{\check{S}}%{S'}
\newcommand{\sno}{n}%%{p}%{m}
\newcommand{\mno}{m} %{n}

\newcommand{\tty}{\tilde{y}_{t,\text{res}}}
\newcommand{\xhat}{\hat{x}}
\newcommand{\betahat}{\hat{\beta}}
\newcommand{\xhatold}{\hat{x}_{t,\text{old}}}

\newcommand{\Nhat}{\hat{N}}
\newcommand{\Dnum}{D_{num}}
\newcommand{\pss}{p^{**,i}}
\newcommand{\fr}{f_{r}^i}

\newcommand{\A}{{\cal A}}
\newcommand{\Z}{{\cal Z}}
\newcommand{\B}{{\cal B}}
\newcommand{\R}{{\cal R}}
\newcommand{\reg}{{\cal G}}
\newcommand{\const}{\mbox{const}}

\newcommand{\trace}{\mbox{trace}}

\newcommand{\hsim}{{\hspace{0.0cm} \sim  \hspace{0.0cm}}}
\newcommand{\he}{{\hspace{0.0cm} =  \hspace{0.0cm}}}

\newcommand{\vect}[2]{\left[\begin{array}{cccccc}
     #1 \\
     #2
   \end{array}
  \right]
  }

\newcommand{\matr}[2]{ \left[\begin{array}{cc}
     #1 \\
     #2
   \end{array}
  \right]
  }
\newcommand{\vc}[2]{\left[\begin{array}{c}
     #1 \\
     #2
   \end{array}
  \right]
  }

\newcommand{\gdot}{\dot{g}}
\newcommand{\Cdot}{\dot{C}}
\newcommand{\re}{\mathbb{R}}
\newcommand{\n}{{\cal N}}  %normal distribution
\newcommand{\N}{{\overrightarrow{\bf N}}}  % normal to contour
\newcommand{\chat}{\tilde{C}_t}
\newcommand{\chati}{\chat^i}

\newcommand{\cmin}{C^*_{min}}
\newcommand{\twi}{\tilde{w}_t^{(i)}}
\newcommand{\twj}{\tilde{w}_t^{(j)}}
\newcommand{\wi}{{w}_t^{(i)}}
\newcommand{\twio}{\tilde{w}_{t-1}^{(i)}}

\newcommand{\tWi}{\tilde{W}_n^{(m)}}
\newcommand{\tWj}{\tilde{W}_n^{(k)}}
\newcommand{\Wi}{{W}_n^{(m)}}
\newcommand{\tWio}{\tilde{W}_{n-1}^{(m)}}

\newcommand{\ds}{\displaystyle}

\newcommand{\SAR}{S$\!$A$\!$R }
\newcommand{\MAR}{MAR}
\newcommand{\MMRF}{MMRF}
\newcommand{\AR}{A$\!$R }
\newcommand{\GMRF}{G$\!$M$\!$R$\!$F }
\newcommand{\DTM}{D$\!$T$\!$M }
\newcommand{\MSE}{M$\!$S$\!$E }
\newcommand{\RCS}{R$\!$C$\!$S }
\newcommand{\uomega}{\underline{\omega}}
\newcommand{\y}{v}
\newcommand{\x}{w}
\newcommand{\lu}{\mu}
\newcommand{\g}{g}
\newcommand{\s}{{\bf s}}
\newcommand{\bft}{{\bf t}}
\newcommand{\refmap}{{\cal R}}
\newcommand{\totrefl}{{\cal E}}
\newcommand{\beq}{\begin{equation}}
\newcommand{\eeq}{\end{equation}}
\newcommand{\bdm}{\begin{displaymath}}
\newcommand{\edm}{\end{displaymath}}
\newcommand{\hatz}{\hat{z}}
\newcommand{\hatu}{\hat{u}}
\newcommand{\tilz}{\tilde{z}}
\newcommand{\tilu}{\tilde{u}}
\newcommand{\hhatz}{\hat{\hat{z}}}
\newcommand{\hhatu}{\hat{\hat{u}}}
\newcommand{\tilc}{\tilde{C}}
\newcommand{\hatc}{\hat{C}}
\newcommand{\tim}{n}

\newcommand{\ssp}{\renewcommand{\baselinestretch}{1.0}}
\newcommand{\defd}{\mbox{$\stackrel{\mbox{$\triangle$}}{=}$}}
\newcommand{\goes}{\rightarrow}
\newcommand{\tends}{\rightarrow}
\newcommand{\defn}{:=} %{\stackrel{\triangle}{=}}
\newcommand{\se}{&=&}
\newcommand{\sdefn}{& \defn  &}
\newcommand{\sle}{& \le &}
\newcommand{\sge}{& \ge &}
\newcommand{\plusminus}{\stackrel{+}{-}}
\newcommand{\Ey}{E_{Y_{1:t}}}
\newcommand{\ey}{E_{Y_{1:t}}}

\newcommand{\equivto}{\mbox{~~~which is equivalent to~~~}}
\newcommand{\nonzero}{i:\pi^n(x^{(i)})>0}
\newcommand{\nonzeroc}{i:c(x^{(i)})>0}

\newcommand{\supn}{\sup_{\phi:\|\phi\|_\infty \le 1}}
\newtheorem{theorem}{Theorem}
\newtheorem{lemma}{Lemma}
\newtheorem{proposition}{Proposition}
\newtheorem{corollary}{Corollary}
\newtheorem{definition}{Definition}
\newtheorem{remark}{Remark}
\newtheorem{example}{Example}
\newtheorem{ass}{Assumption}
\newtheorem{fact}{Fact}
\newtheorem{heuristic}{Heuristic}
\newcommand{\eps}{\epsilon}
\newcommand{\bd}{\begin{definition}}
\newcommand{\ed}{\end{definition}}
\newcommand{\udq}{\underline{D_Q}}
\newcommand{\td}{\tilde{D}}
\newcommand{\epsinv}{\epsilon_{inv}}
\newcommand{\al}{\mathcal{A}}

\newcommand{\bfx} {\bf X}
\newcommand{\bfy} {\bf Y}
\newcommand{\bfz} {\bf Z}
\newcommand{\ddas}{\mbox{${d_1}^2({\bf X})$}}
\newcommand{\ddbs}{\mbox{${d_2}^2({\bfx})$}}
\newcommand{\dda}{\mbox{$d_1(\bfx)$}}
\newcommand{\ddb}{\mbox{$d_2(\bfx)$}}
\newcommand{\xinc}{{\bfx} \in \mbox{$C_1$}}
\newcommand{\eqa}{\stackrel{(a)}{=}}
\newcommand{\eqb}{\stackrel{(b)}{=}}
\newcommand{\eqe}{\stackrel{(e)}{=}}
\newcommand{\leqc}{\stackrel{(c)}{\le}}
\newcommand{\leqd}{\stackrel{(d)}{\le}}

\newcommand{\leqa}{\stackrel{(a)}{\le}}
\newcommand{\leqb}{\stackrel{(b)}{\le}}
\newcommand{\leqe}{\stackrel{(e)}{\le}}
\newcommand{\leqf}{\stackrel{(f)}{\le}}
\newcommand{\leqg}{\stackrel{(g)}{\le}}
\newcommand{\leqh}{\stackrel{(h)}{\le}}
\newcommand{\leqi}{\stackrel{(i)}{\le}}
\newcommand{\leqj}{\stackrel{(j)}{\le}}

\newcommand{\w}{{W^{LDA}}}
\newcommand{\halpha}{\hat{\alpha}}
\newcommand{\hsigma}{\hat{\sigma}}
\newcommand{\slmax}{\sqrt{\lambda_{max}}}
\newcommand{\slmin}{\sqrt{\lambda_{min}}}
\newcommand{\lmax}{\lambda_{max}}
\newcommand{\lmin}{\lambda_{min}}

\newcommand{\da} {\frac{\alpha}{\sigma}}
\newcommand{\chka} {\frac{\check{\alpha}}{\check{\sigma}}}
\newcommand{\sumo}{\sum _{\underline{\omega} \in \Omega}}
\newcommand{\distance}{d\{(\hatz _x, \hatz _y),(\tilz _x, \tilz _y)\}}
\newcommand{\col}{{\rm col}}
\newcommand{\rcs}{\sigma_0}
\newcommand{\CalR}{{\cal R}}
\newcommand{\df}{{\delta p}}
\newcommand{\dq}{{\delta q}}
\newcommand{\dZ}{{\delta Z}}
\newcommand{\pprime}{{\prime\prime}}

\newcommand{\vn}{N}
\newcommand{\diff}{\text{diff}}

\newcommand{\bv}{\begin{vugraph}}
\newcommand{\ev}{\end{vugraph}}
\newcommand{\bi}{\begin{itemize}}
\newcommand{\ei}{\end{itemize}}
\newcommand{\ben}{\begin{enumerate}}
\newcommand{\een}{\end{enumerate}}
\newcommand{\be}{\protect\[}
\newcommand{\ee}{\protect\]}
\newcommand{\bean}{\begin{eqnarray*} }
\newcommand{\eean}{\end{eqnarray*} }
\newcommand{\bea}{\begin{eqnarray} }
\newcommand{\eea}{\end{eqnarray} }
\newcommand{\nn}{\nonumber}
\newcommand{\ba}{\begin{array} }
\newcommand{\ea}{\end{array} }
\newcommand{\ep}{\mbox{\boldmath $\epsilon$}}
\newcommand{\epp}{\mbox{\boldmath $\epsilon '$}}
\newcommand{\Lep}{\mbox{\LARGE $\epsilon_2$}}
\newcommand{\und}{\underline}
\newcommand{\pdif}[2]{\frac{\partial #1}{\partial #2}}
\newcommand{\odif}[2]{\frac{d #1}{d #2}}
\newcommand{\dt}[1]{\pdif{#1}{t}}
\newcommand{\urho}{\underline{\rho}}

\newcommand{\spc}{{\cal S}}
\newcommand{\tspc}{{\cal TS}}

\newcommand{\uv}{\underline{v}}
\newcommand{\us}{\underline{s}}
\newcommand{\uc}{\underline{c}}
\newcommand{\utheta}{\underline{\theta}^*}
\newcommand{\ualpha}{\underline{\alpha^*}}

\newcommand{\uxy}{\underline{x}^*}
\newcommand{\uxyj}{[x^{*}_j,y^{*}_j]}
\newcommand{\arcl}[1]{arclen(#1)}
\newcommand{\one}{{\mathbf{1}}}

\newcommand{\uxyjt}{\uxy_{j,t}}
\newcommand{\E}{\mathbb{E}}

\newcommand{\rhomat}{\left[\begin{array}{c}
                        \rho_3 \ \rho_4 \\
                        \rho_5 \ \rho_6
                        \end{array}
                   \right]}
\newcommand{\deltat}{\tau} %{\Delta t}
\newcommand{\deltatt}{\Delta t_1}
\newcommand{\ceil}[1]{\ulcorner #1 \urcorner}

\newcommand{\xxi}{x^{(i)}}
\newcommand{\txi}{\tilde{x}^{(i)}}
\newcommand{\txj}{\tilde{x}^{(j)}}

\newcommand{\mi}[1]{{#1}^{(m,i)}}

\newcommand{\tx}{\tilde{x}}
\newcommand{\tN}{\tilde{N}}

\title{Modified-CS: Modifying Compressive Sensing for Problems with Partially Known Support
\thanks{N. Vaswani and W. Lu are with the ECE dept. at Iowa State University (email: \{namrata,luwei\}@iastate.edu).
A part of this work appeared in \cite{isit09,icip09}.
%presented at the IEEE Intl. Symp. Info. Theory (ISIT), 2009 \cite{isit09} and a part will be presented at IEEE Intl. Conf. Image Proc. (ICIP), 2009 \cite{icip09}.  conference papers
This research was supported by NSF grants ECCS-0725849 and CCF-0917015. Copyright (c) 2010 IEEE. Personal use of this material is permitted. However, permission to use this material for any other purposes must be obtained from the IEEE by sending a request to pubs-permissions@ieee.org.
}
}

\author{Namrata Vaswani and Wei Lu}

%\authorblockN{Namrata Vaswani and Wei Lu} \\ \authorblockA{ECE Dept., Iowa State University, Ames, IA 50011, USA, Email: \{namrata,luwei\}@iastate.edu}}
%Dept. of Electrical and Computer Engineering\\
%Iowa State University, Ames, IA 50011, USA\\
%Email: namrata@iastate.edu}

\maketitle
%(this uses the empirically observed fact that in a time sequence of signals, the support changes slowly).n $\ell_1$
%whose support contains the smallest number of new additions to $T$, although it may or may not contain all elements of $T$.
\begin{abstract}
We study the problem of reconstructing a sparse signal from a limited number of its linear projections when a part of its support is known, although the known part may contain some errors. The ``known" part of the support, denoted $T$, may be available from prior knowledge. Alternatively, in a problem of recursively reconstructing time sequences of sparse spatial signals, one may use the support estimate from the previous time instant as the ``known" part. The idea of our proposed solution (modified-CS) is to solve a convex relaxation of the following problem:  find the signal that satisfies the data constraint and is sparsest outside of $T$.
We obtain sufficient conditions for exact reconstruction using modified-CS. These are much weaker than those needed for compressive sensing (CS) when the sizes of the unknown part of the support and of errors in the known part are small compared to the support size. An important extension called Regularized Modified-CS (RegModCS) is developed which also uses prior signal estimate knowledge.
Simulation comparisons for both sparse and compressible signals are shown.
%reconstructing simulated sparse signals from random Gaussian measurements and for reconstructing sparsified and actual (approximately sparse) images and image sequences from simulated random Gaussian and partial Fourier measurements.%We show comparisons using simulated sparse signals and simulated random Gaussian and Fourier measurements (simulated MRI) of sparsified and true image sequences. $T$ is large compared to that of  and signal sequences
%\vspace{-0.15in}
\end{abstract}

%{\em Index terms: } modified-cs, compressive sensing, sparse reconstruction, prior knowledge

\vspace{-0.1in}

\setlength{\arraycolsep}{0.05cm}
\newcommand{\signumfn}{\text{sgn}}

\section{Introduction}%estimation using sensor networks
In this work, we study the sparse reconstruction problem from noiseless measurements when a part of the support is known, although the known part may contain some errors. The ``known" part of the support may be available from prior knowledge. For example, consider MR image reconstruction using the 2D discrete wavelet transform (DWT) as the sparsifying basis. If it is known that an image has no (or very little) black background, all (or most) approximation coefficients will be nonzero. In this case, the ``known support" is the set of indices of the approximation coefficients. Alternatively, in a problem of recursively reconstructing time sequences of sparse spatial signals, one may use the support estimate from the previous time instant as the ``known support". This latter problem occurs in various practical applications such as real-time dynamic MRI reconstruction, real-time single-pixel camera video imaging or video compression/decompression. There are also numerous other potential applications where sparse reconstruction for time sequences of signals/images may be needed, e.g. see \cite{igorcarron,rice}.%, and real-time time-varying spatial field sensing \cite{nowaknoise}.% (causal and recursive) igorcarron, elements of the

%\begin{figure}[t!] jung_etal,  \cite{kfcsmri}
%\centerline{
%\subfigure[\small{Plot of $|N_t \setminus N_{t-1}|$ versus $t$}]{
%\includegraphics [height=3.5cm,width=5cm]{slowlychangedelta999.eps}%{slowlychangedelta.eps}
%}
%\hspace{-0.3in}
%\subfigure[\small{Plot of $|N_{t-1} \setminus N_t|$ versus $t$}]{
%\includegraphics [height=3.5cm,width=5cm]{slowlychangedeltae999.eps}%{slowlychangedeltae.eps}
%}
%}
%\caption{\small{Slow changes in the sparsity pattern of the 2D discrete wavelet transform coefficients of a $128 \times 128$ cardiac image sequence and of a $256 \times 256$ vocal tract (larynx) image sequence shown in Fig. \ref{larynxnoiseless}. $N_t$ denotes the $99.9\%$-energy support set (defined in Sec. \ref{exts}). $|N_t|$ varied between 4400-4600 for the larynx sequence and between 1400-1500 for the cardiac sequence. (a) Plot of $|N_t \setminus N_{t-1}|$ (no. of new additions) versus time, $t$. (b) Plot of $|N_{t-1} \setminus N_t|$ (no. of deletions) against $t$. The maximum size of the change (additions or deletions) is less than $2\%$ of the minimum sparsity size in both cases.
%%The larynx sequence shown in Fig. \ref{larynxnoiseless}.  (defined in the first para of Sec. \ref{exts})
%}}
%% As can be seen, the maximum size of the additions is 20 for larynx and 22 for cardiac, i.e. less than $2\%$ of the minimum sparsity size in both cases.  number is 38 for larynx and 19 for cardiac, i.e. less than $2\%$ of the minimum sparsity size in both cases.
%%\vspace{-0.15in} (less than $7\%$)  (less than $9\%$)  (2-level Daubechies-4)
%\label{slowchange}
%\end{figure}

\begin{figure}
\centerline{
\subfigure[Top: larynx image sequence, Bottom: cardiac sequence]{%, Bottom: brain sequence
\label{examples}
\begin{tabular}{c}
\epsfig{file = 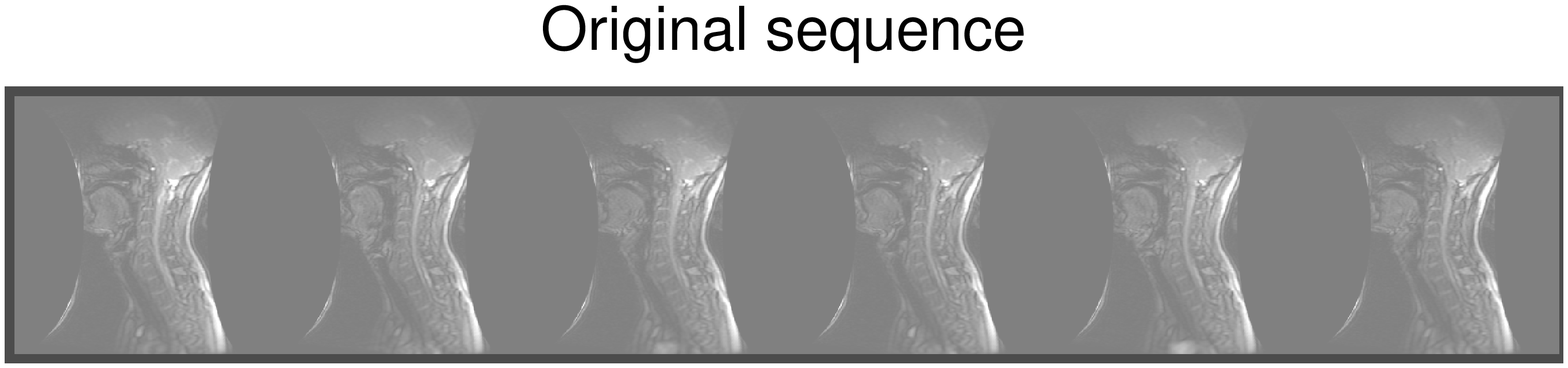,height=2cm} \vspace{-0.2in}  \\ %U:/ModCS/ModCSicip/
\epsfig{file = 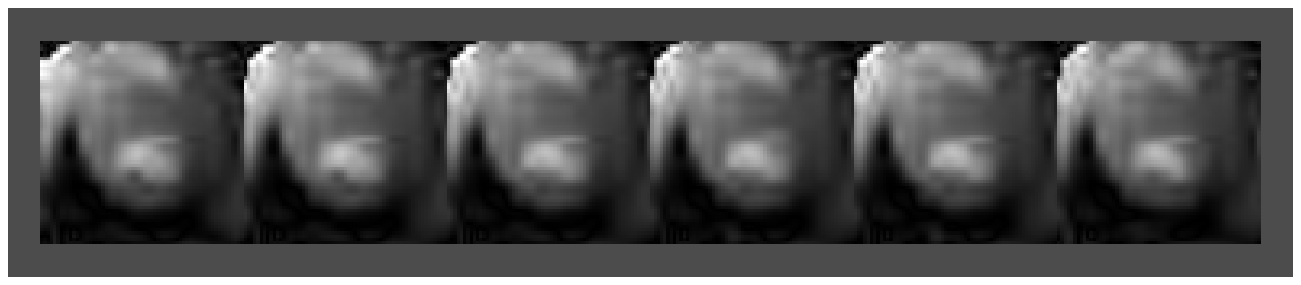, height=1.5cm} %\epsfig{file = U:/KFCS/arxiv_lscs/vowelorg.eps,height=2cm} \vspace{-0.2in}  \\ %U:/ModCS/ModCSicip/
\end{tabular}
}
}
\centerline{
\subfigure[Slow support change plots. Left: additions, Right: removals]{
\label{supp90_99}% taken from I (projection)
\begin{tabular}{cc}
\epsfig{file = 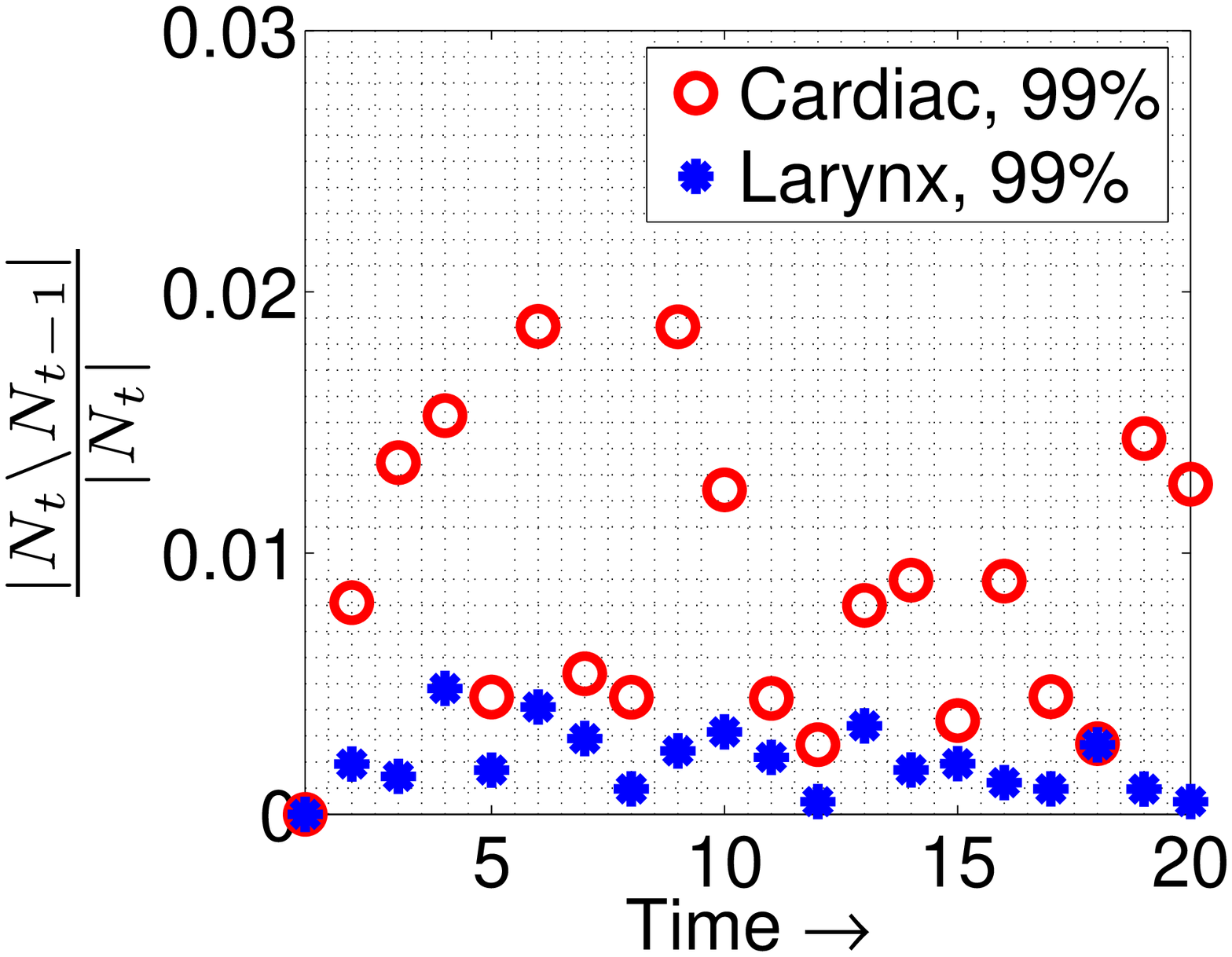,height=2.5cm, width = 4cm} & %C:/Code/KFCS/weilu_fig_jan09/
\epsfig{file = 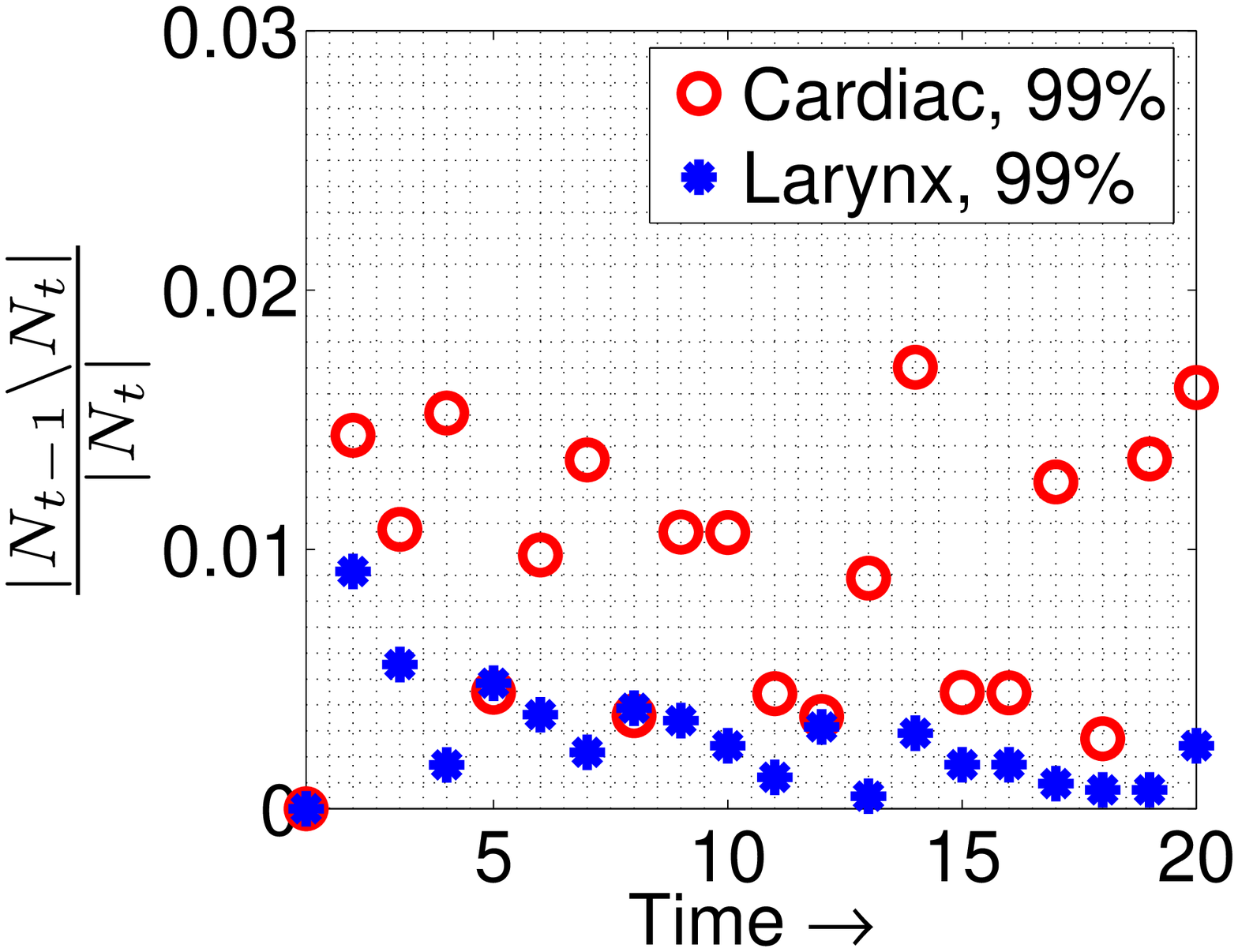,height=2.5cm, width = 4cm}
\end{tabular}
}
}
\vspace{-0.1in}
\caption{\small{ %{\bf Slow support change examples.}
In Fig. \ref{examples}, we show two medical image sequences. In Fig. \ref{supp90_99}, $N_t$ refers to the 99\% energy support of the two-level Daubechies-4 2D discrete wavelet transform (DWT) of these sequences. $|N_t|$ varied between 4121-4183 ($\approx 0.07m$) for larynx and between 1108-1127 ($\approx 0.06m$) for cardiac. We plot the number of additions (left) and the number of removals (right) as a fraction of $|N_t|$. {\em Notice that all changes are less than 2\% of the support size.}
}}
\vspace{-0.2in}
\label{slowchange}
\end{figure}

Sparse reconstruction has been well studied for a while, e.g. see \cite{bpdn,sbl}. Recent work on Compressed Sensing (CS) gives conditions for its exact reconstruction \cite{candes,donoho,decodinglp} and bounds the error when this is not possible \cite{tropp,dantzig}.%the theoretical guarantees on the  without any knowledge of the support

%studied sparse reconstruction from noisy measurements with partly known support. kfcspap0, sparse reconstruction with
%addressed the noisy measurements' version of the problem of
Our recent work on Least Squares CS-residual (LS-CS) \cite{kfcsicip,kfcspap} can be interpreted as a solution to the problem of sparse reconstruction with partly known support. LS-CS replaces CS on the observation by CS on the LS observation residual, computed using the ``known" part of the support. Since the observation residual measures the signal residual which has much fewer {\em large nonzero} components, LS-CS greatly improves reconstruction error when fewer measurements are available. But the exact sparsity size (total number of nonzero components) of the signal residual is equal to or larger than that of the signal. Since the number of measurements required for exact reconstruction is governed by the exact sparsity size, LS-CS is not able to achieve exact reconstruction using fewer noiseless measurements than those needed by CS.%, e.g. see Fig \ref{timeseqsparse}.% (than what is needed for CS)  compared to CS  (previous support estimate)  which is approximately much more sparse

%Also see Fig. \ref{modcscompress}.%  than those needed by CS
%Over time, all the nonzero elements of the signal will change and some new elements also become nonzero. Thus the exact sparsity size of the signal difference will be equal to or larger than that of the signal itself. As a result CS-diff will also not achieve exact reconstruction using fewer noiseless measurements than those needed by CS.
%When a signal changes over time, its coefficients also change. Thus the exact sparsity size of the signal difference will be equal to that of the signal itself. As a result this approach will also not achieve exact reconstruction using fewer noiseless measurements. We demonstrate this in Fig ??. %For compressible signal sequences, if the changes in the signal are very small from $t-1$ to $t$, nothing conclusive can be said, while if the changes are significant, modified-CS will be much better.  than those needed for CS
 %The ``known" support may be available from prior knowledge or for time sequences we can use the estimated support from the previous time instant as the known part. The latter works especially well for problems such as dynamic MRI reconstruction where the sparsity pattern changes very slowly over time [see  Fig. \ref{slowchange}].
%the sizes of the unknown support and of the errors in the known support are small compared to the support size

{\em Exact reconstruction using fewer noiseless measurements than those needed for CS} is the focus of the current work. Denote the ``known" part of the support by $T$. Our proposed solution (modified-CS) solves an $\ell_1$ relaxation of the following problem: find the signal that satisfies the data constraint and is sparsest outside of $T$. We derive sufficient conditions for exact reconstruction using modified-CS. When $T$ is a fairly accurate estimate of the true support, these are much weaker than the sufficient conditions for CS. For a recursive time sequence reconstruction problem, this holds if the reconstruction at $t=0$ is exact and the support changes slowly over time. The former can be ensured by using more measurements at $t=0$, while the latter is often true in practice, e.g. see Fig. \ref{slowchange}.%

%$T \approx N_{t-1}$, where $N_{t-1}$ is the true support at $t-1$.%(previous reconstruction accurate enough). , this is usually true for medical image sequences  (done using CS)  or accurate enough %whose support contains the smallest number of new additions to $T$, although it may or may not contain all elements of $T$.
%
% As shown in Fig. \ref{slowchange}, for medical image sequences, the support changes very slowly over time. If previous support estimation is accurate, i.e.  $T \approx N_{t-1}$, then this means that $T$ is a very accurate estimate of the current true support, $N_t$, and so modified-CS will need much fewer measurements for exact reconstruction.

We also develop an important extension called Regularized Modified-CS which also uses prior signal estimate knowledge. It improves the error when exact reconstruction is not possible.% Detailed simulation comparisons are shown.% %s showing greatly improved performance are also shown.%unknown part of the support (  (or in other words the support set difference from $T$ is smallest) %%The idea of our proposed solution (modified-CS) is to modify CS for problems where part of the support is known, although the known part may have some error.  for both simulated signals and real images as in Fig. \ref{slowchange},  for both sparse and compressible signals and signal sequences
%This will work if the {\em sparsity pattern (support set of the signal) changes slowly over time.}  We have empirically observed that this is indeed true for MRI sequences, see Fig. \ref{slowchange}.

%A part of this work appeared in \cite{isit09}.
A shorter version of this work first appeared in ISIT'09 \cite{isit09}. In parallel and independent work in \cite{hassibi}, Khajehnejad et al have also studied a similar problem to ours but they assume a probabilistic prior on the support. Other related work includes \cite{camsap07}. %Recently (in Feb 2010), we learnt about the older work of von Borries~et~al~\cite{camsap07} which also suggests an approach similar to modified-CS.
Very recent work on causal reconstruction of time sequences includes \cite{reclasso} (focusses on the time-invariant support case) and \cite{dynamicl1} (use past estimates to only speed up the current optimization but not to improve reconstruction error). {\em Except \cite{hassibi}, none of these prove exact reconstruction using fewer measurements and except \cite{camsap07,hassibi}, none of these even demonstrate it.}% noise-free \cite{jung_etal} (causal but batch solution developed for dynamic MRI) rozell: is more about sparse representation, not reconstruction

Other recent work, e.g. \cite{reddy}, applies CS on observation differences to reconstruct the difference signal. While their goal is to only estimate the difference signal, the approach could be easily modified to also reconstruct the actual signal sequence (we refer to this as {\em CS-diff}). But, since all nonzero coefficients of a sparse signal in any sparsity basis will typically change over time, though gradually, and some new elements will become nonzero, thus the exact sparsity size of the signal difference will also be equal to/larger than that of the signal itself. As a result CS-diff will also not achieve exact reconstruction using fewer measurements, e.g. see Fig.\ref{timeseqsparse}.%

In this work, {\em whenever we use the term CS, we are actually referring to basis pursuit (BP)} \cite{bpdn}. As pointed out by an anonymous reviewer, modified-CS is a misnomer and a more appropriate name for our approach should be {\em modified-BP}. %But since we have already used the term in a published conference paper and it has been used by other authors now, we will stick with it.

As pointed out by an anonymous reviewer, modified-CS can be used in conjunction with multiscale CS for video compression \cite{multiscaleCS} to improve their compression ratios. %We can use the reconstructed nonzero wavelet coefficients at scale $j-1$, and the correlation across scales, to define a ``known" part of the support for the problem at scale $j$. Using this, we can replace CS at scale $j$ by modified-CS which should require fewer measurements.% to achieve the same fidelity.This relies on the fact that wavelet coefficients have strong correlations across scales \cite{wavelethmm}. For reconstruction at scale $j$,  (step 3 of algorithm 1 of \cite{multiscaleCS})  \cite{wavelethmm} multiscale CS for video compression  structure

The paper is organized as follows. We give the notation and problem definition below. Modified-CS is developed in Sec. \ref{algo}. We obtain sufficient conditions for exact reconstruction using it in Sec. \ref{exactrecon}. In Sec. \ref{compare}, we compare these with the corresponding conditions for CS and we also do a Monte Carlo comparison of modified-CS and CS. We discuss Dynamic Modified-CS and Regularized Modified CS in Sec. \ref{exts}. Comparisons for actual images and image sequences are given in Sec. \ref{expts} and conclusions and future work in Sec.~\ref{conclusions}.%modified-CS for time sequences (and future work are given

\subsection{Notation}
 We use $'$ for transpose. The notation $\|c\|_k$ denotes the $\ell_k$ norm of the vector $c$. The $\ell_0$ pseudo-norm,  $\|c\|_0$, counts the number of nonzero elements in $c$.  For a matrix, $M$, $\|M\|$ denotes its induced $\ell_2$ norm, i.e. $\|M\| := \max_{c: \|c\|_2=1} \|Mc\|_2$.%spectral norm (

We use the notation $A_T$ to denote the sub-matrix containing the columns of $A$ with indices belonging to $T$. For a vector, the notation $(\beta)_T$ (or $\beta_T$) refers to a sub-vector that contains the elements with indices in $T$. The notation, $[1,\sno]:=[1,2,\dots \sno]$.
 We use $T^c$ to denote the complement of the set $T$ w.r.t. $[1,\sno]$, i.e. $T^c:=[1,\sno] \setminus T$. The set operations, $\cup, \cap$ stand for set union and intersection respectively. Also $T_1 \setminus T_2 : = T_1 \cap T_2^c$ denotes set difference. For a  set $T$, $|T|$ denotes its size (cardinality). But for a scalar, $b$, $|b|$ denotes the magnitude of $b$. %The meaning is clear from context.%have their usual meanings. , \setminus

The $S$-restricted isometry constant \cite{decodinglp}, $\delta_S$, for a matrix, $A$, is defined as the smallest real number satisfying
%\bea
%(1- \delta_S) \|v\|_2^2 \le \|A v\|_2^2 \le (1 + \delta_S) \|v\|_2^2, \ \text{whenever} \ \|v\|_0 \le S \nn
%\eea
%where $\|v\|_k$ denotes the $\ell_k$ norm. We say that $A$ is $S$-approximately-orthonormal if $\delta_{S} < 1$ \cite{decodinglp}. In other words, it is is the smallest real number such that
\bea
(1- \delta_S) \|c\|_2^2 \le \|A_T c\|_2^2 \le (1 + \delta_S) \|c\|_2^2
\label{def_delta}
\eea
for all subsets $T \subset [1,\sno]$ of cardinality $|T| \le S$ and all real vectors $c$ of length $|T|$.
The restricted orthogonality constant \cite{decodinglp}, $\theta_{S_1,S_2}$, is defined as the smallest real number satisfying
\bea
| {c_1}'{A_{T_1}}'A_{T_2} c_2 | \le \theta_{S_1,S_2} \|c_1\|_2 \|c_2\|_2
\label{def_theta}
\eea
for all disjoint sets $T_1, T_2 \subset [1,\sno]$ with $|T_1| \le S_1$, $|T_2| \le S_2$ and  $S_1+S_2 \le \sno$, and for all vectors $c_1$, $c_2$ of length $|T_1|$, $|T_2|$ respectively.
By setting $c_1 \equiv {A_{T_1}}'A_{T_2} c_2$ in (\ref{def_theta}),% it is easy to see that
\bea
\|{A_{T_1}}'A_{T_2}\| \le \theta_{S_1,S_2}
\eea

The notation $X \sim \n(\mu,\Sigma)$ means that $X$ is Gaussian distributed with mean $\mu$ and covariance $\Sigma$ while $\n(x;\mu,\Sigma)$ denotes the value of the Gaussian PDF computed at point $x$.%

\subsection{Problem Definition}
\label{probdef}
We measure an $\mno$-length vector $y$ where
\bea
y := A x
\label{obsmod}
\eea
We need to estimate $x$ which is a sparse $\sno$-length vector with $\sno > \mno$. The support of $x$, denoted $N$, can be split as $N =T \cup \Delta \setminus \Delta_e$ where $T$ is the ``known" part of the support, $\Delta_e:= T \setminus N$ is the error in the the known part and $\Delta:=N \setminus T$ is the unknown part. Thus, $\Delta_e \subseteq T$, $\Delta$, $T$ are disjoint and $|N| = |T| + |\Delta|-|\Delta_e|$.

{\em We use $\sn := |N|$ to denote the size of the (s)upport, $\st:=|T|$ to denote the size of the (k)nown part of the support, $\sde=|\Delta_e|$ to denote the size of the (e)rror in the known part and $\sd=|\Delta|$ to denote the size of the (u)nknown part of the support.} %Notice that $\sn = \st + \sd - \sde$.

We assume that $A$ satisfies the $S$-restricted isometry property (RIP) \cite{decodinglp} for  $S=(\sn + \sde + \sd) = (\st + 2\sd)$. $S$-RIP means that $\delta_S < 1$ where $\delta_S$ is the RIP constant for $A$ defined in (\ref{def_delta}).%%are assumed to be ``approximately orthonormal", i.e.
%Sub-matrices of $A$ containing $S=(\sn + \sde + \sd) = (\st + 2\sd)$ or less columns have nonzero singular values between $\sqrt{1\pm \delta_S}$ with $\delta_S < 1$. In other words $A$ satisfies the $S$-restricted isometry property (RIP) \cite{decodinglp}.%, i.e. $\delta_S < 1$ where $\delta_S$ is the RIP constant for $A$ defined in (\ref{def_delta}).%As we show later (in Corollary \ref{corol1}), this is nearly sufficient for exact reconstruction using modified-CS.  in the notation subsection . $S$-RIP means that the measurement matrix,

%for a natural image which is wavelet sparse,
In a static problem, $T$ is available from prior knowledge. For example, in the MRI problem described in the introduction, let $N$ be the (unknown) set of all DWT coefficients with magnitude above a certain zeroing threshold. Assume that the smaller coefficients are set to zero. Prior knowledge tells us that most image intensities are nonzero and so the approximation coefficients are mostly nonzero. Thus we can let $T$  be the (known) set of indices of all the approximation coefficients. The (unknown) set of indices of the approximation coefficients which are zero form $\Delta_e$. The (unknown) set of indices of the nonzero detail coefficients form $\Delta$.% (these often correspond to regions with a black background)  := N \setminus T  := T \setminus N
% is the set of the nonzero coefficients from the higher subbands. the (unknown) error in the known part,
%But typically there is a small black background in the image, so that only most (not all) lowest subband wavelet coefficients will be nonzero.

For the time series problem, $y \equiv y_t$ and $x \equiv x_t$ with support, $N_t = T \cup \Delta \setminus \Delta_e$, and $T = \hat{N}_{t-1}$ is the support estimate from the previous time instant. %Also, $\Delta:= N_t \setminus T$ and $\Delta_e := T \setminus N_t$. , $t-1$
If exact reconstruction occurs at $t-1$, $T = N_{t-1}$. In this case, $\Delta_e = N_{t-1} \setminus N_t$ is the set of indices of elements that were nonzero at $t-1$, but are now zero (deletions) while $\Delta = N_t \setminus N_{t-1}$ is the newly added coefficients at $t$ (additions). %\footnote{When exact reconstruction does not occur, $\Delta_e$ will include both deletions and the extras from $t-1$, $\Nhat_{t-1} \setminus N_{t-1}$. Similarly, $\Delta$ will include both additions and the misses from $t-1$, $N_{t-1} \setminus \Nhat_{t-1}$.}.
Slow sparsity pattern change over time, e.g. see Fig. \ref{slowchange}, then implies that {\em $\sd \equiv |\Delta|$ and $\sde \equiv |\Delta_e|$ are much smaller than $\sn \equiv |N|$.}% This follows from the empirical observation that sparsity patterns change very slowly over time, e.g. see Fig. \ref{slowchange}.% and the caption of Fig. \ref{timeseqsparse}).

When exact reconstruction does not occur, $\Delta_e$ includes both the current deletions and the extras from $t-1$, $\Nhat_{t-1} \setminus N_{t-1}$. Similarly, $\Delta$ includes both the current additions and the misses from $t-1$, $N_{t-1} \setminus \Nhat_{t-1}$. In this case, slow support change, along with $\Nhat_{t-1} \approx N_{t-1}$, still implies that $\sd \ll \sn$ and $\sde \ll \sn$.%$ being an accurate estimate of $

\section{Modified Compressive Sensing (modified-CS)}
\label{algo}
%Assume for a moment that $\Delta_e$ is empty, i.e. $N = T \cup \Delta$ (there is no error in the known part of the support).
Our goal is to find a signal that satisfies the data constraint given in (\ref{obsmod}) and whose support contains the smallest number of new additions to $T$, although it may or may not contain all elements of $T$.
In other words, we would like to solve %find a $\hat{x}$ which solves
\bea
\min_\beta  \|(\beta)_{T^c}\|_0 \ \text{subject to} \ y = A \beta
\label{l0seqcs}
\eea
If $\Delta_e$ is empty, i.e. if $N = T \cup \Delta$, then the solution of (\ref{l0seqcs}) is also the sparsest solution whose support contains $T$.

As is well known, minimizing the $\ell_0$ norm is a combinatorial optimization problem \cite{natarajan}. We propose to use the same trick that resulted in CS \cite{bpdn,candes,donoho,tropp}. We replace the $\ell_0$ norm by the $\ell_1$ norm, which is the closest norm to $\ell_0$ that makes the optimization problem convex, i.e. we solve %Thus at each time $t$, we solve
\bea
\min_\beta  \|(\beta)_{T^c}\|_1 \ \text{subject to} \ y = A \beta
\label{l1seqcs}
\eea
Denote its output by $\xhat$. If needed, the support can be estimated as% of $\xhat$
\bea
\Nhat := \{i \in [1,\sno]: (\xhat)_i^2 > \alpha \}
\label{supp_est}
\eea
where $\alpha \ge 0$ is a zeroing threshold. If exact reconstruction occurs, $\alpha$ can be zero. We discuss threshold setting for cases where exact reconstruction does not occur in Sec. \ref{dynmodcs}.

\newcommand{\tw}{\tilde{w}}

\section{Exact Reconstruction Result}% using Fewer Observations}
\label{exactrecon}
We first analyze the $\ell_0$ version of modified-CS in Sec. \ref{l0exactrec}. We then give the exact reconstruction result for the actual $\ell_1$ problem in Sec. \ref{l1exactrec}.  In Sec. \ref{lemmas}, we give the two key lemmas that lead to its proof and we explain how they lead to the proof. The complete proof is given in the Appendix. The proof of the lemmas is given in Sec. \ref{lemmaproof}.% \ref{thm1proof}and we give the key idea of how they lead to the proof
\\

Recall that $\st=|T|$, $\sd=|\Delta|$, $\sde=|\Delta_e|$ and $\sn=|N|$.

\subsection{Exact Reconstruction Result: $\ell_0$ version of modified-CS}
\label{l0exactrec}
Consider the $\ell_0$ problem, (\ref{l0seqcs}). Using a rank argument similar to \cite[Lemma 1.2]{decodinglp} we can show the following. The proof is given in the Appendix.

\begin{proposition}
%Consider the problem of reconstructing a sparse vector, $x$, with support $T \cup \Delta$, from  $y:=Ax$.  $T$ is the known part of the support and $\Delta$ is the unknown part.%Consider reconstructing a sparse vector $x$ with support $T \cup \Delta$ from  $y:=Ax$ by solving (\ref{l0seqcs}).
Given a sparse vector, $x$, with support, $N = T \cup \Delta \setminus \Delta_e$, where $\Delta$ and $T$ are disjoint and $\Delta_e \subseteq T$.
Consider  reconstructing it from $y:=Ax$ by solving (\ref{l0seqcs}). $x$ is the unique minimizer of (\ref{l0seqcs}) if $\delta_{\st + 2\sd} < 1$ ($A$ satisfies the $(\st+2\sd)$-RIP).%Recall that $\st=|T|$ and $\sd=|\Delta|$.
\label{l0exact}
\end{proposition}

Using $\st = \sn+\sde-\sd$, %where $\sn=|N|$ and $\sde=|\Delta_e|$,  the fact that
this is equivalent to $\delta_{\sn + \sde + \sd} < 1$. Compare this with \cite[Lemma 1.2]{decodinglp} for the $\ell_0$ version of CS. It requires $\delta_{2\sn} < 1$ which is much stronger when $\sd \ll \sn$ and $\sde \ll \sn$, as is true for time series problems.%stronger, if $\sde + \sd < \sn$. It is

\subsection{Exact Reconstruction Result: modified-CS}
\label{l1exactrec}
Of course we do not solve (\ref{l0seqcs}) but its $\ell_1$ relaxation, (\ref{l1seqcs}). Just like in CS, the sufficient conditions for this to give exact reconstruction will be slightly stronger. In the next few subsections, we prove the following result. %and compare it with the corresponding CS result \cite[Theorem 1.3]{decodinglp}.

\begin{theorem}[Exact Reconstruction]
Given a sparse vector, $x$, whose support, $N = T \cup \Delta \setminus \Delta_e$, where $\Delta$ and $T$ are disjoint and $\Delta_e \subseteq T$. %In particular, $x$ is nonzero on the set $\Delta$ and $x$ is zero outside $T \cup \Delta$.
Consider reconstructing it from $y:=Ax$ by solving (\ref{l1seqcs}). $x$ is the unique minimizer of  (\ref{l1seqcs}) if
\ben
\item  $\delta_{\st+\sd} < 1$ and $\delta_{2\sd} + \delta_{\st} + \theta_{\st,2\sd}^2 < 1$ and %$\delta_{\st + 2\sd} < 1$ and
\label{cond1}
\\

\item $a_{\st}(2\sd,\sd) + a_{\st}(\sd,\sd) < 1$ where
\label{cond2}
\bea
a_{\st}(S,\Sp) \sdefn \frac{\theta_{\Sp,S} + \frac{\theta_{\Sp,\st} \ \theta_{S,\st}}{1 - \delta_{\st}}}{ 1-\delta_S - \frac{\theta_{S,\st}^2}{1 - \delta_{\st}} }
\label{def_a0}
\eea
\een
The above conditions can be rewritten using $\st = \sn + \sde - \sd$.%Recall that $\st=|T|$ and $\sd=|\Delta|$. Since $|T|=|N|+|\Delta_e|-|\Delta|$, thus
\\
\label{thm1}
\end{theorem}
%with support

%&& \le \frac{\theta_{\sd,2\sd} + \frac{\theta_{\sd,\st}  \theta_{2\sd,\st}}{1 - \delta_{\st}} + \theta_{\sd,\sd} + \frac{\theta_{\sd,\st}^2}{1 - \delta_{\st}}  }{ 1-\delta_{2\sd} - \frac{\theta_{2\sd,\st}^2}{1 - \delta_{\st}} } \nn

To understand the second condition better and relate it to the corresponding CS result, let us simplify it.
$
 a_{\st}(2\sd,\sd) + a_{\st}(\sd,\sd)
 \le \frac{ \theta_{\sd,2\sd} + \theta_{\sd,\sd} +  \frac{\theta_{2\sd,\st}^2 + \theta_{\sd,\st}^2}{1 - \delta_{\st}} }{ 1-\delta_{2\sd} - \frac{\theta_{2\sd,\st}^2}{1 - \delta_{\st}} }.
$
Simplifying further, a sufficient condition for $a_{\st}(2\sd,\sd) +  a_{\st}(\sd,\sd) < 1$ is
$
\theta_{\sd,2\sd} + \theta_{\sd,\sd} +  \frac{2\theta_{2\sd,\st}^2 + \theta_{\sd,\st}^2}{1 - \delta_{\st}} + \delta_{2\sd} < 1 $. %+ \frac{\theta_{2\sd,\st}^2}{1 - \delta_{\st}}
% or equivalently, $\delta_{2\sd} + \theta_{\sd,2\sd} + \theta_{\sd,\sd} +  \frac{\theta_{\sd,\st}^2 + \theta_{2\sd,\st}}{1 - \delta_{\st}} +   \frac{\theta_{2\sd,\st}^2}{1 - \delta_{\st}} < 1 $
Further, a sufficient condition for this is $\theta_{\sd,\sd}+ \delta_{2\sd} + \theta_{\sd,2\sd}  + \delta_{\st}  + \theta_{\sd,\st}^2  + 2\theta_{2\sd,\st}^2  < 1$.

To get a condition only in terms of $\delta_S$'s, use the fact that $ \theta_{S,\Sp} \le \delta_{S+\Sp}$ \cite{decodinglp}. A sufficient condition is $2\delta_{2\sd} + \delta_{3\sd} + \delta_{\st} + \delta_{\st + \sd}^2 +  2\delta_{\st + 2\sd}^2 < 1$. Further, notice that if $\sd \le \st$ and if $\delta_{\st + 2\sd} < 1/5$, then $2\delta_{2\sd} + \delta_{3\sd} + \delta_{\st} + \delta_{\st + \sd}^2 +  2\delta_{\st + 2\sd}^2 < 4 \delta_{\st + 2\sd} + \delta_{\st + 2\sd}(3\delta_{\st + 2\sd}) \le (4+3/5)\delta_{\st + 2\sd} <  23/25< 1$.
\\
\begin{corollary}[Exact Reconstruction]
Given a sparse vector, $x$, whose support, $N = T \cup \Delta \setminus \Delta_e$, where $\Delta$ and $T$ are disjoint and $\Delta_e \subseteq T$. Consider reconstructing it from $y:=Ax$ by solving (\ref{l1seqcs}).

\bi

\item $x$ is the unique minimizer of (\ref{l1seqcs}) if $\delta_{\st+\sd} < 1$ and
\bea
(\delta_{2\sd} + \theta_{\sd,\sd} + \theta_{\sd,2\sd})  + (\delta_{\st}  + \theta_{\st,\sd}^2  + 2\theta_{\st,2\sd}^2)  < 1
\label{modcscond1}
\eea

\item This, in turn, holds if
$$2\delta_{2\sd} + \delta_{3\sd} + \delta_{\st} + \delta_{\st + \sd}^2 + 2 \delta_{\st + 2\sd}^2 < 1.$$

\item This, in turn, holds if $\sd \le \st$  and
$$\delta_{\st + 2\sd} < 1/5.$$

%\item By setting $\st = \sn+\sde-\sd$, we can obtain the above conditions in terms of $\sn, \sde, \sd$ only.
\ei
These conditions can be rewritten by substituting $\st = \sn + \sde - \sd$.% if needed.
%Recall that $|T|=|N|+|\Delta_e|-|\Delta|$, i.e. $\st = \sn + \sde - \sd$.%Recall that $\st=|T|$ and $\sd=|\Delta|$. %Recall that $\st=|T|$, $\sd=|\Delta|$, $\sde=|\Delta_e|$ and $\sn=|N|$.
\label{corol1}
\end{corollary}

Compare (\ref{modcscond1}) to the sufficient condition for CS given in \cite{decodinglp}:% given in
\bea
\delta_{2\sn} + \theta_{\sn,\sn} + \theta_{\sn,2\sn} < 1
\label{cscond1}
\eea
%As shown in Fig. \ref{slowchange}, usually $\sd \ll \sn$, $\sde \ll \sn$ and $\sd \approx \sde$ (which means that $\st \approx \sn$). Under this assumption, compare (\ref{modcscond1}) with (\ref{cscond1}). The first bracket of (\ref{modcscond1}) will be small compared to the left hand side (LHS) of (\ref{cscond1}), particularly when $\sn/\mno$ is larger. Also, if $\theta_{\st,2\sd} < 1/2$ (requires $\sn/\mno$ to not be too large), then each term of the second bracket will also be smaller than the LHS of (\ref{cscond1}). In fact, if $\sd$ is small enough, the second and third terms of this bracket will be small compared to the second and third terms of the LHS of (\ref{cscond1}). Thus, for a certain range of values of $\sn/\mno$,  if $\sd, \sde$ are small enough, the LHS of (\ref{modcscond1}) will be small compared to that of (\ref{cscond1}). Since $\delta_S$, $\theta_{S,S'}$ are non-increasing in $\mno$, this means that (\ref{modcscond1}) can hold for smaller values of $\mno$ than (\ref{cscond1}) can, i.e. {\em exact reconstruction with modified-CS can be guaranteed for smaller values of $\mno$ than what is needed for CS.} A detailed comparison is done in Sec. \ref{compare}. %if $\sd, \sde$ are small enough,

%=\sn + \sde - \sd
As shown in Fig. \ref{slowchange}, usually $\sd \ll \sn$, $\sde \ll \sn$ and $\sd \approx \sde$ (which means that $\st \approx \sn$). Consider the case when the number of measurements, $\mno$, is smaller than what is needed for exact reconstruction for a given support size, $\sn$, but is large enough to ensure that $\theta_{\st,2\sd} < 1/2$. Under these assumptions, compare (\ref{modcscond1}) with (\ref{cscond1}). Notice that (a) the first bracket of the left hand side (LHS) of (\ref{modcscond1}) will be small compared to the LHS of (\ref{cscond1}). The same will hold for the second and third terms of its second bracket compared with the second and third terms of (\ref{cscond1}). The first term of its second bracket, $\delta_{\st}$, will be smaller than the first term of (\ref{cscond1}), $\delta_{2\sn}$. Thus, for a certain range of values of $\mno$, the LHS of (\ref{modcscond1}) will be smaller than that of (\ref{cscond1}) and it may happen that (\ref{modcscond1}) holds, but (\ref{cscond1}) does not hold. For example, if $\mno < 2\sn$, (\ref{cscond1}) {\em will not hold}, but if $\sn + \sd + \sde < \mno < 2\sn$, {\em (\ref{modcscond1}) can hold} if $\sd, \sde$ are small enough. A detailed comparison is done in Sec. \ref{compare}.
%

%Also, if $\theta_{\st,2\sd} < 1/2$ (requires $\sn/\mno$ to not be too large), then each term of the second bracket will also be smaller than the LHS of (\ref{cscond1}). In fact, if $\sd$ is small enough, the second and third terms of this bracket will be small compared to the second and third terms of the LHS of (\ref{cscond1}). Thus, for a certain range of values of $\sn/\mno$,  if $\sd, \sde$ are small enough, the LHS of (\ref{modcscond1}) will be small compared to that of (\ref{cscond1}). Since $\delta_S$, $\theta_{S,S'}$ are non-increasing in $\mno$, this means that (\ref{modcscond1}) can hold for smaller values of $\mno$ than (\ref{cscond1}) can, i.e. {\em exact reconstruction with modified-CS can be guaranteed for smaller values of $\mno$ than what is needed for CS.} A detailed comparison is done in Sec. \ref{compare}. %if $\sd, \sde$ are small enough,

\subsection{Proof of Theorem \ref{thm1}: Main Lemmas and Proof Outline}
\label{lemmas}
The idea of the proof is motivated by that of \cite[Theorem 1.3]{decodinglp}.
Suppose that we want to minimize a convex function $J(\beta)$ subject to $A \beta = y$ and that $J$ is differentiable. The Lagrange multiplier optimality condition requires that there exists a Lagrange multiplier, $w$, s.t.
$\nabla J(\beta) - A'w = 0$. Thus for $x$ to be a solution we need $A'w = \nabla J(x)$. In our case, $J(x) = \|x_{T^c}\|_1 = \sum_{j \in T^c} |x_j|$. Thus $(\nabla J(x))_j = 0$ for $j \in T$ and $(\nabla J(x))_j = \signumfn(x_j)$ for $j \in \Delta$. For $j \notin T \cup \Delta$, $x_j = 0$. Since $J$ is not differentiable at 0, we require that $(A'w)_j = {A_j}'w =  w'A_j$ lie in the subgradient set of $J(x_j)$ at 0, which is the set $[-1, 1]$ \cite{boyd}. In summary, we need a $w$ that satisfies
\bea
w'A_j \se 0 \ \text{if}  \ j \in T, \ \ w'A_j = \signumfn(x_j) \  \text{if} \ j \in \Delta, \ \text{and} \nn \\
|w'A_j| \sle 1, \  \text{if} \ j \notin T \cup \Delta
\label{reqw}
\eea
Lemma \ref{wcond} below shows that by using (\ref{reqw}) but with $|w'A_j| \le 1$ replaced by $|w'A_j| < 1$ for all $j \notin T \cup \Delta$, we get a set of sufficient conditions for $x$ to be {\em the unique} solution of (\ref{l1seqcs}).
\begin{lemma}
The sparse signal, $x$, with support as defined in Theorem \ref{thm1}, and with $y:=Ax$, is the unique minimizer of (\ref{l1seqcs}) if $\delta_{\st + \sd} < 1$
%$\delta_{\max(\st,\sd)} < 1$ %  $\delta_{\st + 2\sd} < 1$
and if we can find a vector $w$ satisfying
\ben
\item $w'A_j = 0 \ \text{if}  \ j \in T$
\label{zero}
\item  $w'A_j = \signumfn(x_j) \  \text{if} \ j \in \Delta$
\label{sgn}
\item $|w'A_j| < 1, \  \text{if} \ j \notin T \cup \Delta$
\label{subgrad}
\een
Recall that $\st=|T|$ and $\sd=|\Delta|$.
\label{wcond}
\end{lemma}
The proof is given in the next subsection.

Next we give Lemma \ref{wbnd} which constructs a $\tw$ which satisfies ${A_T}'\tw = 0$ and ${A_{T_d}}'\tw = c$ for any set $T_d$ disjoint with $T$ of size $|T_d| \le S$ and  for any given vector $c$ of size $|T_d|$. It also bounds $|{A_j}'\tw|$ for all $j \notin T \cup T_d \cup E$ where $E$ is called an ``exceptional set". We prove Theorem \ref{thm1} by applying Lemma \ref{wbnd} iteratively to construct a $w$ that satisfies the conditions of Lemma \ref{wcond} under the assumptions of Theorem \ref{thm1}.

\begin{lemma} % $\delta_{\st + S} < 1$ and
Given the known part of the support, $T$, of size $\st$. Let $S$, $\Sp$ be such that $\st+S+\Sp \le \sno$ and $\delta_{S} + \delta_{\st} + \theta_{\st,S}^2 < 1$. Let $c$ be a vector supported on a set $T_d$, that is disjoint with $T$, of size $|T_d| \le S$. Then there exists a vector $\tw$ and an exceptional set, $E$, disjoint with $T \cup T_d$, s.t.
\bea
{A_j}'\tw \se 0, \ \forall \ j \in T  \nn \\
\label{T_Td}
{A_j}'\tw \se c_j, \ \forall \ j \in T_d  \\
|E| & < & \Sp \nn \\
\|{A_E}'\tw\|_2 \sle a_{\st}(S,\Sp) \|c\|_2 \nn \\ %\ \text{and} \nn \\
|{A_j}'\tw| \sle \frac{a_{\st}(S,\Sp)}{\sqrt{\Sp}} \|c\|_2  \ \forall j \notin T \cup T_d \cup E \ \  \text{and} \nn \\
\|\tw\|_2 \sle K_{\st}(S) \|c\|_2
\label{Ebnds}
\eea
where $a_{\st}(S,\Sp)$ is defined in (\ref{def_a0}) and
\bea
K_{\st}(S) \sdefn \frac{ \sqrt{1 + \delta_S} }{1-\delta_S - \frac{\theta_{S,\st}^2}{1 - \delta_{\st}} }
\label{def_K0}
\eea
\label{wbnd}
\end{lemma}
The proof is given in the next subsection.
%a_{\st}(S,\Sp) \sdefn \frac{\theta_{\Sp,S} + \frac{\theta_{\Sp,\st} \ \theta_{S,\st}}{1 - \delta_{\st}}}{ 1-\delta_S - \frac{\theta_{S,\st}^2}{1 - \delta_{\st}} } \nn \\ %where $a(S,\Sp,\st)$ is defined in (\ref{def_a}).
\\

{\em Proof Outline of Theorem \ref{thm1}. }
To prove Theorem \ref{thm1}, apply Lemma \ref{wbnd} iteratively, in a fashion similar to that of the proof of  \cite[Lemma 2.2]{decodinglp} (this proof had some important typos).
The main idea is as follows. At iteration zero, apply Lemma \ref{wbnd} with $T_{d} \equiv \Delta$ (so that $S \equiv \sd$), $c_j \equiv \signumfn(x_j) \ \forall \ j \in \Delta$, and $\Sp \equiv \sd$, to get  a $w_{1}$ and an exceptional set $T_{d,1}$, of size less than $\sd$, that satisfy the above conditions. At iteration $\iter > 0$, apply Lemma \ref{wbnd} with $T_{d} \equiv \Delta \cup T_{d,\iter}$ (so that $S \equiv 2\sd$), $c_j \equiv 0 \ \forall \ j \in \Delta$, $c_j \equiv {A_j}' w_\iter \ \forall \ j \in T_{d,\iter}$ and $\Sp \equiv \sd$ to get a  $w_{\iter+1}$ and an exceptional set $T_{d,\iter+1}$, of size less than $\sd$. Lemma \ref{wbnd} is applicable in the above fashion because condition \ref{cond1} of Theorem \ref{thm1} holds. Define $w : = \sum_{\iter=1}^\infty (-1)^{\iter-1} w_\iter$. We then argue that if condition \ref{cond2} of Theorem \ref{thm1} holds, $w$ satisfies the conditions of Lemma \ref{wcond}. Applying  Lemma \ref{wcond}, the result follows. %Notice that the $n$ used in this proof should not be confused with $n$ used in the rest of the paper for the number of measurements.
We give the entire proof in the Appendix.%
%We can apply the above lemma iteratively to make the size of the exceptional set $E$ smaller and smaller and to create a $w$ that satisfies the conditions of Lemma \ref{wcond} if $a_{\st}(\sd,\sd) + a_{\st}(2\sd,\sd) < 1$. Combining with Lemma \ref{wcond}, we get the result of Theorem \ref{thm1}.  Lemma \ref{wbnd}
%For completion, we give the entire proof in the Appendix. %$a_{\st}(\sd,\sd) + a_{\st}(2\sd,\sd) < 1$ and if its denominator is positive % to find a $w$ that satisfies the conditions of Lemma \ref{wcond}.

\subsection{Proofs of Lemmas \ref{wcond} and \ref{wbnd}}
\label{lemmaproof}
We prove the lemmas from the previous subsection here. Recall that $\st=|T|$ and $\sd=|\Delta|$.

\subsubsection{Proof of Lemma \ref{wcond}}
The proof is motivated by \cite[Section II-A]{decodinglp}.
 %Standard convex arguments give that there is at least one minimizer of (\ref{l1seqcs}).
There is clearly at least one element in the feasible set of (\ref{l1seqcs}) - $x$ - and hence there will be at least one minimizer of (\ref{l1seqcs}). %We need to prove that $x$ is a minimizer  of (\ref{l1seqcs}) and any minimizer, $\beta$,  of (\ref{l1seqcs}) is equal to $x$.
Let $\beta$ be a minimizer of (\ref{l1seqcs}). We need to prove that if the conditions of the lemma hold, it is equal to $x$. For any minimizer, $\beta$,
\bea
\|(\beta)_{T^c}\|_1 \le \|(x)_{T^c}\|_1 := \sum_{j \in \Delta} |x_j|
\label{ineq1}
\eea
Recall that $x$ is zero outside of $T \cup \Delta$, $T$ and $\Delta$ are disjoint, and $x$ is always nonzero on the set $\Delta$. Take a $w$ that satisfies the three conditions of the lemma. Then,%
\bea
\|(\beta)_{T^c}\|_1 \se \sum_{j \in \Delta} |x_j + (\beta_j - x_j)| + \sum_{j \notin T \cup \Delta} |\beta_j| \nn \\%\sge \sum_{j \in \Delta} \signumfn(x_j) (x_j + (\beta_j - x_j)) + \sum_{j \notin T \cup \Delta} |\beta_j| \nn \\
\sge \sum_{j \in \Delta} |x_j + (\beta_j - x_j)| + \sum_{j \notin T \cup \Delta} w'A_j \beta_j  \nn \\
\sge \sum_{j \in \Delta} \signumfn(x_j) (x_j + (\beta_j - x_j)) + \sum_{j \notin T \cup \Delta} w'A_j \beta_j  \nn \\
%\se \sum_{j \in \Delta}  |x_j| + \sum_{j \in \Delta} w'A_j (\beta_j - x_j) + \sum_{j \notin T \cup \Delta} w'A_j \beta_j \nn \\
\se \sum_{j \in \Delta}  |x_j| + \sum_{j \in \Delta} w'A_j (\beta_j - x_j) + \sum_{j \notin T \cup \Delta} w'A_j \beta_j  \nn \\ && + \sum_{j \in T} w'A_j (\beta_j - x_j) \nn \\
\se \|x_{T^c}\|_1 + w'(A \beta - A x) = \|x_{T^c}\|_1
%\se \sum_{j \in \Delta}  |x_j| +  w' ( \sum_{j \in T \cup \Delta}A_j (\beta_j - x_j) + \sum_{j \notin T \cup \Delta} A_j \beta_j   ) \nn \\
%\se \sum_{j \in \Delta}  |x_j| +  w' (A \beta - A_{T \cup \Delta} (x)_{T \cup \Delta}) \nn \\
%\se \sum_{j \in \Delta}  |x_j| +  w' (y  - y) = \sum_{j \in \Delta}  |x_j| = \|x_{T^c}\|_1
\label{ineq2}
\eea
%The 2nd row follows from the condition \ref{subgrad} on $w$. The 4th row follows from condition \ref{sgn}. The 5th row follows from condition \ref{zero}. The last row follows because $x$ is supported on $T \cup \Delta$ only.

Now, the only way (\ref{ineq2}) and (\ref{ineq1}) can hold simultaneously is if all inequalities in (\ref{ineq2}) are actually equalities. %Thus we need $\sum_{j \in \Delta} w'A_j (\beta_j - x_j) + \sum_{j \notin T \cup \Delta} w'A_j \beta_j = 0$. Consider the second term. Since we need
Consider the first inequality. Since $|w'A_j|$ is strictly less than 1 for all $j  \notin T \cup \Delta$, the only way $\sum_{j \notin T \cup \Delta} |\beta_j| = \sum_{j \notin T \cup \Delta} w'A_j \beta_j$  is if $\beta_j = 0$ for all $j  \notin T \cup \Delta$.

Since both $\beta$ and $x$ solve (\ref{l1seqcs}), $y= Ax = A \beta$. Since $\beta_j = 0 = x_j$ for all $j  \notin T \cup \Delta$, this means that $y = A_{T \cup \Delta} (\beta)_{T \cup \Delta}  = A_{T \cup \Delta} (x)_{T \cup \Delta}$ or that $A_{T \cup \Delta} ((\beta)_{T \cup \Delta} -  (x)_{T \cup \Delta})=0$. Since $\delta_{\st + \sd} < 1$, $A_{T \cup \Delta}$ is full rank and so the only way this can happen is if $(\beta)_{T \cup \Delta} =  (x)_{T \cup \Delta}$. Thus any minimizer, $\beta = x$, i.e. $x$ is the unique minimizer of (\ref{l1seqcs}).% This proves the claim.
$\blacksquare$%

\subsubsection{Proof of Lemma \ref{wbnd}}
The proof of this lemma is significantly different from that of the corresponding lemma in \cite{decodinglp}, even though the form of the final result is similar.

Any $\tw$ that satisfies ${A_T}'\tw = 0$ will be of the form
\bea
\tw = [I - A_{T}({A_{T}}'A_{T})^{-1}{A_{T}}'] \gamma : = M \gamma
\eea
We need to find a $\gamma$ s.t. ${A_{T_d}}'\tw = c$, i.e. ${A_{T_d}}'M \gamma = c$. Let $\gamma = M'A_{T_d} \eta$. Then
$\eta = ({A_{T_d}}'M M'A_{T_d} )^{-1} c = ({A_{T_d}}'M A_{T_d} )^{-1} c$. This follows because $M M' = M^2 = M$ since $M$ is a projection matrix.
Thus,
\bea
\tw = MM'A_{T_d} ({A_{T_d}}'M A_{T_d} )^{-1} c = M A_{T_d} ({A_{T_d}}'M A_{T_d} )^{-1} c \ \ \ \
\label{def_tw}
\eea
Consider any set $\Tdp$ with $|\Tdp| \le \Sp$ disjoint with $T \cup T_d$. Then
\bea
\|{A_{\Tdp}}'\tw \|_2 %\se \|{A_{\Tdp}}' M A_{T_d} ({A_{T_d}}'M A_{T_d} )^{-1} c \|_2 \nn \\
                  \sle \|{A_{\Tdp}}' M A_{T_d}\| \ \|({A_{T_d}}'M A_{T_d} )^{-1}\| \ \|c\|_2  \ \ \
\label{eq1}
\eea
Consider the first term from the right hand side (RHS) of (\ref{eq1}).
\bea
\|{A_{\Tdp}}' M A_{T_d}\| \sle \|{A_{\Tdp}}'A_{T_d}\| + \|{A_{\Tdp}}' A_{T}({A_{T}}'A_{T})^{-1}{A_{T}}' A_{T_d}\| \nn \\
                          \sle \theta_{\Sp,S} + \frac{\theta_{\Sp,\st} \ \theta_{S,\st}}{1 - \delta_{\st}}
\label{eq2}
\eea
%The second bound is valid since $\delta_{\st} \le \delta_{S} + \delta_{\st} + \theta_{\st,S}^2 < 1$.
Consider the second term from the RHS of (\ref{eq1}). Since ${A_{T_d}}'M A_{T_d}$ is non-negative definite,
\bea
\|({A_{T_d}}'M A_{T_d} )^{-1}\| \se \frac{1}{\lambda_{\min}({A_{T_d}}'M A_{T_d} )}
%\se  \frac{1}{\lambda_{\min}({A_{T_d}}'A_{T_d}  - {A_{T_d}}' A_{T}({A_{T}}'A_{T})^{-1}{A_{T}}'A_{T_d} )}
\eea
Now, ${A_{T_d}}'M A_{T_d} = {A_{T_d}}'A_{T_d}  - {A_{T_d}}' A_{T}({A_{T}}'A_{T})^{-1}{A_{T}}'A_{T_d}$ which is the difference of two symmetric non-negative definite matrices. Let $B_1$ denote the first matrix and $B_2$ the second one. Use the fact that $\lambda_{\min}(B_1 - B_2) \ge \lambda_{\min}(B_1) + \lambda_{\min}(-B_2) =  \lambda_{\min}(B_1) - \lambda_{\max}(B_2)$ where $\lambda_{\min}(.), \lambda_{\min}(.)$ denote the minimum, maximum eigenvalue.
%
%For symmetric matrices, $\lambda_{\min}(B_1 + B_2) \ge \lambda_{\min}(B_1) + \lambda_{\min}(B_2)$ where $\lambda_{\min}(.)$ denotes the minimum eigenvalue. Let $B_1 := {A_{T_d}}'A_{T_d}$ and $B_2 := -{A_{T_d}}' A_{T}({A_{T}}'A_{T})^{-1}{A_{T}}'A_{T_d}$.
%This is the difference of two non-negative definite matrices. It is easy to see that if $B_1$ and $B_2$ are two non-negative definite matrices, then $\lambda_{\min}(B_1 - B_2) \ge \lambda_{\min}(B_1) - \lambda_{\max}(B_2)$.
%
Since $\lambda_{\min}(B_1) \ge (1-\delta_S)$ and $\lambda_{\max}(B_2) = \|B_2\| \le \frac{\|({A_{T_d}}' A_{T})\|^2}{1 - \delta_{\st}} \le \frac{\theta_{S,\st}^2}{1 - \delta_{\st}}$, thus
\bea
\|({A_{T_d}}'M A_{T_d} )^{-1}\| \sle \frac{1}{1-\delta_S - \frac{\theta_{S,\st}^2}{1 - \delta_{\st}} }
\label{eq3}
\eea
as long as the denominator is positive. It is positive because we have assumed that $\delta_{S} + \delta_{\st} + \theta_{\st,S}^2 < 1$.
Using (\ref{eq2}) and (\ref{eq3}) to bound (\ref{eq1}), we get that for any set $\Tdp$ with $|\Tdp| \le \Sp$,
\bea
\|{A_{\Tdp}}'\tw \|_2 \sle \frac{\theta_{\Sp,S}    + \frac{\theta_{\Sp,\st} \ \theta_{S,\st}}{1 - \delta_{\st}}}{ 1-\delta_S - \frac{\theta_{S,\st}^2}{1 - \delta_{\st}} } \|c\|_2  = a_{\st}(S,\Sp) \|c\|_2 \ \ \ \ \ \ \
 \label{def_a}
\eea
where $a_{\st}(S,\Sp)$ is defined in (\ref{def_a0}).
Notice that $a_{\st}(S,\Sp)$ is non-decreasing in $\st$, $S$, $\Sp$. % all its arguments:
Define an exceptional set, $E$, as%
\bea
E :=\{ j \in (T \cup T_d)^c : |{A_j}'\tw| > \frac{a_{\st}(S,\Sp) }{\sqrt{\Sp}} \|c\|_2 \}
\eea
Notice that $|E|$ must obey $|E| < \Sp$ since otherwise we can contradict (\ref{def_a}) by taking $\Tdp \subseteq E$.

Since $|E| < \Sp$ and $E$ is disjoint with $T \cup T_d$, (\ref{def_a}) holds for $\Tdp \equiv E$, i.e. $\|{A_{E}}'\tw \|_2 \le a_{\st}(S,\Sp) \|c\|_2$. Also, by definition of $E$, $|{A_j}'\tw| \le  \frac{a_{\st}(S,\Sp) }{\sqrt{\Sp}} \|c\|_2$, for all $j \notin T \cup T_d \cup E$.
Finally,%(\ref{def_K0}) holds because
\bea
\|\tw\|_2 \sle \| M A_{T_d} ({A_{T_d}}'M A_{T_d} )^{-1} \| \ \|c\|_2 \nn \\
\sle \| M\| \ \|A_{T_d} \| \ \| ({A_{T_d}}'M A_{T_d} )^{-1} \| \  \|c\|_2   \nn \\
\sle  \frac{ \sqrt{1 + \delta_S} }{1-\delta_S - \frac{\theta_{S,\st}^2}{1 - \delta_{\st}} } \|c\|_2 = K_{\st}(S) \|c\|_2
\label{def_K}
\eea
since $\|M\| =  1$ (holds because $M$ is a projection matrix). Thus, all equations of (\ref{Ebnds}) hold. Using (\ref{def_tw}), (\ref{T_Td}) holds. %This proves the lemma.
$\blacksquare$%$M=M^2$ and so $\|M\| = \|M^2\|$)

\begin{figure*}[t!]
\centering{
\subfigure[\small{Plots of $\rho_{CS}$ defined in (\ref{defrhocs})}]{ %$= g_{\sno/\mno}(2\sn)+ g_{\sno/\mno}(3\sn)$}]{
\includegraphics [width=5.5cm,height=3.6cm]{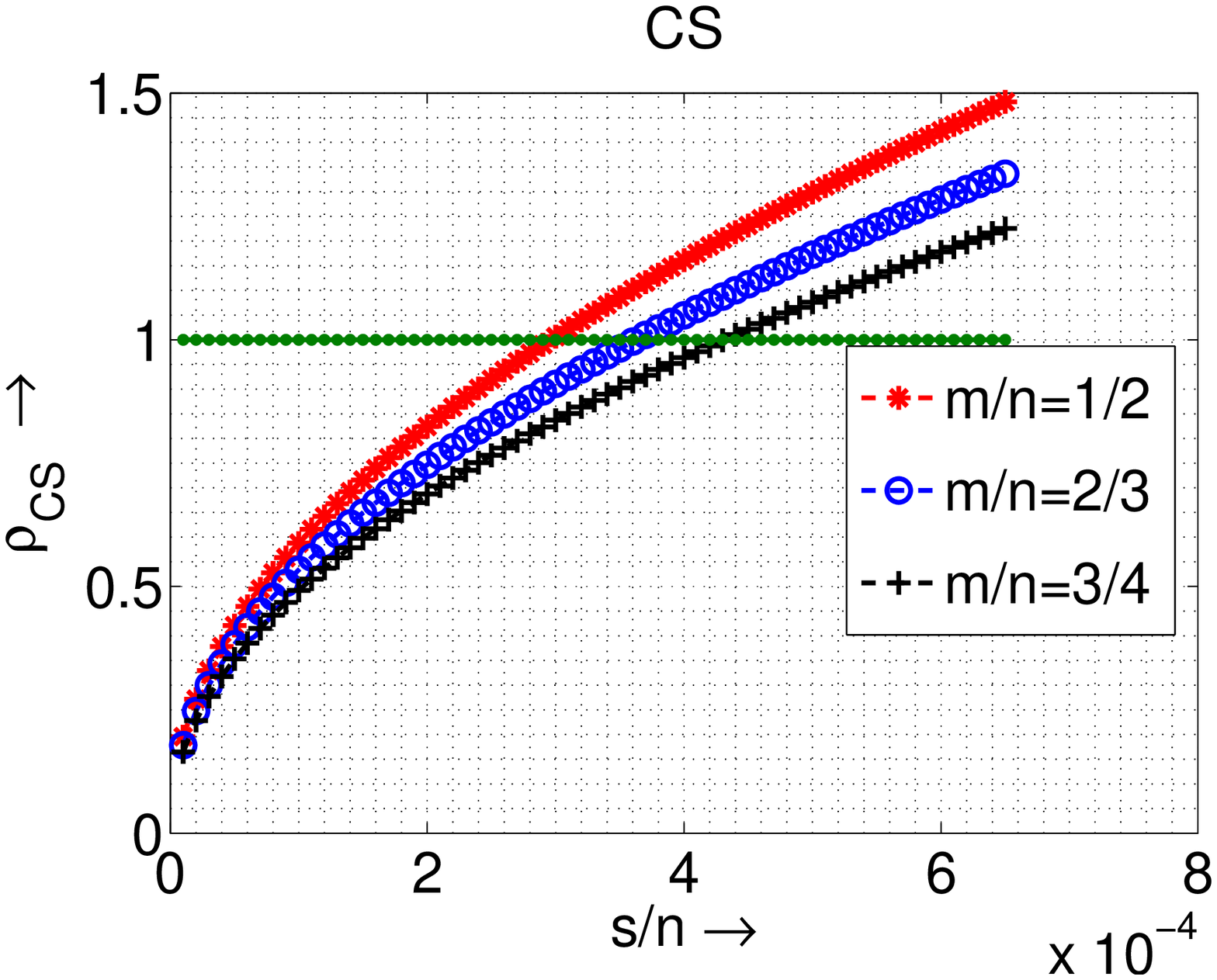}
}
\subfigure[\small{Plots of $\rho_{CS,2}$ defined in (\ref{defrhocs})}]{ %$= g_{\sno/\mno}(2\sn)$}]{
\includegraphics [width=5.5cm,height=3.6cm]{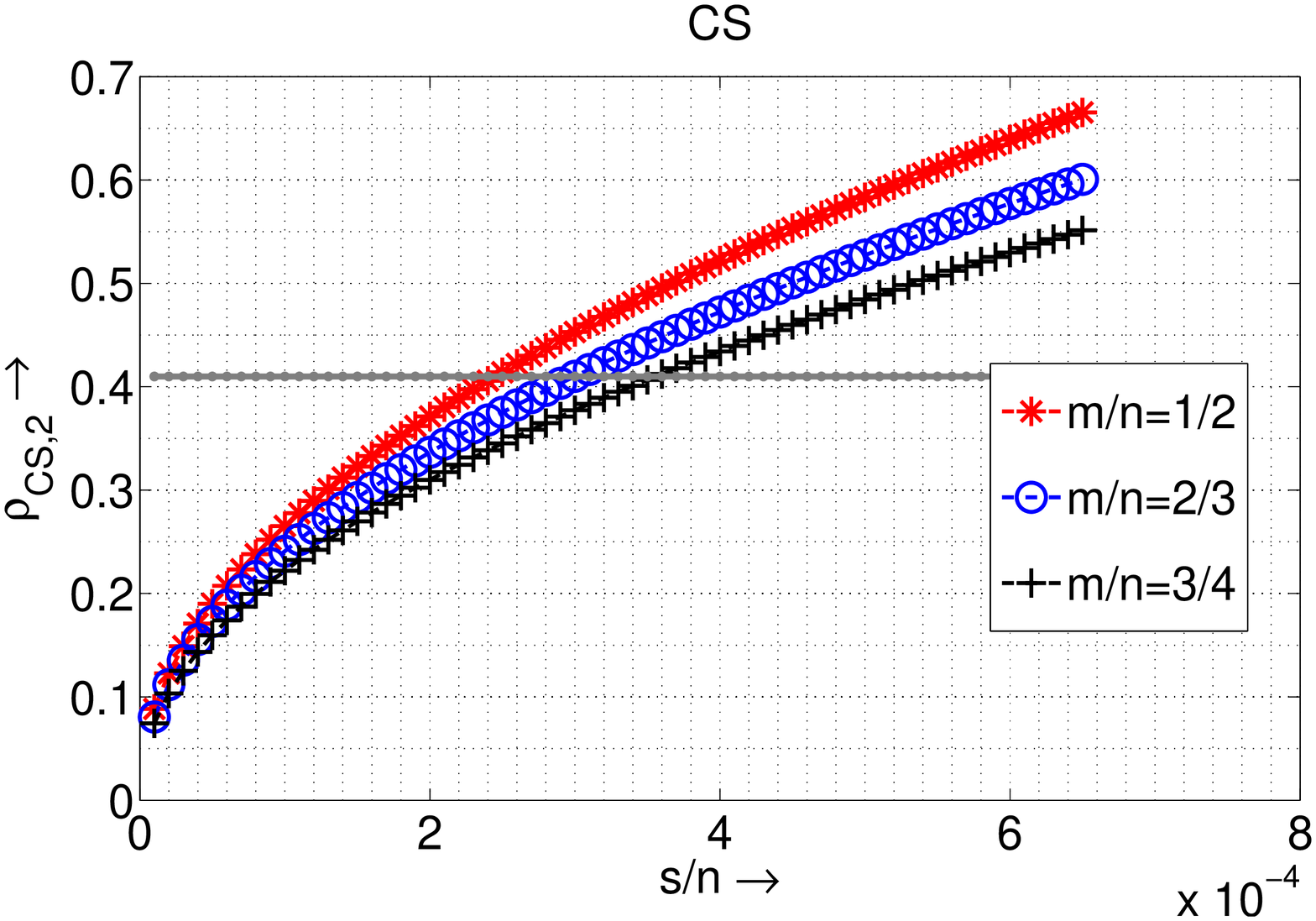}
}
%\hspace{-0.35in}
\subfigure[\small{Plots of $\rho_{modCS}$ defined in (\ref{defrhomodcs})}]{
\includegraphics [width=5.5cm,height=3.6cm]{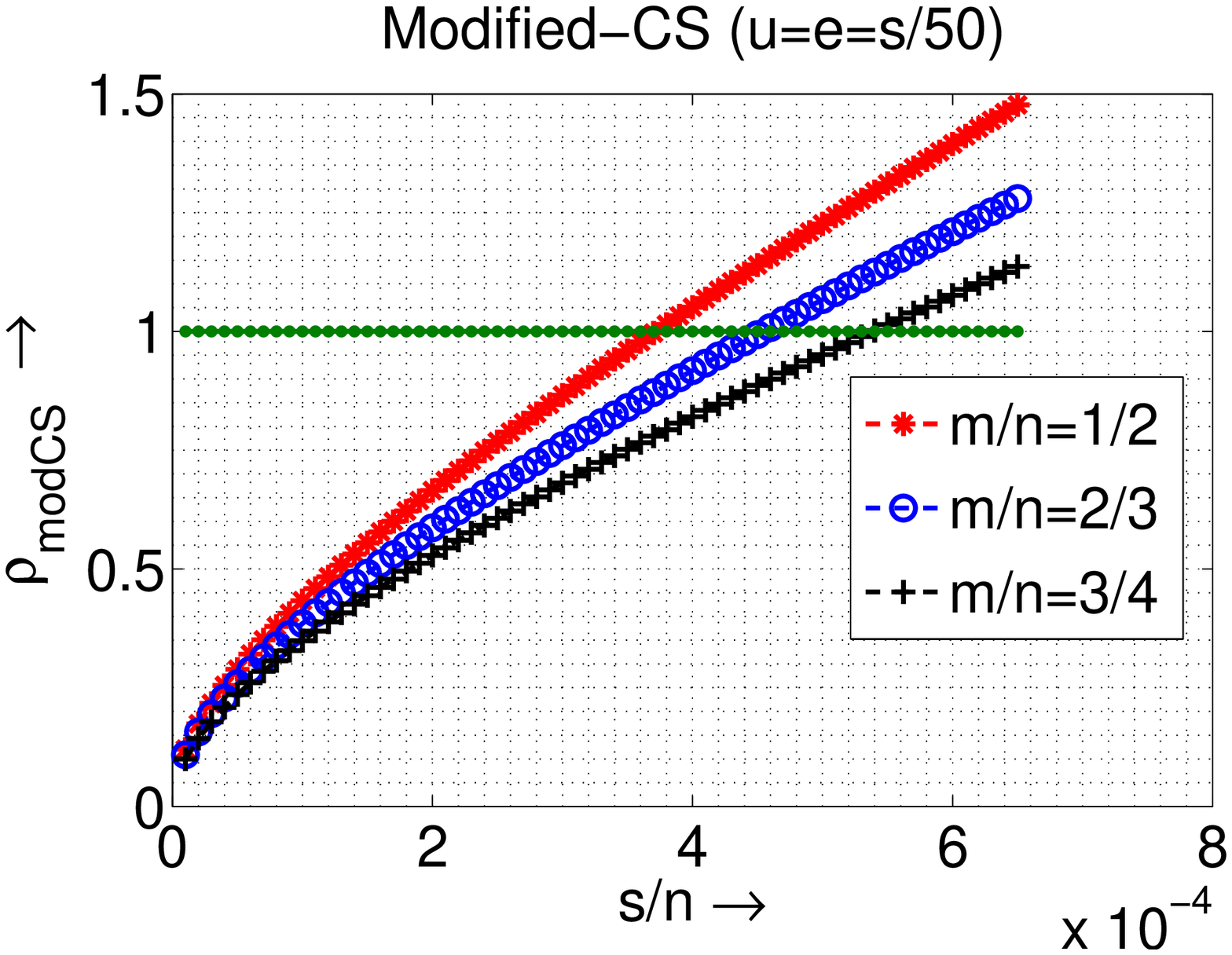}}
}
\caption{\small{Plots of $\rho_{CS}$ and $\rho_{CS,2}$ (in (a) and (b)) and $\rho_{modCS}$ (in (c)) against $\sn/\sno$ for 3 different values of $\mno/\sno$. For  $\rho_{modCS}$, we used $\sd=\sde = \sn/50$. Notice that,  for any given $\mno/\sno$, the maximum allowed sparsity, $\sn/\sno$, for $\rho_{modCS}<1$ is larger than that for which either $\rho_{CS}<1$ or $\rho_{CS,2} < \sqrt{2}-1$. Also, both are much smaller than what is observed in simulations.% ?? CHANGE n to $\mno$ and m to $\sno$ in the FIGURES??
%(a) Upper bound function $\rho_{cs,n/m}(\sn,\sd)$ for $\sn=50\sd$, (b)Upper bound function $\rho_{modcs,n/m}(\st,\sd)$ for $\sn=50\sd=50\sde$. The maximum allowed sparsity for $\rho_{modcs,n/m}(\st,\sd)<1$ is larger than that for $\rho_{cs,n/m}(\sn,\sd)<1$ for given $\mno/\sno$.
}}
%\vspace{-0.15in}
\label{upperbound}
\end{figure*}

\section{Comparison of CS and Modified-CS}
\label{compare}
In Theorem \ref{thm1} and Corollary \ref{corol1}, we derived sufficient conditions for exact reconstruction using modified-CS. In Sec. \ref{highprob}, we compare the sufficient conditions for modified-CS with those for CS. In Sec. \ref{montecarlo}, we use Monte Carlo to compare the probabilities of exact reconstruction for both methods.% In Sec. \ref{noisycompare} we do a brief comparison for noisy measurements.% when $A$ is random Gaussian  modified-CS and CS compare the sufficient conditions for modified-CS to give exact reconstruction with those for CS using the high probability upper bound on $\delta_S$ \cite{decodinglp}

\subsection{Comparing sufficient conditions}
\label{highprob}
We compare the sufficient conditions for modified-CS and for CS, expressed only in terms of $\delta_S$'s. Sufficient conditions for an algorithm serve as a designer's tool to decide the number of measurements needed for it and in that sense comparing the two sufficient conditions is meaningful.

For modified-CS, from Corollary \ref{corol1}, the sufficient condition in terms of only  $\delta_S$'s  is $2\delta_{2\sd}+\delta_{3\sd}+\delta_{\st}+\delta_{\st+\sd}^2+2\delta_{\st+2\sd}^2<1$. Using $\st=\sn+\sde-\sd$,
%where $\st=|T|$, $\sd=|\Delta|$, $\sde=|\Delta_e|$ and $\sn=|N|$,
this becomes
\bea
2\delta_{2\sd}+\delta_{3\sd}+\delta_{\sn+\sde-\sd}+\delta_{\sn+\sde}^2+2\delta_{\sn+\sde +\sd}^2 < 1.
\label{modcscond2}
\eea
For CS, two of the best (weakest) sufficient conditions that use only $\delta_S$'s are given in \cite{candes_rip,foucart_lai} and \cite{dantzig}.
Between these two, it is not obvious which one is weaker. Using \cite{candes_rip} and \cite{dantzig}, CS achieves exact reconstruction if either
\bea
\delta_{2\sn} < \sqrt{2}-1  \text{~~or~~}  \delta_{2\sn}+\delta_{3\sn}<1.
\label{CScond2}
\eea

To compare (\ref{modcscond2}) and (\ref{CScond2}), we use $\sd = \sde = 0.02 \sn$ which is typical for time series applications (see Fig. \ref{slowchange}). One way to compare them is to use $\delta_{cr} \le c \delta_{2r}$ \cite[Corollary 3.4]{cosamp} to get the LHS's of both in terms of a scalar multiple of $\delta_{2\sd}$. Thus, (\ref{modcscond2}) holds if $\delta_{\sn+\sde +\sd} < 1/2 \ \text{and} \ \delta_{2\sd} < 1/132.5$.
Since $\delta_{\sn+\sde +\sd} = \delta_{52\sd}< 52 \delta_{2\sd}$, the second condition implies the first, and so only $\delta_{2\sd} < 1/132.5$ is sufficient. On the other hand, (\ref{CScond2}) holds if $\delta_{2\sd} < 1/241.5$ {\em which is clearly  stronger.}

%Finally use $\sd \approx \sde \approx 0.02 \sn$ which is a typical value for $\sd,\sde$ for time series applications. for $\sd,\sde$

Alternatively, we can compare (\ref{modcscond2}) and (\ref{CScond2}) using the high probability upper bounds on $\delta_S$ as in \cite{decodinglp}.
Using  \cite[Eq 3.22]{decodinglp}, for an $\mno \times \sno$ random Gaussian matrix, with high probability (w.h.p.), $\delta_S< g_{\sno/\mno}(\frac{S}{\sno})$, where
\bea
g_{\sno/\mno} \left(\frac{S}{\sno} \right) \sdefn -1+ \left[ 1+f \left(\frac{S}{\sno}, \frac{\sno}{\mno} \right) \right]^2, \ \text{where}  \nn \\
f \left(\frac{S}{\sno}, \frac{\sno}{\mno} \right) \sdefn \sqrt{\frac{\sno}{\mno}} \left(\sqrt{\frac{S}{\sno}}+\sqrt{2H \left(\frac{S}{\sno} \right)} \right), \nn
\eea
and binary entropy $H(r):=-r\log r-(1-r)\log(1-r)$ for $0 \le r \le 1$. Thus, w.h.p., modified-CS achieves exact reconstruction from random-Gaussian measurements if %$\rho_{modCS} < 1$, where %(\sn,\sd,\sde)
\bea
\label{defrhomodcs}
&& \rho_{modCS} := 2g_{\sno/\mno} \left(\frac{2\sd}{\sno} \right) + g_{\sno/\mno} \left(\frac{3\sd}{\sno} \right) + g_{\sno/\mno} \left(\frac{\sn+\sde-\sd}{\sno} \right)  \nn \\ &&
+  g_{\sno/\mno} \left(\frac{\sn+\sde}{\sno} \right)^2+ 2g_{\sno/\mno} \left(\frac{\sn+\sde +\sd}{\sno} \right)^2 < 1.
\eea
Similarly, from  (\ref{CScond2}), w.h.p., CS achieves exact reconstruction from random-Gaussian measurements if either
\bea
&& \rho_{CS}:= g_{\sno/\mno} \left(\frac{2\sn}{\sno} \right) + g_{\sno/\mno} \left(\frac{3\sn}{\sno} \right) < 1 \text{~or~} \nn \\
&& \rho_{CS,2} := g_{\sno/\mno} \left(\frac{2\sn}{\sno} \right) < \sqrt{2}-1.
\label{defrhocs}
\eea

%\ \ \text{where}   \\ && g_{\sno/\mno}(\frac{S}{\sno}) \defn  -1+[1+f(\frac{S}{\sno}, \frac{\sno}{\mno})]^2
%$\rho_{CS} < 1$, where %Let $g(S):=-1+[1+f(\frac{S}{\sno})]^2$. (\sn)  \cite{candes_rip,dantzig}

%Notice that both $\rho_{CS}$ and $\rho_{modCS}$ also depend on $\sno$ and $\mno$. using the maximum value obtained
In Fig. \ref{upperbound}, we plot $\rho_{CS}$, $\rho_{CS,2}$ and  $\rho_{modCS}$ against $\sn/\sno$ for three different choices of $\mno/\sno$.  For $\rho_{modCS}$, we use $\sd = \sde = 0.02\sn$ (from Fig. \ref{slowchange}). %and thus $\sd/\sno = \sde/\sno = 0.02 \sn/\sno$.
As can be seen, the maximum allowed sparsity, i.e. the maximum allowed value of $\sn/\sno$, for which either $\rho_{CS}< 1$ or $\rho_{CS,2} < \sqrt{2}-1$ is smaller than that for which $\rho_{modCS} < 1$. Thus, for a given number of measurements, $\mno$,  {\em w.h.p., modified-CS will give exact reconstruction from random-Gaussian measurements, for larger sparsity sizes, $\sn/\sno$, than CS would.} As also noted in \cite{decodinglp}, in all cases, the maximum allowed $\sn/\sno$ is much smaller than what is observed in simulations, because of the looseness of the bounds. For the same reason, the difference between CS and modified-CS is also not as significant.% (since the bound is very loose).  as it is in simulations
%Also, from the plots, the condition $\rho_{CS} < 1$ \cite{dantzig} is better (weaker) than $\rho_{CS,2} < \sqrt{2}-1$ \cite{candes_rip}.%of $\sn/\sno$  from the figures much

%\begin{remark}
%By modifying the proof approach of \cite{candes_rip}, it is possible to also get sufficient conditions for modified-CS that compare directly with those from \cite{candes_rip} or \cite{dantzig}. The former is done by Jacques \cite{arxiv}. One can show that modified-CS achieves exact reconstruction if either
%\bea
%\delta_{\sn+\sde + \sd} < \sqrt{2}-1 \text{~~or~~} \delta_{\sn+\sde + \sd} + \delta_{\sn+\sde + 2\sd} < 1
%\label{cond2}
%\eea
%This can be directly compared with the corresponding conditions for CS given in (\ref{CScond2}). Clearly, if $\sde \ll \sn$ and $\sd \ll \sn$, (\ref{cond2}) is significantly weaker than (\ref{CScond2}).
%\end{remark}

\subsection{Comparison using Monte Carlo}
\label{montecarlo}

%The above comparison provides very small maximum values for the allowed support size, $\sn$, for a given $\mno$, $\sno$. Also, in the above we are only comparing sufficient conditions.
So far we only compared sufficient conditions. The actual allowed $\sn$ for CS may be much larger. To actually compare exact reconstruction ability of modified-CS with that of CS, we thus need Monte Carlo. We use the following procedure to obtain a Monte Carlo estimate of the probability of exact reconstruction using CS and modified-CS, for a given $A$ (i.e. we average over the joint distribution of $x$ and $y$ given $A$).%
\ben
\item Fix signal length, $\sno = 256$ and its support size, $\sn= 0.1\sno=26$.
 Select $\mno$, $\sd$ and $\sde$.%= |\Delta| =|\Delta_e|
\item Generate the  $\mno\times \sno$ random-Gaussian matrix, $A$ (generate an $\mno \times \sno$ matrix with independent identically distributed (i.i.d.) zero mean Gaussian entries and normalize each column to unit $\ell_2$ norm)\footnote{As pointed out by an anonymous reviewer, we actually do not need to normalize each column to unit norm.  As proved in \cite{RIPsimpleproof}, a matrix with i.i.d. zero mean Gaussian entries with variance $1/n$ will itself satisfy the RIP. If the variance is not $1/n$, there will just be a scaling factor in the RIP. This does not affect reconstruction performance in any way.}.
% A matrix with i.i.d. zero mean entries and any variance $v$ will also satisfy a scaled version of the RIP and that again will not change its incoherence.}.

\item Repeat the following $\text{tot}=500$ times
\label{gendata}
\ben
\item  Generate the support, $N$, of size $\sn$, uniformly at random from $[1,\sno]$.
\item Generate $(x)_N \sim \n(0,100 I)$. Set $(x)_{N^c} = 0$.

\item Set $y:=Ax$.
 %(each element i.i.d. Gaussian with zero mean and variance $100$).
 %Generate the nonzero elements of $x$ on the support $N$ with i.i.d Gaussian distributed entries with zero mean and variance 100. Compute $y:=Ax$.

\item Generate $\Delta$ of size $\sd$ uniformly at random from the elements of $N$.%the unknown part of support,

\item Generate $\Delta_e$ of size $\sde$, uniformly at random from the elements of $[1,\sno] \setminus N$.%the error in known part of support,

\item Let $T = N \cup \Delta_e \setminus \Delta$. Run modified-CS, i.e. solve (\ref{l1seqcs})). Call the output $\xhat_{modCS}$.

\item Run CS, i.e. solve (\ref{l1seqcs}) with $T$ being the empty set. Call the output $\xhat_{CS}$.%econstruct using

\een
%\item Repeat 500 times for each $\sd$,$\sde$ and fixed $\sn$.

\item Estimate the probability of exact reconstruction using modified-CS by counting the number of times $\xhat_{modCS}$ was equal to $x$ (``equal" was defined as $\|\xhat_{modCS}-x\|_2/\|x\|_2 < 10^{-5}$) and dividing by $\text{tot}=500$.
\label{probcompute}

\item Do the same for CS using $\xhat_{CS}$.% for each $\mno$, $\sd$, $\sde$

\item Repeat for various values of $\mno$,  $\sd$ and $\sde$.
\een

 %=0.16m, 0.19m, 0.25m, 0.3m, 0.4\sno$.
 % (the minimum number of measurements needed for CS to work)
We set $\sno=256$ and $\sn=0.1\sno$ and we varied $\mno$ between $0.16\sno = 1.6\sn$ and $0.4\sno = 4\sn$. For each $\mno$, we varied $\sd = |\Delta|$ between $0.04\sn$ to $\sn$ and $\sde=|\Delta_e|$ between $0$ to $0.4\sn$. We tabulate our results in Table \ref{prob1}. {\em The case $\sd = \sn$ and  $\sde=0$ corresponds to CS.} Notice that when $\mno$ is just $0.19\sno = 1.9\sn < 2\sn$, modified-CS achieves exact reconstruction more than 99.8\% of the times if $\sd \le 0.08\sn$ and $\sde \le 0.08\sn$. % This probability is 95\% if $\sd \le 0.12\sn$.
In this case, CS has {\em zero} probability of exact reconstruction. With $\mno=0.3\sno=3\sn$, CS has a very small (14\%) chance of exact reconstruction. On the other hand, modified-CS works almost all the time for $\sd \le 0.2\sn$ and $\sde \le 0.4\sn$. CS needs at least $\mno=0.4\sno = 4\sn$ to work reliably.%with $\mno=0.3\sno$,

The above simulation was done in a fashion similar to that of \cite{decodinglp}. It does not compute the $\mno$ required for Theorem \ref{thm1} to hold. Theorem \ref{thm1} says that if $\mno$ is large enough for a given $\sn$, $\sd$, $\sde$, so that the two conditions given there hold, modified-CS will {\em always} work. But all we show above is that {\em (a) for certain large enough values of $\mno$, the Monte Carlo estimate of the probability of exact reconstruction using modified-CS is one (probability computed by averaging over the joint distribution of $x$ and $y$); and (b) when $\sd$, $\sde \ll \sn$, this happens for much smaller values of $\mno$ with modified-CS than with CS.}% compared to $\sn$

%Either for CS or for Modified-CS, finite sample Monte Carlo simulations usually ignore pathological cases, i.e. cases that occur with zero/low probability. As pointed out by an anonymous reviewer,
As pointed out by an anonymous reviewer, Monte Carlo only computes expected values (here, expectation of the indicator function of the event that exact reconstruction occurs) and thus, it ignores the pathological cases which occur with zero probability \cite{dossal_peyre,dossal}. %only computes the probability of exact reconstruction and hence ignores pathological (zero probability)  cases. Finite sample Monte Carlo also usually ignores low probability cases. with probability one
%Notice that finite sample Monte Carlo ignores pathological cases, i.e. cases which occur with low/zero probability. the probability of exact reconstruction. This issue has been discussed in detail in \cite{dossal_peyre,dossal}. probability or expected chance of exact reconstruction)
In \cite{dossal_peyre}, the authors give a greedy pursuit algorithm to find these pathological cases for CS, i.e. to find the sparsest vector $x$ for which CS does not give exact reconstruction. The support size of this vector then gives an upper bound on the sparsity that CS can handle. Developing a similar approach for modified-CS is a useful open problem.%comes close to 1 almost always n efficient

\begin{table*}[t!]
\caption{\small{Probability of exact reconstruction for modified-CS. Recall that $\sd=|\Delta|$, $\sde=|\Delta_e|$ and $\sn=|N|$.
Notice that $\sd=\sn$ and $\sde=0$ corresponds to CS.}} %$\st=|T|$,
\label{prob1}
\centering
\subtable[$\mno=0.16 \sno$]{%, CS: 0
\begin{tabular}{|c|c|c|c|c|}
  \hline
  % after \\: \hline or \cline{col1-col2} \cline{col3-col4} ...
  \backslashbox{$\sd$}{$\sde$} & 0 & $0.08\sn$ & $0.24\sn$ & $0.40\sn$ \\
  \hline
 % $0$ & 1 & 1 & 1 &1 \\
  $0.04\sn$ & 0.9980 & 0.9900 & 0.8680 & 0.4100 \\
  \hline
  $0.08\sn$ & 0.8880 & 0.8040 & 0.3820 & 0.0580 \\
  \hline
%  $0.12\sn$ & 0.6000 & 0.4400 & 0.1340 & 0.0060 \\
%  \hline
%  $0.20\sn$ & 0.1360 & 0.0560 & 0.0040 & 0 \\
%  \hline
%  $0.35\sn$ & 0.0020 & 0.0020 & 0 & 0 \\
%  \hline
%  $0.50\sn$ & 0 & 0 & 0 & 0 \\
%  \hline
  $\sn$  &  {\bf (CS)} 0.0000 &  &  &  \\  %& 0 & 0 & 0\\
  \hline
\end{tabular}}
\subtable[$\mno=0.19 \sno$]{%, CS: 0
\begin{tabular}{|c|c|c|c|c|}
  \hline
  % after \\: \hline or \cline{col1-col2} \cline{col3-col4} ...
  \backslashbox{$\sd$}{$\sde$} & 0 & $0.08\sn$ & $0.24\sn$ & $0.40\sn$ \\
   \hline
  $0.08\sn$ & 0.9980 & 0.9980 & 0.9540 & 0.7700 \\
  \hline
  $0.12\sn$ & 0.9700 & 0.9540 & 0.7800 & 0.4360 \\
  \hline
  $\sn$ &  {\bf (CS)} 0.0000 &  &  &  \\ %& 0 & 0 & 0 \\
  \hline
\end{tabular}}
\subtable[$\mno=0.25 \sno$]{%, CS: 0
\begin{tabular}{|c|c|c|c|c|}
  \hline
  % after \\: \hline or \cline{col1-col2} \cline{col3-col4} ...
  \backslashbox{$\sd$}{$\sde$} & 0 & $0.08\sn$ & $0.24\sn$ & $0.40\sn$ \\
  \hline
%  $0$ & 1 & 1 & 1 &1 \\
  $0.04\sn$ & 1 & 1 & 1 & 1 \\
  \hline
  $0.20\sn$ & 1 & 1 & 0.9900 & 0.9520 \\
  \hline
  $0.35\sn$ & 0.9180 & 0.8220 & 0.6320 & 0.3780 \\
  \hline
  $0.50\sn$ & 0.4340 & 0.3300 & 0.1720 & 0.0600 \\
  \hline
  $\sn$ &  {\bf (CS)} 0.0020 &  &  &  \\ %& 0 & 0 & 0 \\
  \hline
\end{tabular}}
%\end{table*}
%\begin{table}
%\caption{\small{Probability of exact reconstruction for modified-CS}}
%\label{prob2}
\centering
\subtable[$\mno=0.30 \sno$]{%, CS: 0.16
\begin{tabular}{|c|c|c|c|c|}
  \hline
  % after \\: \hline or \cline{col1-col2} \cline{col3-col4} ...
  \backslashbox{$\sd$}{$\sde$} & 0 & $0.08\sn$ & $0.24\sn$ & $0.40\sn$ \\
  \hline
%  $0$ & 1 & 1 & 1 &1 \\
  $0.04\sn$ & 1 & 1 & 1 & 1 \\
  \hline
  $0.20\sn$ & 1 & 1 & 1 & 1 \\
  \hline
  $0.35\sn$ & 1 & 1 & 0.9940 & 0.9700 \\
  \hline
  $0.50\sn$ & 0.9620 & 0.9440 & 0.8740 & 0.6920 \\
  \hline
  $\sn$   &  {\bf (CS)} 0.1400 &  &  &  \\ %& 0.1 & 0.03 & 0.02\\
  \hline
\end{tabular}}
\subtable[$\mno=0.40 \sno$]{
\begin{tabular}{|c|c|c|c|c|}
  \hline
  % after \\: \hline or \cline{col1-col2} \cline{col3-col4} ...
  \backslashbox{$\sd$}{$\sde$} & 0 & $0.40\sn$ \\
  \hline
%  %$0$ & 1 & 1 \\ % & 1 &1 \\
 $0.04\sn$ & 1 & 1 \\ %& 1 & 1 \\
 \hline
  $0.20\sn$ & 1 & 1 \\ %& 1 & 1 \\
  \hline
  $0.35\sn$ & 1 & 1 \\ %& 1 & 1 \\
  \hline
  $0.50\sn$ & 1 & 1 \\ %& 1 & 1 \\
  \hline
  $\sn$ & {\bf (CS)} 0.9820 &  \\ %&  &  \\ %& 0.98 & 0.96 & 0.94\\
  \hline
\end{tabular}}
%\subtable[$\mno=0.40 \sno$]{%, CS: 0.98
%\begin{tabular}{|c|c|c|c|c|}
%  \hline
%  % after \\: \hline or \cline{col1-col2} \cline{col3-col4} ...
%  \backslashbox{$\sd$}{$\sde$} & 0 & $0.08\sn$ & $0.24\sn$ & $0.40\sn$ \\
%%  \hline
%  %  $0$ & 1 & 1 & 1 &1 \\
%%  $0.04\sn$ & 1 & 1 & 1 & 1 \\
%%  \hline
%%  $0.20\sn$ & 1 & 1 & 1 & 1 \\
%%  \hline
%%  $0.35\sn$ & 1 & 1 & 1 & 1 \\
%  \hline
%  $0.50\sn$ & 1 & 1 & 1 & 1 \\
%  \hline
%  $\sn$ & {\bf (CS)} 0.9820  &  &  &  \\ %& 0.98 & 0.96 & 0.94\\
%  \hline
%\end{tabular}}
\end{table*}
%%%%%%%%%%%%%%%%%%%%%%%%%%%%%%%%%%%%%%%%%%%%%%%%%%%%%

\subsection{Robustness to noise}
\label{noisycompare}
Using an anonymous reviewer's suggestion, we studied the robustness of modified-CS to measurement noise. Of course notice that in this case the true signal, $x$, does not satisfy the data constraint. Thus it is not clear if (\ref{l1seqcs}) will even be feasible. A correct way to approach noisy measurements is to relax the data constraint as is done for CS in \cite{bpdn} or \cite{candes_rip}. This is done for modified-CS in our recent work \cite{modcsicassp10} and also in \cite{arxiv}.% Similarly for CS, i.e. (\ref{l1seqcs}) with $T$ empty.  BPDN  as is done for CS in \cite{bpdn} or \cite{candes_rip}  and compared it with that of CS

In practice though, at least with random Gaussian measurements and small enough noise, (\ref{l1seqcs}) did turn out to be feasible, i.e. we were able find a solution, in all our simulations. %, both for modified-CS and for CS ($T$=empty). $ is i.i.d. zero mean Gaussian noise with variance  of our Monte Carlo test
We used $\sno = 256$, $\sn= 0.1\sno$, $u=e=0.08s$ and $\mno=0.19 \sno$. We ran the simulation as in step \ref{gendata} of the previous subsection with the following change. The measurements were generated as $y:=Ax+w$ where $w \sim \n(0, \sigma_w^2I)$. We varied $\sigma_w^2$ and compared the normalized root mean squared error  (N-RMSE) of modified-CS with that of CS in Table \ref{noisytable}. N-RMSE is computed as $\sqrt{\E[\|x-\hat{x}\|_2^2]/\E[\|x\|_2^2]}$ where $\E[.]$ denotes the expected value computed using Monte Carlo.
Recall that $x_N \sim \n(0,100 I)$.
When the noise is small enough, modified-CS has small error. CS has large error in all cases since $\mno$ is too small for it.% (this is for $\mno=1.9\sn$ being small). In large noise, both have large error.for all values of $\sigma_w^2$  ($\sigma_w^2 < 0.1$)

%Clearly modified-CS has much smaller error when $\sigma_w^2 < 1$. independent identically distributed (

\vspace{-0.1in}
\begin{table}[h]
\caption{Reconstruction error (N-RMSE) from noisy measurements.}
%$\mno=0.19 \sno$,$\sd=0.08\sn$,$\sde=0.08\sn$}
\vspace{-0.1in}
\begin{center}
\begin{tabular}{|c|c|c|c|c|c|c|}
  \hline
  % after \\: \hline or \cline{col1-col2} \cline{col3-col4} ...
  $\sigma^2_w$ & 0.001 & 0.01 & 0.1 & 1 & 10 \\ %& 100\\
  \hline
  CS & 0.7059 & 0.7011 & 0.7243 & 0.8065 & 1.1531 \\ %& 2.4888 \\
  \hline
  Modified-CS & 0.0366 & 0.0635 & 0.1958 & 0.5179 & 1.3794 \\ %& 4.3204\\
  \hline
\end{tabular}
\end{center}
\vspace{-0.2in}
\label{noisytable}
\end{table}

\section{Extensions of Modified-CS}
\label{exts}
We now discuss some key extensions - dynamic modified-CS, regularized modified-CS (RegModCS) and dynamic RegModCS. RegModCS is useful when exact reconstruction does not occur - either $\mno$ is too small for exact reconstruction or the signal is compressible. The dynamic versions are for recursive reconstruction of a time sequence of sparse signals.

Before going further we define the $b\%$-energy support.%{\em clarify the meaning of support for compressible signals}.
\bd[$b\%$-energy support or $b\%$-support]
For sparse signals, clearly the support is $N:=\{i \in [1,\sno]: x_i^2 > 0 \}$. For compressible signals, we misuse notation slightly and let $N$ be the {\em $b\%$-energy support}, i.e. $N:=\{i \in [1,\sno]: x_i^2 > \zeta \}$, where $\zeta$ is the largest real number for which $N$ contains at least $b$\% of the signal energy, e.g. $b=99$ in Fig. \ref{slowchange}.
\ed
%use $N$ to denote the $\gamma$-support, i.e. $N:=\{i \in [1,\sno]: x_i^2 > \gamma \}$. We let $\gamma$ be the largest possible value for which $N$ contains at least $b$\% of the signal energy, e.g. $b=99.9$ in Fig. \ref{slowchange}. We also refer to this as the $b\%$-energy support.%(for an $\alpha>0$) In most places $p=99.9$.

\subsection{Dynamic Modified-CS: Modified-CS for Recursive Reconstruction of Signal Sequences}
\label{dynmodcs}
The most important application of modified-CS is for recursive reconstruction of time sequences of sparse or compressible signals. To apply it to time sequences, at each time $t$, we solve (\ref{l1seqcs}) with $T = \Nhat_{t-1}$ where $\Nhat_{t-1}$ is the support estimate from $t-1$ and is computed using (\ref{supp_est}). At $t=0$ we can either initialize with CS, i.e. set $T$ to be the empty set, or with modified-CS with $T$ being the support available from prior knowledge, e.g. for wavelet sparse images, $T$ could be the set of indices of the approximation coefficients.
The prior knowledge is usually not very accurate and thus at $t=0$ one will usually need more measurements i.e. one will need to use $y_0 = A_0 x_0$ where $A_0$ is an $\mno_0 \times \sno$ measurement matrix with $\mno_0 > \mno$. The full algorithm is summarized in Algorithm \ref{modcsalgo}.%as explained in Sec. \ref{algo} $y_0$ may be obtained from  with $\mno_0 > \mno$

%{\bf Setting the support estimation threshold, $\alpha$. } %  (as mentioned earlier)
{\bf Threshold Selection. }
If $\mno$ is large enough for exact reconstruction, the support estimation threshold, $\alpha$, can be set to zero. In case of very accurate reconstruction, if we set $\alpha$ to be equal/slightly smaller than the magnitude of the smallest element of the support, it will ensure zero misses and fewest false additions. As $\mno$ is reduced further (error increases), $\alpha$ should be increased further to prevent too many false additions. %reconstruction  (if that is roughly known)
For compressible signals, one should do the above but with ``support" replaced by the $b\%$-support. %, i.e. $\alpha$ should be equal/slightly smaller than the magnitude of the smallest element of the $b\%$-support.
For a given $\mno$, $b$ should be chosen to be just large enough so that the elements of the $b\%$-support can be exactly reconstructed.%$\alpha$ should be to be equal/slightly smaller than the magnitude of the smallest element of the $b\%$-energy support.
%Alternatively, here again one can use the above approach.% of \cite{kfcspap}. $\mno$ is large enough for And $b$ should be chosen so that, with the given $\mno$,

Alternatively, one can use the approach proposed in \cite[Section II]{kfcspap}. First, only detect additions to the support using a small threshold (or keep adding largest elements into $T$ and stop when the condition number of $A_T$ becomes too large); then compute an LS estimate on that support and then use this LS estimate to perform support deletion, typically, using a larger threshold. If there are few misses in the support addition step, the LS estimate will have lower error than the output of modified-CS, thus making deletion more accurate.%, even with a larger threshold.%The reason for doing this is that if there are few misses, the just large enough to ensure that $A_T$ is well-conditioned

\begin{algorithm}[h!]
\caption{{\bf \small Dynamic Modified-CS}}% of Sparse/Compressible Signal Sequences  ModCS for Time Sequence Reconstruction
%{\em Use $\alpha > 0$ for compressible signal sequences or for sparse sequences when $\mno$ is not enough for exact reconstruction. Use $\alpha \approx 0$ otherwise.} \\
%Select $\alpha$ as explained earlier in Sec. \ref{algo}. For compressible signals, replace the support by the $99.9\%$ support.
At $t=0$, compute $\hat{x}_{0}$ as the solution of $\min_{\beta}  \|(\beta)_{T^c}\|_1, \ \text{s.t.} \ y_0 = A_0 \beta$, where $T$ is either empty or is available from prior knowledge. Compute $\Nhat_0 = \{i \in [1,\sno] : (\xhat_{0})_i^2 > \alpha \}$.
%
% $\min_{\beta}  \|\beta\|_1, \ \text{s.t.} \ y_0 = A_0 \beta$. Here $A_0$ is an $\mno_0 \times \sno$ measurement matrix with $\mno_0> \mno$. Compute $\Nhat_0 = \{i \in [1,\sno] : (\xhat_{0})_i^2 > \alpha \}$.
%Choose $\mno_0$ large enough s.t. CS achieves exact (small error) reconstruction for sparse (compressible) signal sequences.  do CS, i.e.
For $t>0$, do
\ben

\item {\em Modified-CS. } Let $T = \Nhat_{t-1}$. Compute $\xhat_{t}$ as the solution of $\min_\beta  \|(\beta)_{T^c}\|_1, \ \text{s.t.} \ y_t = A \beta$. %Here $A$ is an $\mno \times \sno$ measurement matrix. Choose $\mno$ large enough s.t. modified-CS achieves exact (small error) reconstruction for sparse (compressible) signal sequences. Typically $\mno$ is much smaller than $\mno_0$.
\label{step1noiseless}

\item {\em Estimate the Support. } $\Nhat_t=\{i \in [1,\sno] : (\xhat_{t})_i^2 > \alpha \}$.

\item Output the reconstruction $\hat{x}_{t}$.
\een
Feedback $\Nhat_t$, increment $t$, and go to step \ref{step1noiseless}.
\label{modcsalgo}
\end{algorithm}

\subsection{RegModCS: Regularized Modified-CS}

%In Algorithm \ref{modcsalgo}, we only use the estimate of the support from the previous time instant to improve reconstruction error at the current time. The next step is to also use knowledge of the reconstructed large coefficients $(\xhat_{t-1})_T$, where $T:=\Nhat_{t-1}$, to further reduce reconstruction error. Due to temporal correlations, these also change slowly over time \cite{kfcspap,kfcsmri}.

So far we only used prior knowledge about the support to reduce the $\mno$ required for exact reconstruction or to reduce the error in cases where exact reconstruction is not possible. If we also know something about how the signal along $T$ was generated, e.g. we know that the elements of $x_T$ were generated from some distribution with mean $\mu_T$, we can use this knowledge\footnote{Because of error in $T$, this knowledge is also not completely correct.} to reduce the reconstruction error by solving%$x_T$ was generated from some distribution with mean $\mu$, we can use this information to reduce the reconstruction error as follows.
\bea
\min_\beta \|(\beta)_{T^c}\|_1 +  \gamma \|(\beta)_T - \mu_T \|_2^2 \ \ \text{s.t.} \ \ y = A \beta
\label{regmodcs}
\eea
We call the above Regularized Modified-CS or RegModCS. Denote its output by $\xhat_{reg}$.

%\subsubsection{Comparing RegModCS with Modified-CS for Sparse Signals} $ from a Gaussian distribution with mean $  step \ref{probcompute} of
We ran a Monte Carlo simulation to compare Modified-CS with RegModCS for sparse signals.
We fixed $\sno = 256$, $\sn=26\approx 0.1\sno$, $\sd = \sde = 0.08s$. We used $\mno=0.16\sno, 0.12\sno, 0.11\sno$ in three sets of simulations done in a fashion similar to that of Sec. \ref{montecarlo}, but with the following change. %The matrix $A$ and in each run, the sets $N,\Delta,\Delta_e$ were generated as in Sec. \ref{montecarlo}. %was run as in step \ref{gendata} of Sec. \ref{montecarlo} with the following change.
In each run of a simulation, we generated each element of $\mu_{N \setminus \Delta}$ to be i.i.d. $\pm 1$ with probability (w.p.) 1/2 and each element of $\mu_\Delta$ and of $\mu_{\Delta_e}$ to be i.i.d. $\pm 0.25$ w.p. 1/2. We generated $x_N \sim \n(\mu_N,0.01I)$ and we set $x_{N^c}=0$. We set $y:=Ax$.  We tested RegModCS with various values of $\gamma$ ($\gamma=0$ corresponds to modified-CS). We used $\text{tot}=50$. The results are tabulated in Table \ref{regmodcscomparetable}. We computed the exact reconstruction probability as in Sec. \ref{montecarlo} by counting the number of times $\xhat_{reg}$ equals $x$ and normalizing. As can be seen, RegModCS does not improve the exact reconstruction probability, in fact it can reduce it. This is primarily because the elements of $(\xhat_{reg})_{\Delta_e}$ are often nonzero, though small\footnote{But if we use $\xhat_{reg}$ to first estimate the support using a small threshold, $\alpha$, and then estimate the signal as ${A_{\Nhat}}^\dag y$, this probability does not decrease as much and in fact it even increases when $\mno$ is smaller.}.
But, it significantly reduces the reconstruction error, particularly when $\mno$ is small.% and if $\gamma$ is chosen appropriately.  by $\text{tot}$
% If $\mno$ is reduced to 12\% or 11\%, neither gives exact reconstruction, but RegModCS has much smaller error than that of Modified-CS.
%the reduction in error for RegModCS is much more significant.% ($\gamma=0.5$ is the best).%We set $\mu_{N^c}=0$. , in fact it reduces it  = T \setminus N  = T \setminus N

\begin{table}[h!]
\caption{\small{Comparing probability of exact reconstruction (prob) and reconstruction error (error) of RegModCS with different $\gamma$'s. $\gamma=0$ corresponds to modified-CS.}}%We used $\sno=256$, $\sn=26$, $\mno=0.16 \sno$, $\sd=0.08\sn$, $\sde=0.08\sn$}}
%\vspace{-3mm}
\begin{center}
\subtable[$\mno = 0.16 \sno$]{
\begin{tabular}{|c|c|c|c|c|c|c|c|c|}%c|}|c
  \hline
  % after \\: \hline or \cline{col1-col2} \cline{col3-col4} ...
  $\gamma$ & 0 {\bf (modCS)} & 0.001 & 0.05  & 0.1 & 0.5 & 1 \\ %& 5\\ 0.001 &
  \hline
  prob & 0.76  & 0.76  & 0.74   & 0.74 & 0.70 & 0.34 \\ %& 0 \\ & 0.8
  \hline
  error & 0.0484   &  0.0469  & 0.0421 &0.0350 & 0.0273 &0.0286 \\ %& 0.0714\\ & 0.0532
  \hline
\end{tabular}
}
\subtable[$\mno = 0.12 \sno$]{
\begin{tabular}{|c|c|c|c|c|c|c|c|c|}%c|}|c
  \hline
  % after \\: \hline or \cline{col1-col2} \cline{col3-col4} ...
  $\gamma$ & 0  {\bf (modCS)} & 1 \\ %& 0.001  & 0.05  & 0.1 & 0.5
  \hline
  prob & 0.04  & 0 \\ % & 0.04    & 0.02 & 0 & 0
  \hline
  error & 0.2027   &0.0791 \\ % & 0.1899  & 0.1593 & 0.1404 &0.0893
  \hline
\end{tabular}
}
\subtable[$\mno = 0.11 \sno$]{
\begin{tabular}{|c|c|c|c|c|c|c|c|c|}%c|}|c
  \hline
  % after \\: \hline or \cline{col1-col2} \cline{col3-col4} ...
  $\gamma$ & 0  {\bf (modCS)}  & 1 \\ %& 0.001  & 0.05  & 0.1 & 0.5
  \hline
  prob & 0  & 0  \\ %& 0  & 0 & 0 & 0
  \hline
  error & 0.3783 & 0.0965 \\ %&   0.3583 &  0.2029& 0.1702& 0.1093
  \hline
\end{tabular}
}
\end{center}
\label{regmodcscomparetable}
\end{table}

%\subsubsection{MAP Interpretation}
\subsection{Setting $\gamma$ using an MAP interpretation of RegModCS}
One way to select $\gamma$ is to interpret the solution of (\ref{regmodcs}) as a maximum a posteriori (MAP) estimate under the following prior model and under the observation model of (\ref{obsmod}).
Given the prior support and signal estimates, $T$ and $\mu_T$, assume that $x_T$ and $x_{T^c}$ are mutually independent and%, and the prior signal knowledge
\bea
p(x_T | T, \mu_T) \se \n(x_T; \mu_T,\sigma_p^2I), \nn \\  %:= \frac{1}{(\sqrt{2\pi}\sigma_p)^{|T|}} e^{-\frac{\|x_T-\mu_T\|_2^2}{2\sigma_p^2}}, \nn \\
 p( x_{T^c} | T, \mu_T) \se  \left(\frac{1}{2b_p} \right)^{|T^c|} e^{-\frac{\|x_{T^c}\|_1}{b_p}},
%iidLap(0,b_p):=  (1/2b_p)^{|T^c|}e^{-\frac{\|x_{T^c}\|_1}{b_p}}  \prod_{i \in T^c} (\frac{1}{2b_p}) e^{-\frac{|x_i|}{b_p}}  =
\label{priormod}
\eea
%Thus, we assume that, given $T,\mu_T$,
i.e. all elements of $x$ are mutually independent; each element of $T^c$ is zero mean Laplace distributed with parameter $b_p$; and the $i^{th}$ element of $T$ is Gaussian with mean $\mu_i$ and variance $\sigma_p^2$. Under the above model, if $\gamma = b_p/2\sigma_p^2$ in (\ref{regmodcs}), then, clearly, its solution, $\xhat_{reg}$, will be an MAP solution. %Conditions under which $\xhat_{reg}$ will be {\em the} unique MAP solution are being studied in ongoing work.% (unique minimizer of (\ref{regmodcs})),  in ongoing work

Given i.i.d. training data, the maximum likelihood estimate (MLE) of $b_p$, $\sigma_p^2$ can be easily computed in closed form.% \cite{falsealarm}.
%In the absence of training data, one could use alternating minimization or ideas similar to those of Sparse Bayesian Learning \cite{sbl} to also estimate $b_p,\sigma_p^2$ just from $y$.%If needed, the same can be done for $T, \mu$ also.% as well as $T, \mu$ alternating minimization or  (and if needed $T, \mu$ also)

%Notice that RegModCS may not (???) reduce the number or measurements, $\mno$, required for exact reconstruction. But, if $\sigma_p^2$ is small enough, and $\gamma$ is appropriately chosen, it should reduce the reconstruction error in situations where exact reconstruction does not occur (i.e. sparse signals and $\mno$ is too small for exact reconstruction or compressible signals).

\subsection{Dynamic Regularized Modified-CS (RegModCS)} %for Recursive Reconstruction of Signal Sequences}
To apply RegModCS to time sequences, we solve (\ref{regmodcs}) with $T = \Nhat_{t-1}$ and $\mu_T = (\xhat_{t-1})_T$. Thus, we use Algorithm \ref{modcsalgo} with step \ref{step1noiseless} replaced by
%In other words, we solve
\bea
\min_\beta \|(\beta)_{\Nhat_{t-1}^c}\|_1 + \gamma \|(\beta)_{\Nhat_{t-1}} - (\xhat_{t-1})_{\Nhat_{t-1}} \|_2^2 \  \text{s.t.} \  y_t = A \beta \ \ \ \
\label{regmodcs2}
\eea
%with $T = \Nhat_{t-1}$.
and in the last step of Algorithm \ref{modcsalgo}, we feed back $\xhat_{t}$ and $\Nhat_t$.%both $\Nhat_t$ and

In Appendix \ref{appendix_algos}, we give the conditions under which the solution of (\ref{regmodcs2}) becomes a causal MAP estimate. %We also give the expression for computing the ML estimates of the parameters of the signal model from a training sequence of signals.
To summarize that discussion, if we set $\gamma = b_p/2\sigma_p^2$ where $b_p, \sigma_p^2$ are the parameters of the signal model given in Appendix \ref{appendix_algos}, and if we assume that the previous signal is perfectly estimated from $y_0,\dots y_{t-1}$ with the estimate being zero outside $\Nhat_{t-1}$ and equal to $(\xhat_{t-1})_{\Nhat_{t-1}}$ on it, then the solution of (\ref{regmodcs2}) will be the causal MAP solution under that model.

In practice, the model parameters are usually not known. But, if we have a training time sequence of signals, we can compute their MLEs using (\ref{mle}), also given in Appendix \ref{appendix_algos}.%. The MLE expression is given in  $[{(\xhat_{t-1})_{\Nhat_{t-1}}}', {0_{{\Nhat_{t-1}}^c}}']'$

\section{Reconstructing Sparsified/True Images from Simulated Measurements} %and for noisy measurements as $\sqrt{\E[\|x_t-\hat{x}_t\|_2^2]/\E[\|x_t\|_2^2]}$,
\label{expts}
We simulated two applications: CS-based image/video compression (or single-pixel camera imaging) and static/dynamic MRI. The measurement matrix was $A=H\Phi$ where $\Phi$ is the sparsity basis of the image  and $H$ models the measurement acquisition. All operations are explained by rewriting the image as a 1D vector. We used $\Phi=W'$ where $W$ is an orthonormal matrix corresponding to a 2D-DWT for a 2-level Daubechies-4 wavelet.
For video compression (or single-pixel imaging), {\em $H$ is a random Gaussian matrix, denoted $G_r$}, (i.i.d. zero mean Gaussian $\mno \times \sno$ matrix with columns normalized to unit $\ell_2$ norm). For MRI, {\em $H$ is a partial Fourier matrix, i.e. $H = MF$} where $M$ is an $\mno \times \sno$ mask which contains a single 1 at a different randomly selected location in each row and all other entries are zero and $F$ is the matrix corresponding to the 2D discrete Fourier transform (DFT).% $H=MF$ is thus a partial Fourier matrix.%commonly referred to as

N-RMSE, defined here as $\|x_t-\hat{x}_t\|_2/\|x_t\|_2$, is used to compare the reconstruction performance. We first used the sparsified and then the true image and then did the same for image sequences. In all cases, the image was sparsified by computing its 2D-DWT, retaining the coefficients from the 99\%-energy support while setting others to zero and taking the inverse DWT. We used the 2-level Daubechies-4 2D-DWT as the sparsifying basis.
We compare modified-CS and RegModCS with simple CS, CS-diff \cite{reddy} and LS-CS \cite{kfcspap}.% (CS at each time instant)

\begin{figure}[t]
\centerline{
\subfigure[$H=G_r$, $\mno_0$=$0.5\sno$, $\mno$=$0.16\sno$]{
\label{timeseqsparse_gauss}
\includegraphics [width=4.5cm,height=3cm]{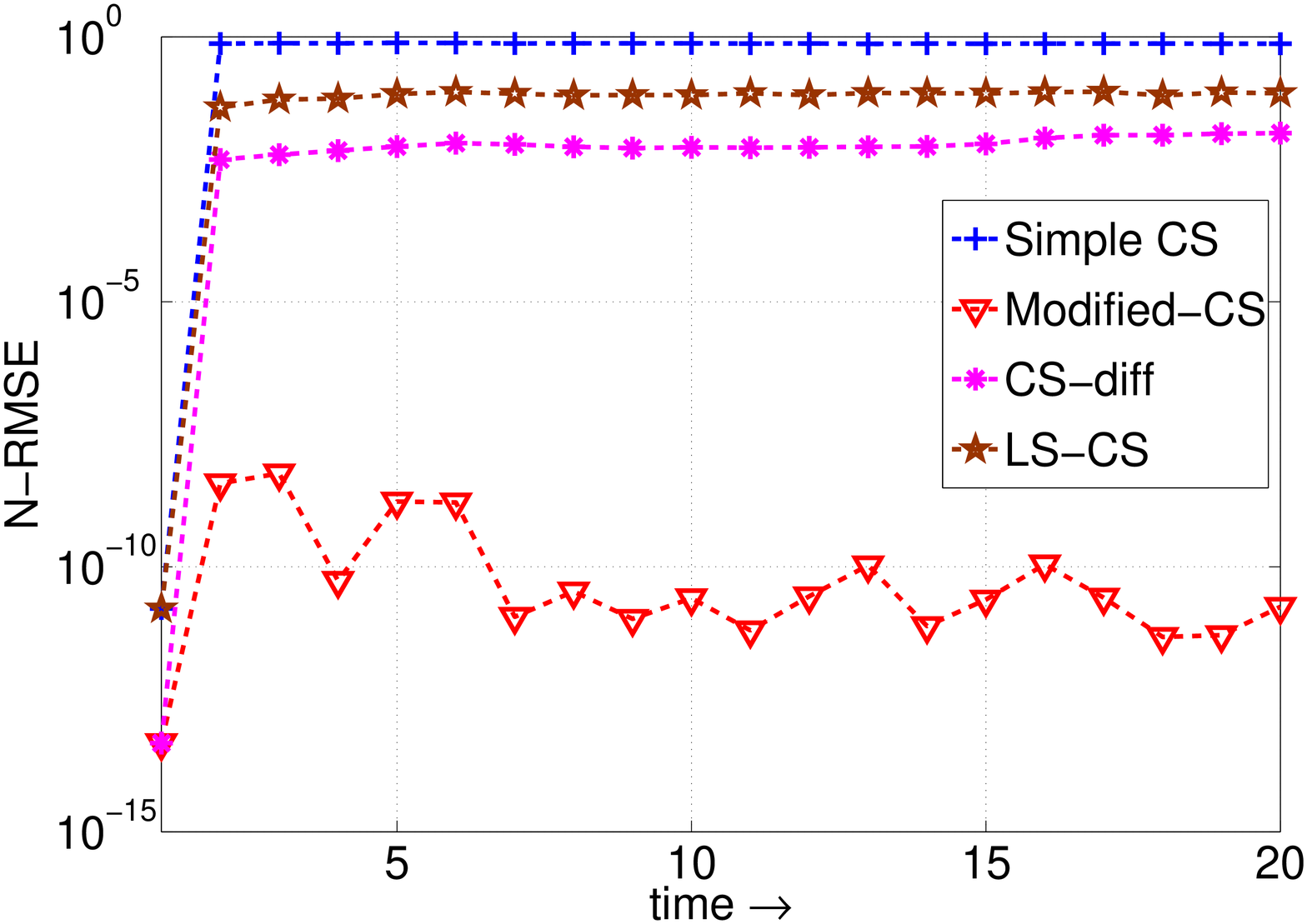}
}
\hspace{-0.15in}
\subfigure[$H=MF$, $\mno_0$=$0.5\sno$, $\mno$=$0.16\sno$]{
\label{timeseqsparse_fourier}
\includegraphics [width=4.5cm,height=3cm]{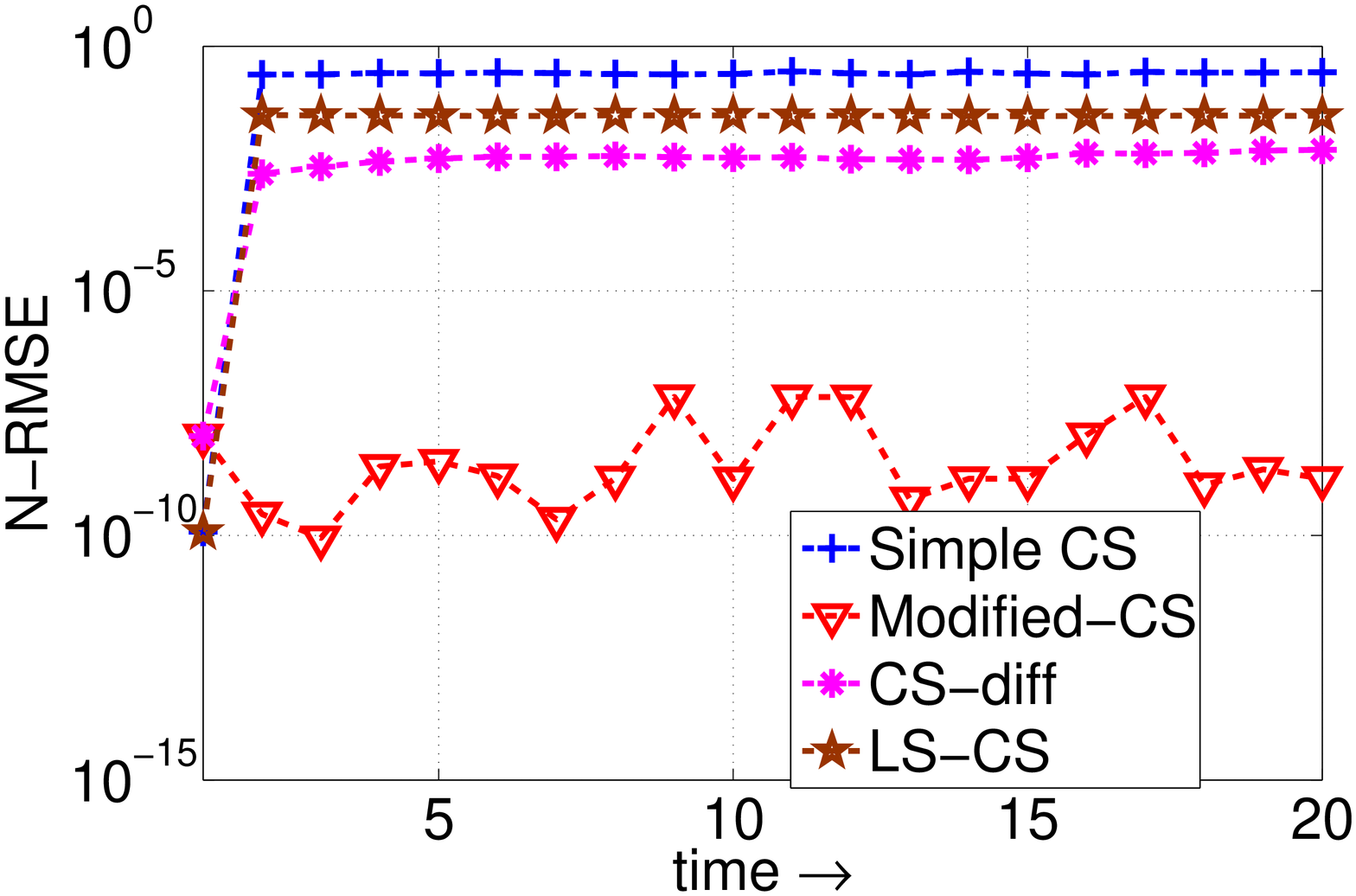}
}
%\subfigure[Noisy meas, $H=G_r$, $\mno_0=0.5\sno$, $\mno=0.16\sno$]{
%\label{timeseqsparse_noisy}
%\includegraphics [width=5.75cm,height=4cm]{cardiac16Gaussiannoisycsmodcs30timesMC.eps}%{cardiac16Gaussiannoisymodcs.eps}
%}
}
\caption{\small{Reconstructing the {\em sparsified} $32 \times 32$ cardiac image sequence.  $\sn \approx 0.1 \sno $, $\sd \approx 0.01\sno$, $\sde \approx 0.005\sno$. (a) $H=G_r$, (b) $H=MF$. Similar results were also obtained for the larynx sequence. These are shown in \cite[Fig. 3]{icip09} (not repeated here due to lack of space).%
%(c) Noisy measurements, Video Compression.
%($H=G_r$, $\mno_0=0.5 \sno$, $\mno=0.16\sno=164$ for $t>1$). (b) MRI ($H=MF$,  $\mno_0=0.5 \sno$, $\mno = 0.16\sno=164$ for $t>1$). (c) Noisy measurements, video compression.
%Simple CS is referred to as CS in the figures.
%Modified-CS achieved exact reconstruction, while simple CS (referred to as CS in the figure) has very large ($85-90\%$) error.
%Modified-CS achieved exact reconstruction, while simple CS (referred to as CS in the figure) has very large ($15-25\%$) error.
%random Gaussian measurements  partial Fourier measurements
}}
%\vspace{-0.2in}
\label{timeseqsparse}
%\label{sparsified}
\end{figure}

For solving the minimization problems given in (\ref{l1seqcs}) and (\ref{regmodcs}), we used CVX, \url{http://www.stanford.edu/~boyd/cvx/}, for smaller sized problems ($\sno < 4096$). All simulations of Sec. \ref{compare} and all results of Table \ref{statictable} and Figs. \ref{timeseqsparse}, \ref{modcscompress} used CVX. For bigger signals/images, (i) the size of the matrix $A$ becomes too large to store on a PC (needed by most existing solvers including the ones in CVX) %\footnote{As suggested by an anonymous reviewer, the Sparco toolbox, \url{http://www.cs.ubc.ca/labs/scl/sparco/} addresses this issue in part.}
and (ii) direct matrix multiplications take too much time. For bigger images and structured matrices like DFT times DWT, we wrote our own solver for (\ref{l1seqcs}) by using a modification of the code in L1Magic \cite{l1magic}. We show results using this code on a $256 \times 256$ larynx image sequence ($\sno=65536$) in Fig. \ref{larynxnoiseless}. This code used the operator form of primal-dual interior point method. With this, one only needs to store the sampling mask which takes $O(\sno)$ bits of storage and one uses FFT and fast DWT to perform matrix-vector multiplications in $O(\sno \log \sno)$ time instead of $O(\sno^2)$ time. In fact for a $b \times b$ image the cost difference is $O(b^2 \log b)$ versus $O(b^4)$.
All our code, for both small and large problems, is posted online at \url{http://www.ece.iastate.edu/~namrata/SequentialCS.html}. This page also links to more experimental results.

 %We show most comparisons for a small block of the image sequence mostly because of computational complexity.

%, while for all other results, we used CVX.
 %As suggested by an anonymous reviewer, the issue of storing a large matrix can also possibly be addressed by using the Sparco toolbox (which stores a random matrix using a single seed, ??WEI - CHECK THIS LAST PART PLEASE?? although from a quick reading we think it contains code only for the standard sparse reconstruction algorithms and does not contain code to solve general LPs which is what we need to implement mod-CS).
%never needs to store the matrix (just

\subsection{Sparsified and True (Compressible) Single Image}
We first evaluated the single image reconstruction problem for a sparsified image. The image used was a $32 \times 32$ cardiac image (obtained by decimating the full $128 \times 128$ cardiac image shown in Fig. \ref{slowchange}), i.e.  $\sno=1024$. %We used the 2-level Daubechies-4 2D-DWT as the sparsifying basis. It was sparsified by retaining the 99\% support of the sparse basis coefficients.
Its support size $\sn=107 \approx 0.1 \sno$. We used the set of indices of the approximation coefficients as the known part of the support, $T$. Thus, $\st=|T|=64$ and so $\sd=|\Delta| \ge 43$ which is a significantly large fraction of $\sn$. We compare the N-RMSE in Table \ref{statictable}. Even with such a large unknown support size, modified-CS achieved exact reconstruction from 29\% random Gaussian and 19\% partial Fourier measurements. CS error in these cases was 34\% and 13\% respectively.%using Gaussian using Fourier measurements.% measurements
% Notice that $195 < 2\sn = 214$, which is the minimum $\mno$ necessary for exact reconstruction using CS for a $\sn$-sparse vector.  and show the reconstructed images in Fig. \ref{static}
%from only $\mno = 0.19\sno=195$ partial Fourier measurements ($H=MF$) and $\mno=0.29\sno=297$ random Gaussian measurements ($H=G_r$)

%\subsubsection{True (Compressible) Single Image}
We also did a comparison for actual cardiac and larynx images (which are only approximately sparse). The results are tabulated in Table \ref{statictable}. Modified-CS works better than CS, though not by much since $|\Delta|$ is a large fraction of $|N|$. Here $N$ refers to the $b\%$ support for any large $b$, e.g. $b=99$.% This is because $\sd \ge 43$ is a significant fraction of $\sn=107$ (size of 99\%-energy support).% In simulations for image sequences, where $\sd$ is small compared to $\sn$, modified-CS did much better than CS even for compressible images (see Figs. \ref{modcscompress},  \ref{larynxnoiseless}). , both for the cardiac and the larynx image shown in Fig. \ref{slowchange}
%In this case, actually decreases since the signal is not strictly sparse and this causes the performance of modified-CS to decay towards CS.

\begin{table}[h]
\centering{
\caption[$\mno=0.29\sno$]{\small{Reconstruction Error (N-RMSE)}}
\label{statictable}
\begin{tabular}{|c|c|c|c|}
  \hline
  % after \\: \hline or \cline{col1-col2} \cline{col3-col4} ...
                                 & Sparsified & True & True \\  %\backslashbox{Algorithm}{Image type}
                                 & Cardiac  & Cardiac & Larynx \\
  \hline
    CS ($H=G_r$, $\mno=0.29\sno$ ) & 0.34 & 0.36 & 0.090 \\
  \hline
    Mod-CS ($H=G_r$, $\mno=0.29\sno$) & 0 & 0.14 & 0.033 \\
  \hline
    CS ($H=MF$, $\mno = 0.19\sno$) & 0.13 & 0.12 & 0.097 \\
  \hline
    Mod-CS ($H=MF$, $\mno = 0.19\sno$) & 0 & 0.11 & 0.025 \\ %0.114 \\
  \hline
\end{tabular}
}
\end{table}

%\begin{table*}[h]
%\centering{
%\caption[$\mno=0.19\sno$]{\small{Reconstruction Error (N-RMSE) for Larynx}}
%\label{staticlarynxtable}
%\begin{tabular}{|c|c|c|c|}
%  \hline
%  % after \\: \hline or \cline{col1-col2} \cline{col3-col4} ...
%  \backslashbox{Algorithm}{Image type} & 99.9\% Sparsified Image & 99\% Sparsified Image & True Image \\
%  \hline
%    CS ($H=G_r$, $\mno=0.19\sno$ ) & 0.848 & 0.842 & 0.849 \\
%  \hline
%    Modified-CS ($H=G_r$, $\mno=0.19\sno$) & 0 & 0 & 0.048 \\
%  \hline
%    CS ($H=MF$, $\mno = 0.19\sno$) & 0.019 & 0.040 & 0.097 \\
%  \hline
%    Modified-CS ($H=MF$, $\mno = 0.19\sno$) & 0.018 & 0.036 & 0.025 \\ %0.114 \\
%  \hline
%\end{tabular}
%}
%\end{table*}

\subsection{Sparsified Image Sequences}
We compared modified-CS with simple CS (CS at each time instant), CS-diff and LS-CS \cite{kfcspap} for the sparsified $32 \times 32$ cardiac sequence in Fig. \ref{timeseqsparse}. Modified-CS was implemented as in Algorithm \ref{modcsalgo}. At $t=0$, the set $T$ was empty and we used 50\% measurements.
For this sequence, $|N_t| \approx 0.1 \sno = 107$, $\sd=|\Delta| \le 10 \approx  0.01\sno$ and $\sde=|\Delta_e| \le 5 \approx 0.005\sno$.  Since $\sd \ll |N_t|$ and $\sde \ll |N_t|$,  modified-CS achieves exact reconstruction with as few as 16\% measurements at $t>0$. Fig. \ref{timeseqsparse_gauss} used $H=G_r$ (compression/single-pixel imaging) and Fig. \ref{timeseqsparse_fourier} used $H=MF$ (MRI). %Both used $\mno_0 = 0.5\sno$ and $m = 0.16 \sno$.
 As can be seen, simple CS has very large error. CS-diff and LS-CS also have significantly nonzero error since the exact sparsity size of both the signal difference and the signal residual is equal to/larger than the signal's sparsity size. {\em Modified-CS error is $10^{-8}$ or less (exact for numerical implementation)}. Similar conclusions were also obtained for the sparsified larynx sequence, see \cite[Fig. 3]{icip09}. This is not repeated here due to lack of space.

%Fig. \ref{timeseqsparse_noisy} was discussed in Sec. \ref{noisycompare}.%

% The error (square root of normalized MSE) for both Gaussian and partial Fourier measurements is plotted in Fig. \ref{timeseqsparse}.
%should use much less measurements than CS requires. We show the reconstruction results in Fig. \ref{timeseqsparse}. From the figures, simple CS  %(referred to as CS in the figure) has very large error for both Gaussian and Fourier measurements while modified-CS gives exact reconstruction. This strongly supports our statement claimed in the above sections.  done at each time instant  at each time instant

 %Modified-CS was implemented as in Algorithm \ref{modcsalgo}. %For RegModCS, we used Algorithm \ref{modcsalgo}, but in step \ref{step1noiseless} we computed $\xhat_{reg,t}$ by solving (\ref{regmodcs2}). %Also, we fed back both $\Nhat_t$ and $(\xhat_{reg,t})_{\Nhat_t}$. %At $t=0$, modified-CS, RegModCS and LS-CS {\em used $T$ to be the set of indices of the approximation coefficients.}

\subsection{True (Compressible) Image Sequences} %compared modified-CS and RegModCS with simple CS, CS-diff and LS-CS
\begin{figure}[t!]
\centerline{
\subfigure[\small{$H$=$G_r$, $\mno_0$=$0.19\sno$, $\mno$=$0.19\sno$}]{%\vspace{-0.3in}
\label{modcscompress_19}
\includegraphics [width=4.65cm,height=3.5cm]{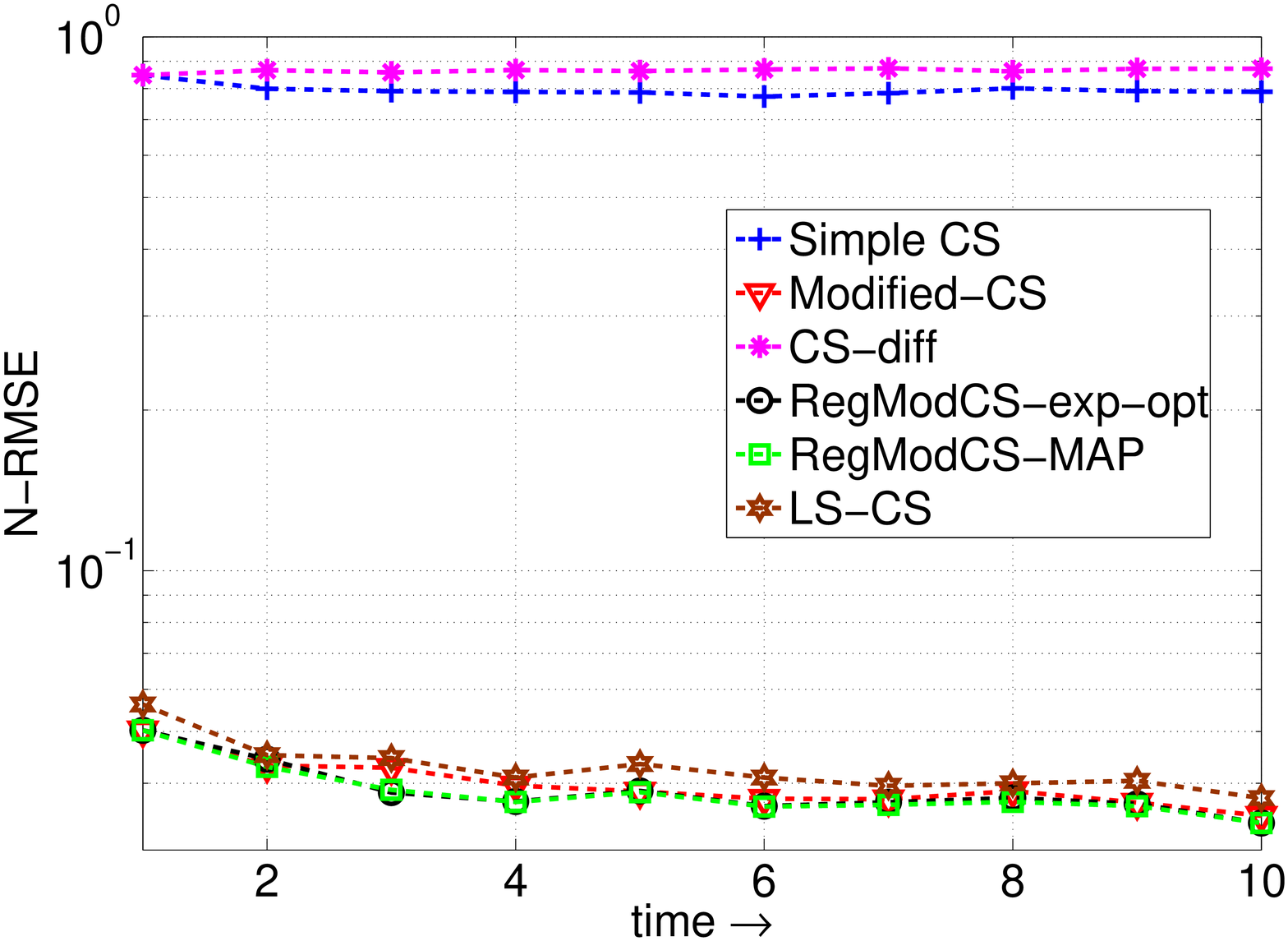}
}
%\subfigure[\small{$H=G_r$, $\mno_0=0.19\sno$, $\mno=0.08\sno$}]{
%\label{modcscompress_8}
%\includegraphics [width=5.75cm,height=4cm]{larynxblock19to8Gaussian7algo_new.eps}
%}
\subfigure[\small{$H$=$G_r$, $\mno_0$=$0.19\sno$, $\mno$=$0.06\sno$}]{
\label{modcscompress_6}
\includegraphics [width=4.65cm,height=3.5cm]{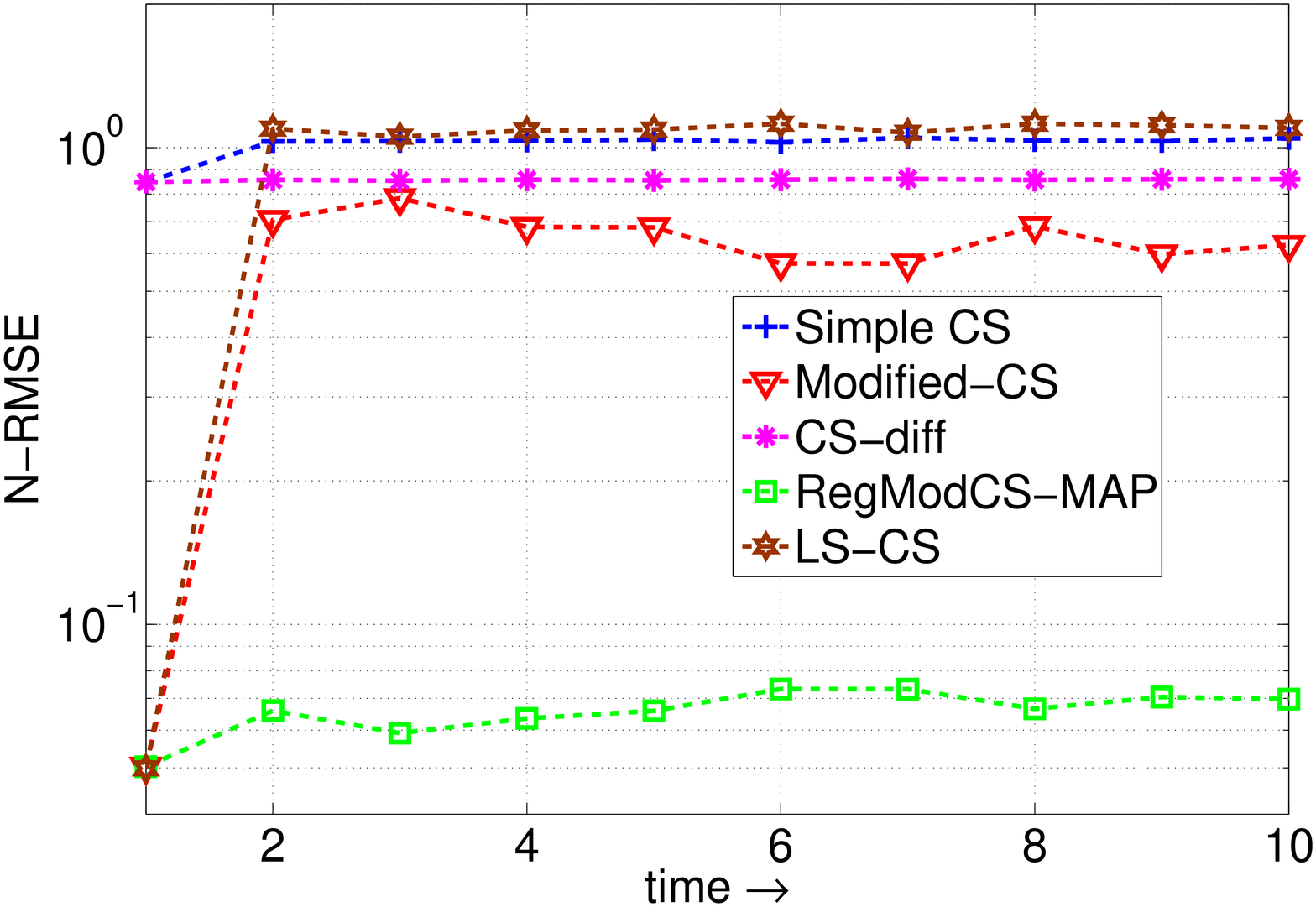}}
}
\caption{\small{Reconstructing a $32 \times 32$ block of the {\em actual (compressible)} larynx sequence from random Gaussian measurements. $\sno=1024$, 99\%-energy support size, $\sn \approx 0.07 \sno$, $\sd \approx 0.001 \sno$ and $\sde \approx 0.002 \sno$. Modified-CS used $\alpha = 50^2$ when $\mno=0.19 \sno$ and increased it to $\alpha = 80^2$ when $\mno= 0.06\sno$.%0.08\sno, shown in Fig. \ref{larynxnoiseless}
}}
\label{modcscompress}
\end{figure}

\begin{figure}[h!]
\centering{
%\subfigure[\small{Reconstruction error}]{
%\includegraphics [width=8cm]{larynxFTWT16obsnoiseless.eps}
%}
\subfigure[\small{Reconstructed sequence. $H$=$MF$. $\mno$=$0.19\sno$,  $\mno_0$=$0.5 \sno$.}]{
\label{larynxnoiseless_recon}
\includegraphics [width=8cm]{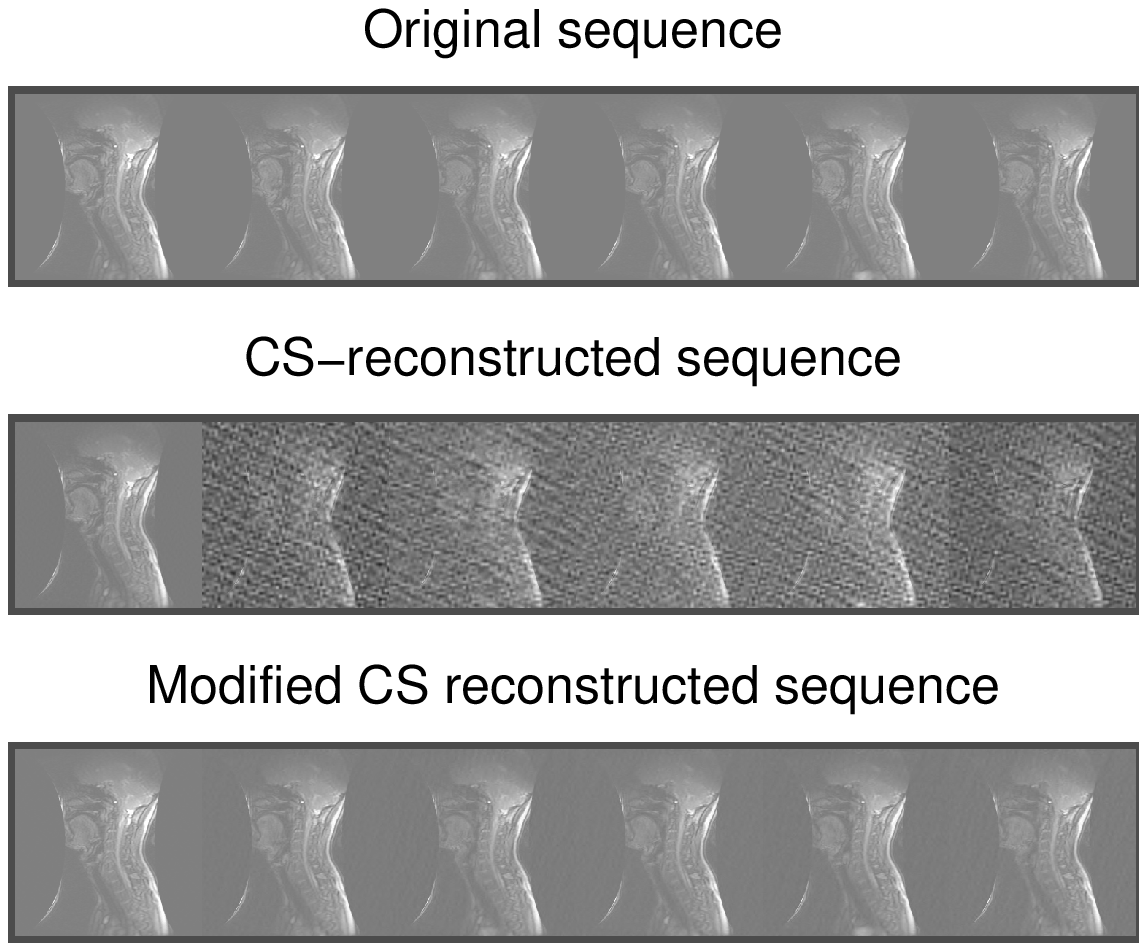}
}
}
\centerline{
\subfigure[\small{$H$=$MF$, $\mno_0$=$0.2\sno$, $\mno$=$0.19\sno$}]{
\label{larynxnoiseless_plots_20}
\includegraphics [width=4.7cm,height=3.1cm]{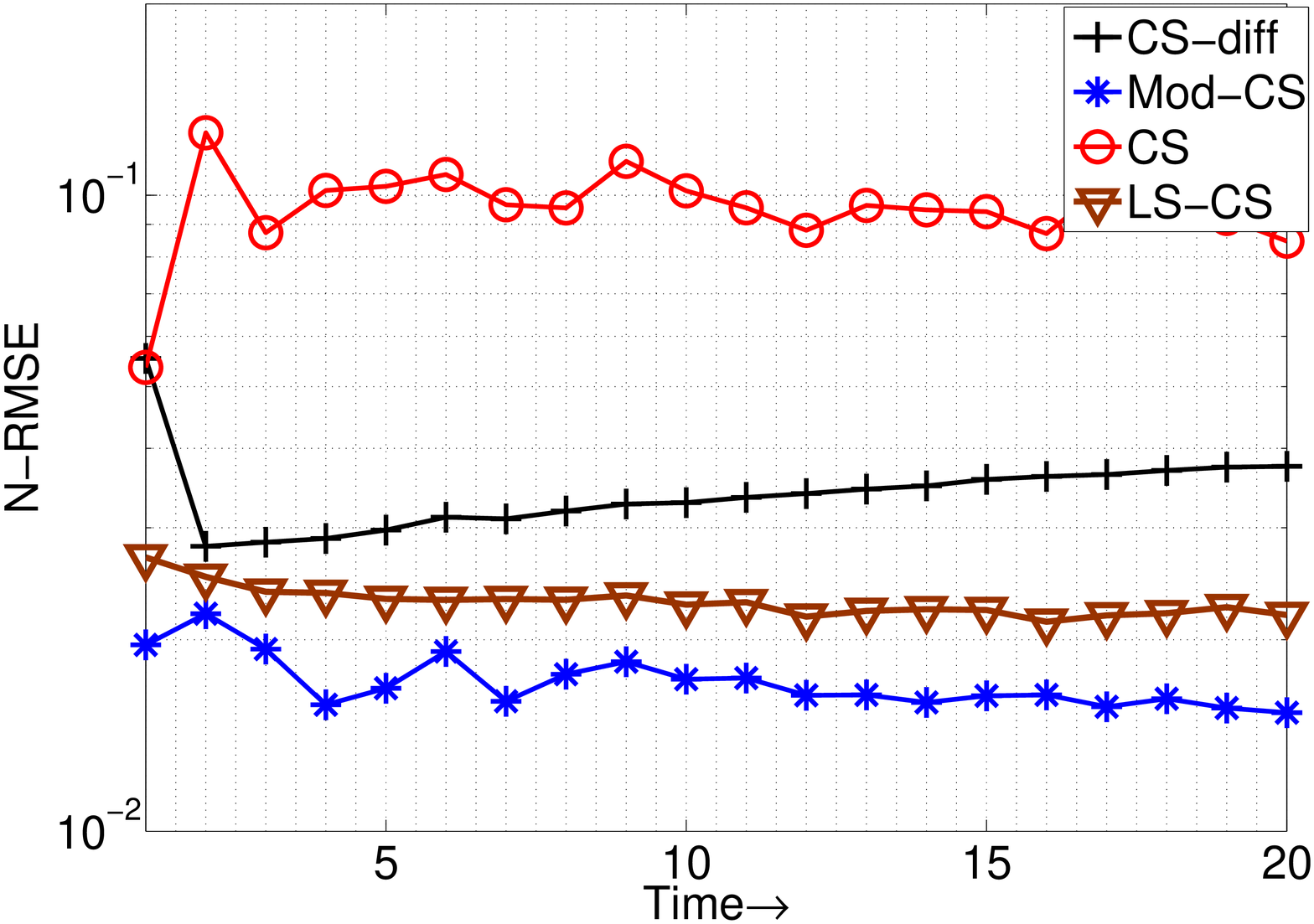} %& \hspace{-0.3in}
}
\hspace{-0.2in}
\subfigure[\small{$H$=$MF$, $\mno_0$=$0.19\sno$, $\mno$=$0.19\sno$}]{
\label{larynxnoiseless_plots_19}
\includegraphics [width=4.7cm,height=3.1cm]{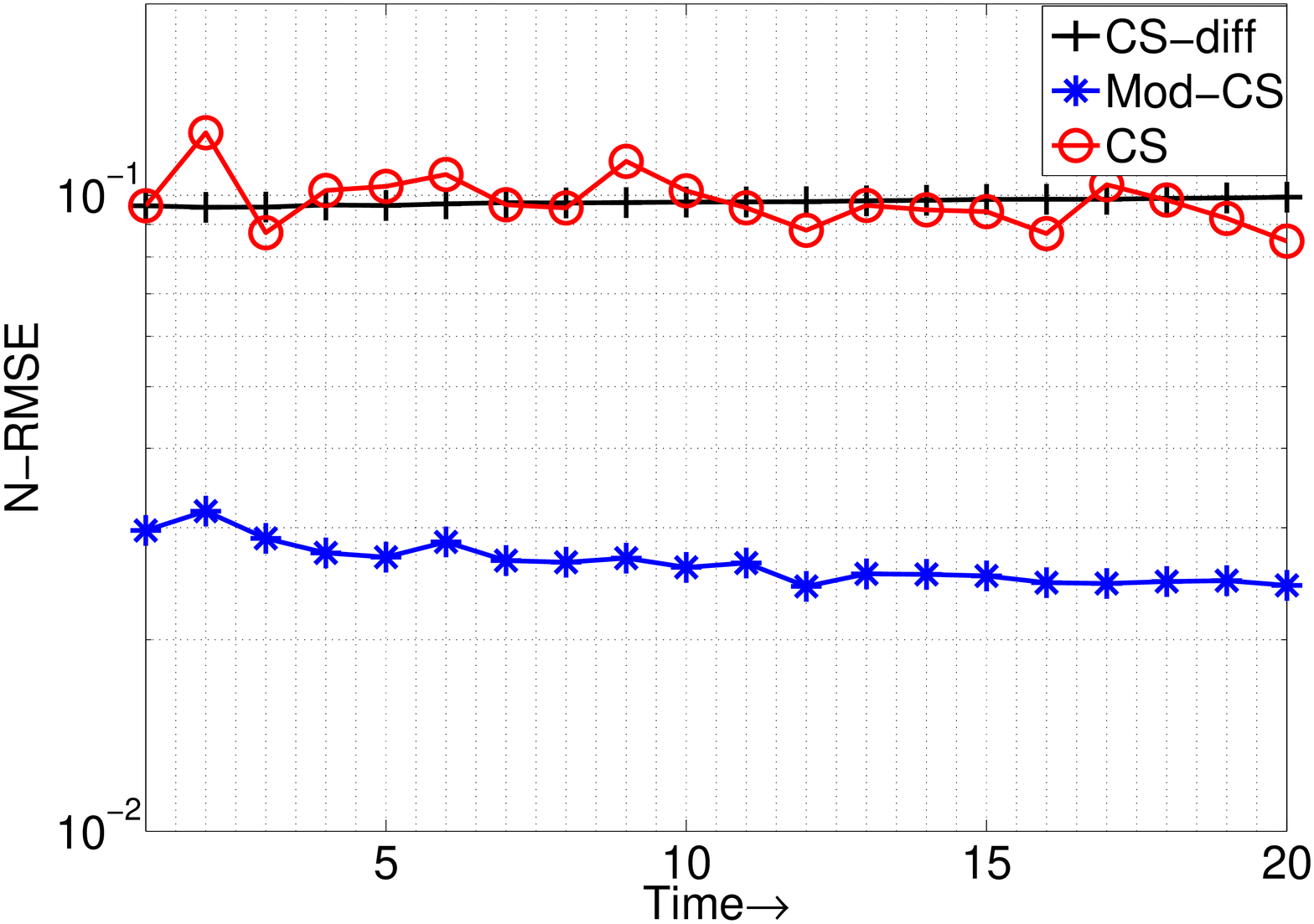}
}
}
\vspace{-0.1in}%At $t=0$, (a) used $\mno_0 = 0.5 \sno$, (b), left used $\mno_0 = 0.2 \sno$ and (b), right used $\mno_0 = 0.19 \sno$.
\caption{\small{Reconstructing the 256x256 {\em actual (compressible)} vocal tract (larynx) image sequence from {\em simulated MRI} measurements, i.e. $H = MF$. All three figures used $\mno=0.19\sno$ for $t>0$ but used different values of $\mno_0$. Image size, $\sno = 256^2=65536$. 99\% energy support, $|N_t| \approx 0.07\sno$; $\sd\approx 0.001\sno$. In Fig. \ref{larynxnoiseless_recon}, modified-CS used $\alpha=10^2$ which is the smallest magnitude element in the 99\% support.%N-RMSE for the sequence shown in (a) was 8-11\% for CS but was stable at 2\% or lesser for modified-CS. N-RMSE of CS-diff (not shown) also started at 2\% but kept increasing slowly.
}}
\vspace{-0.2in}
\label{larynxnoiseless}
\end{figure}

Finally we did the comparison for actual image sequences which are only compressible. We show results on the larynx (vocal tract) image sequence of Fig. \ref{slowchange}. For Fig. \ref{modcscompress}, we used a $32 \times 32$ block of it with random Gaussian measurements. For Fig. \ref{larynxnoiseless} we used the entire $256 \times 256$ image sequence with partial Fourier measurements. {\em At $t=0$, modified-CS, RegModCS and LS-CS used $T$ to be the set of indices of the approximation coefficients.}

%, \ref{modcscompress_8},  0.08\sno,
For the subfigures in Fig. \ref{modcscompress}, we used $H = G_r$ (random Gaussian) and $\mno_0 =0.19 \sno$. Fig. \ref{modcscompress_19} and \ref{modcscompress_6} used $\mno = 0.19 \sno, 0.06\sno$ respectively.
At each $t$, RegModCS-MAP solved (\ref{regmodcs2}) with $b_p,\sigma_p^2$ estimated using (\ref{mle}) from a few frames of the sequence treated as training data. The resulting $\gamma = \hat{b}_p/2\hat{\sigma_p^2}$ was 0.007. RegModCS-exp-opt solved (\ref{regmodcs}) with $T= \Nhat_{t-1}$, $\mu_T = (\xhat_{reg,t-1})_{T}$ and we experimented with many values of $\gamma$ and chose the one which gave the smallest error. Notice from Fig. \ref{modcscompress_19} that RegModCS-MAP gives MSEs which are very close to those of RegModCS-exp-opt. %and thus picking $\gamma = \hat{b}_p/2\hat{\sigma_p^2}$ gives almost the best MSE.

%Notice that {\em modifiedCS and RegModCS significantly outperform CS and CS-diff. In most cases, both also outperform LS-CS.} Secondly,  RegModCS always outperforms all the others, with the difference being largest when $\mno$ is smallest, i.e. in Fig. \ref{modcscompress_6}. Here the advantage of RegModCS over modified-CS is really seen.

Fig. \ref{larynxnoiseless} shows reconstruction of the full larynx sequence using $H = MF$, $\mno = 0.19 \sno$ and three choices of $\mno_0$. In  \ref{larynxnoiseless_recon}, we compare the reconstructed image sequence using modified-CS with that using simple CS. The error (N-RMSE) was 8-11\% for CS, while it was stable at 2\% or lesser for modified-CS. Since $\mno_0$ is large enough for CS to work, the N-RMSE of CS-diff (not shown) also started at a small value of 2\% for the first few frames, but kept increasing slowly over time. In \ref{larynxnoiseless_plots_20}, \ref{larynxnoiseless_plots_19}, we show N-RMSE comparisons with simple CS, CS-diff and LS-CS. In the plot shown, the LS-CS error is close to that of modified-CS because we implemented LS estimation using conjugate gradient and did not allow the solution to converge (forcibly ran it with a reduced number of iterations). Without this tweeking, LS-CS error was much higher, since the computed initial LS estimate itself was inaccurate.%
 %If LS-CS is implemented  has almost the same error was not stable, i.e. it
% and with modified-LS-CS (modified-CS on LS-residual) for smaller values of  $\mno_0$. Modified-LS-CS had the least error.% Similarly one can also develop modified-CS-diff. %At $t=0$, modified-CS, LS-CS and modified-LS-CS used $T$ to be the set of indices of the approximation coefficients.% to improve upon both CS-diff and modified-CS This used $\mno_0 = 0.5 \sno$.

%$|N_t| \approx ??$, $|\Delta| \approx ??$ and $|\Delta_e| \approx ??$.
%
%$\sno=1024$, $99\%$ energy support, $|N_t|\approx 0.1 \sno$, $\sd\approx 0.01\sno$,  $\sde\approx 0.005\sno$.  When $\mno=0.19\sno$ is used for $t>1$, both modified-CS and modified-CS-compressible have similar error which is much lower than that of simple CS. When a smaller $\mno$ is used, modified-CS-compressible has much lower error than that of modified-CS and both are still much lower than simple CS. In all cases, we used  $\mno=0.5\sno$ used for $t=1$.% (referred to as CS in the figure)
%(c) and (d): MRI ($H = MF$). Reconstruction error using  $\mno=0.19\sno=195$ and  $\mno = 0.15\sno=154$ partial Fourier measurements respectively. Same trend as for $H = G_r$.(a) and (b): Video compression ($H = G_r$). Reconstruction error using  $\mno=0.19\sno=195$ and $\mno = 0.12\sno=123$ random Gaussian measurements respectively.

Notice from both Figs. \ref{modcscompress} and \ref{larynxnoiseless}, that {\em modifiedCS and RegModCS significantly outperform CS and CS-diff. In most cases, both also outperform LS-CS.} RegModCS always outperforms all the others, with the difference being largest when $\mno$ is smallest, i.e. in Fig. \ref{modcscompress_6}. %Here the advantage of RegModCS over modified-CS is really seen.
In Figs. \ref{modcscompress} and \ref{larynxnoiseless_plots_19}, CS-diff performs so poorly, in part, because the initial error at $t=0$ is very large (since we use only $\mno_0=0.19\sno$). As a result the difference signal at $t=1$ is not compressible enough, making its error large and so on. But even when $\mno_0$ is larger and the initial error is small, e.g. in Fig. \ref{larynxnoiseless_plots_20}, CS-diff is still the worst and its error still increases over time, though more slowly.%s increases more gradually over time.%is not as large, .% since it is unstable
\section{Conclusions and Future Directions}
\label{conclusions}
We studied the problem of reconstructing a sparse signal from a limited number of its linear projections when the support is partly known (although the known part may contain some errors). Denote the known support by $T$. Modified-CS solves an $\ell_1$ relaxation of the following problem: find the signal that is sparsest outside of $T$ and that satisfies the data constraint. We derived sufficient conditions for exact reconstruction using modified-CS. These are much weaker than those for CS when the sizes of the unknown part of the support and of errors in the known part are small compared to the support size.
An important extension, called RegModCS, was developed that also uses prior signal estimate knowledge. Simulation results showing greatly improved performance of modified-CS and RegModCS using both random Gaussian and partial Fourier measurements were shown. %on both sparse and compressible signals and image sequences.%
%both for a single signal/image and for image sequences are shown  and improves reconstruction error when $\mno$ is too small for exact reconstruction  (or approximately sparse)

The current work does not bound the error either under noisy measurements or for compressible signals or for the TV norm. The former is done in \cite{modcsicassp10,weilu_isit} for modified-CS and RegModCS respectively, and, in parallel, also in \cite{arxiv} for modified-CS. A more important question for recursive reconstruction of signal sequences from noisy measurements, is the stability of the error over time (i.e. how to obtain a time-invariant and small bound on the error over time). This is studied in ongoing work \cite{modcs_stab}. The stability of RegModCS over time is a much more difficult and currently open question. This is due to its dependence on both the previous support and the previous signal estimates.%, i.e. one needs to obtain conditions under which a time-invariant and small error bound holds
%The stability of noisy dynamic modified-CS over time is also being studied \cite{modcs_stab}, while that of RegModCS is an open question. Another related open question is how to obtain weaker conditions under which the dynamic RegModCS estimate is a causal MAP estimate and is {\em unique}.

A key application of our work is for recursive reconstruction of time sequences of (approximately) sparse signals, e.g. for real-time dynamic MRI. As pointed out by an anonymous reviewer, many MRI problems minimize the total variation (TV) norm. The modified-CS idea can be applied easily for the TV norm as follows. Let $T$ contain the set of pixel indices whose spatial gradient magnitude was nonzero at the previous time (or should be nonzero based on some other available prior knowledge). Minimize the TV norm of the image along all pixels not in $T$ subject to the data constraint.
Also, by designing homotopy methods, similar to those in \cite{dynamicl1} for CS, one can efficiently handle sequentially arriving measurements and this can be very useful for MRI applications.%in practice.  for Modified-CS and RegModCS, like  for ModCS or RegModCS
\appendix
Recall that $\st=|T|$, $\sd=|\Delta|$, $\sde=|\Delta_e|$ and $\sn=|N|$.

\subsection{Proof of Proposition \ref{l0exact}}
\label{prop1proof}
The proof follows by contradiction. Suppose that we can find two different solutions $\beta_1$ and $\beta_2$ that satisfy $y = A\beta_1 = A \beta_2$ and have the same $\ell_0$ norm, $\sd$,  along $T^c$. Thus $\beta_1$ is nonzero along $T$ (or a subset of it) and some set $\Delta_1$ of size $\sd$ while $\beta_2$ is nonzero along $T$ (or a subset of it) and some set $\Delta_2$ also of size $\sd$. The sets $\Delta_1$ and $\Delta_2$ may or may not overlap. Thus $A (\beta_1 - \beta_2) = 0$. Since $(\beta_1 - \beta_2)$ is supported on $T \cup \Delta_1 \cup \Delta_2$, this is equivalent to $A_{T \cup \Delta_1 \cup \Delta_2} (\beta_1 - \beta_2)_{T \cup \Delta_1 \cup \Delta_2} = 0$. But if $\delta_{\st+2\sd} < 1$, $A_{T \cup \Delta_1 \cup \Delta_2}$ is full rank and so the only way this can happen is if $\beta_1 - \beta_2 = 0$, i.e $\beta_1 = \beta_2$.

Therefore there can be only one solution with $\ell_0$ norm $\sd$ along $T^c$ that satisfies that data constraint. Since $x$ is one such solution, any other solution has to be equal to $x$. $\blacksquare$

\subsection{Proof of Theorem \ref{thm1}}  % to make the size of the exceptional set $E$ smaller and smaller
\label{thm1proof}
%We use $\iter$ here to denote the iteration number. %It should not be confused with the $\mno$ used in the rest of the paper to denote the number of measurements.

We construct a $w$ that satisfies the conditions of Lemma \ref{wcond} by applying Lemma \ref{wbnd} iteratively as follows and defining $w$ using (\ref{wdef}) below. At iteration zero, we apply Lemma \ref{wbnd} with $T_{d} \equiv \Delta$ (so that $S \equiv \sd$), $c_j \equiv \signumfn(x_j) \ \forall \ j \in \Delta$ (so that $\|c\|_2 = \sqrt{\sd}$), and with $\Sp \equiv \sd$. Lemma \ref{wbnd} can be applied because $\delta_{\sd} + \delta_{\st} + \theta_{\st,\sd}^2  < 1$ (follows from condition \ref{cond1} of the theorem). From Lemma \ref{wbnd}, there exists a $w_1$ and an exceptional set $T_{d,1}$, disjoint with $T \cup \Delta$, of size less than $\Sp=\sd$, s.t.
\bea
{A_j}' w_1 \se 0,  \ \forall  \ j \in T \nn \\
{A_j}' w_1 \se \signumfn(x_j), \ \forall \ j \in \Delta \nn \\
|T_{d,1}| &<&  \sd \nn \\
\|{A_{T_{d,1}}}'w_1\|_2 \sle a_{\st}(\sd,\sd) \sqrt{\sd}  \nn \\
|{A_j}'w_1| \sle  a_{\st}(\sd,\sd),  \ \forall j \notin T \cup \Delta \cup T_{d,1} \nn \\
\|w_1\|_2 \sle  K_{\st}(\sd) \sqrt{\sd}
%\frac{a(\sd,\sd,\st)}{\sqrt{\sd}} \sqrt{\sd}
\label{iter_0}
\eea
At iteration $\iter$, apply Lemma \ref{wbnd} with $T_{d} \equiv \Delta \cup T_{d,\iter}$ (so that $S \equiv 2\sd$), $c_j \equiv 0 \ \forall \ j \in \Delta$, $c_j \equiv {A_j}' w_\iter \ \forall \ j \in T_{d,\iter}$ and $\Sp \equiv \sd$. Call the exceptional set $T_{d,\iter+1}$.
Lemma \ref{wbnd} can be applied because $\delta_{2\sd} + \delta_{\st} + \theta_{\st,2\sd}^2 < 1$  (condition \ref{cond1} of the theorem).
 From Lemma \ref{wbnd}, there exists a $w_{\iter+1}$ and an exceptional set $T_{d,\iter+1}$, disjoint with $T \cup  \Delta \cup T_{d,\iter}$, of size less than $\Sp=\sd$, s.t.
\bea
{A_j}' w_{\iter+1} \se 0  \ \forall \ j \in T \nn \\
{A_j}' w_{\iter+1} \se 0, \ \forall \ j \in \Delta \nn \\
{A_j}' w_{\iter+1} \se {A_j}' w_\iter, \ \forall \ j \in T_{d,\iter} \nn \\
|T_{d,\iter+1}| &<& \sd \nn \\
\|{A_{T_{d,\iter+1}}}'w_{\iter+1}\|_2 \sle a_{\st}(2\sd,\sd)\|{A_{T_{d,\iter}}}'w_\iter\|_2  \nn \\
|{A_j}'w_{\iter+1}| \sle \frac{a_{\st}(2\sd,\sd)}{\sqrt{\sd}} \|{A_{T_{d,\iter}}}'w_\iter\|_2  \nn \\ && \forall j \notin T \cup \Delta \cup T_{d,\iter} \cup T_{d,\iter+1} \nn \\
\|w_{\iter+1}\|_2 \sle  K_{\st}(2\sd) \|{A_{T_{d,\iter}}}'w_\iter\|_2
\label{iter_n}
\eea
Notice that $|T_{d,1}| < \sd$ (at iteration zero) and $|T_{d,\iter+1}| < \sd$ (at iteration $\iter$) ensures that $|\Delta \cup T_{d,\iter}| < S=2\sd$ for all $\iter \ge 1$.%

The last three equations of (\ref{iter_n}), combined with the fourth equation of (\ref{iter_0}), simplify to%Assume that $a(2\sd,\sd,\st) \le 1$. Then t
\bea
\|{A_{T_{d,\iter+1}}}'w_{\iter+1}\|_2 \sle a_{\st}(2\sd,\sd)^\iter a_{\st}(\sd,\sd) \sqrt{\sd} \nn \\
\label{iter_n_2}
|{A_j}'w_{\iter+1}| \sle a_{\st}(2\sd,\sd)^\iter a_{\st}(\sd,\sd),  \nn \\ && \forall j \notin T \cup \Delta \cup T_{d,\iter} \cup T_{d,\iter+1} \\
\|w_{\iter+1}\|_2 \sle  K_{\st}(2\sd) a_{\st}(2\sd,\sd)^{\iter-1} a_{\st}(\sd,\sd) \sqrt{\sd} \nn \\
\label{wnbnd}
%K(2\sd,\st)^n K(\sd,\st) \sqrt{\sd}
\eea
We can define%Since $a_{\st}(2\sd,\sd) < 1$, w
\bea
w : = \sum_{\iter=1}^\infty (-1)^{\iter-1} w_\iter
\label{wdef}
\eea
Since $a_{\st}(2\sd,\sd) < 1$, $\|w_\iter\|_2$ approaches zero with $\iter$, and so the above summation is absolutely convergent, i.e. $w$ is well-defined. %(follows from condition \ref{cond2} of the theorem)

From the first two equations of (\ref{iter_0}) and (\ref{iter_n}),
\bea
{A_j}' w \se  0,  \ \forall  \ j \in T \nn \\
{A_j}' w \se {A_j}' w_1 =  \signumfn(x_j), \ \forall \ j \in \Delta
\label{wcond12}
\eea
Consider ${A_j}' w = {A_j}' \sum_{\iter=1}^\infty (-1)^{\iter-1} w_\iter$ for some $j \notin T \cup \Delta$. If for a given $\iter$, $j \in T_{d,\iter}$, then ${A_j}' w_{\iter} = {A_j}' w_{\iter+1}$ (gets canceled by the $\iter+1^{th}$ term). If $j \in T_{d,\iter-1}$, then ${A_j}' w_{\iter} = {A_j}' w_{\iter-1}$ (gets canceled by the $\iter-1^{th}$ term). Since $T_{d,\iter}$ and $T_{d,\iter-1}$ are disjoint, $j$ cannot belong to both of them.
%If for some other $\tilde{\iter}$, $j \in T_{d,\tilde{\iter}-1}$, then ${A_j}' w_{\tilde{\iter}} = {A_j}' w_{\tilde{\iter}-1}$ (gets canceled by the $\tilde{\iter}-1^{th}$ term). Also, since $T_{d,\iter}$ and $T_{d,\iter-1}$ are disjoint, $j$ cannot belong to both of them.
Thus,
\bea
{A_j}' w =  \sum_{\iter: j \notin T_{d,\iter} \cup T_{d,\iter-1} } (-1)^{\iter-1} {A_j}'w_\iter , \ \forall  j \notin T \cup \Delta
\eea
Consider a given $\iter$ in the above summation. Since $j \notin T_{d,\iter} \cup T_{d,\iter-1} \cup T \cup \Delta$, we can use (\ref{iter_n_2}) to get $|{A_j}'w_\iter| \le  a_{\st}(2\sd,\sd)^{\iter-1} a_{\st}(\sd,\sd)$. Thus, for all $j \notin T \cup \Delta$,%the second-last equation of
\bea
| {A_j}' w | \sle \sum_{\iter: j \notin T_{d,\iter} \cup T_{d,\iter-1} } a_{\st}(2\sd,\sd)^{\iter-1} a_{\st}(\sd,\sd) \nn \\
%\ \ \ \ \ \
%, \  j \notin T \cup \Delta
%\eea
%Since  $a_{\st}(2\sd,\sd) < 1$ (follows from condition \ref{cond2} of the theorem), this simplifies to
%\bea | {A_j}' w |
\sle \frac{a_{\st}(\sd,\sd)}{1 - a_{\st}(2\sd,\sd)} %, \ \forall  j \notin T \cup \Delta
\eea
%The second inequality holds because $a_{\st}(2\sd,\sd) < 1$.
Since $a_{\st}(2\sd,\sd) + a_{\st}(\sd,\sd) < 1$ (condition \ref{cond2} of the theorem),
\bea
| {A_j}' w | < 1, \ \forall  j \notin T \cup \Delta
\label{wcond3}
\eea
Thus, from (\ref{wcond12}) and (\ref{wcond3}),  we have found a $w$ that satisfies the conditions of Lemma \ref{wcond}. From condition \ref{cond1} of the theorem, $\delta_{\st+\sd} < 1$. Applying Lemma \ref{wcond}, the claim follows. $\blacksquare$% exact reconstruction
%if $a(2\sd,\sd,\st) + a(\sd,\sd,\st) < 1$ then,   and using $\delta_{S+\Sp} \le \theta_{S+\Sp} + \delta_{\max(S,\Sp)}$ \cite{decodinglp} 
%%\item the observation model is $y_t=Ax_t$ and $y_t$ is conditionally independent of $y_{t-1}, y_{t-2} \dots y_0$ given $x_t$,
%%and $iidLap(\cdot,\cdot)$ are
%\label{yxmarkov}
%
%\item the random process $x_t$ satisfies the Markov chain property ($x_t$ is conditionally independent of the past given $x_{t-1}$),
%\label{xtmarkov}

%, i.e.
%\ben
%\item $x_t$ is conditionally independent of $x_{t-1}, x_{t-2}, \dots x_0$, given $x_{t-1}$, and
%\item $y_t$ is conditionally independent of $x_{t-1}, x_{t-2} \dots x_0$, given $x_t$% $y_{t-1}, y_{t-2} \dots y_0$,
%\een

\subsection{Causal MAP Interpretation of Dynamic RegModCS}
\label{appendix_algos}

The solution of (\ref{regmodcs2}) becomes a causal MAP estimate under the following assumptions. Let $p(X|Y)$ denote the conditional PDF of $X$ of given $Y$ and let $\delta(X)$ denote the Dirac delta function.
%Also, let $N_t$ be the support or $\alpha$-support of $x_t$, i.e.
%\bea
%N_t:=\{i : |(x_t)_i| > \alpha \}
%\eea
%with $\alpha \ge 0$.
Assume that
\ben
\item the random processes $\{x_t\}, \{y_t\}$ satisfy the hidden Markov model property;
$p(y_t|x_t) = \delta(y_t - A x_t)$ (re-statement of the observation model); and
%$p(x_t|x_{t-1}) = p( (x_t)_{N_{t-1}} | x_{t-1} ) p( (x_t)_{N_{t-1}^c} | x_{t-1} )$ with
\bea
&& p(x_t|x_{t-1}) = p( (x_t)_{N_{t-1}} | x_{t-1} ) p( (x_t)_{N_{t-1}^c} | x_{t-1} ), \text{where} \nn \\
&& p( (x_t)_{N_{t-1}} | x_{t-1} ) = \n( (x_t)_{N_{t-1}}; (x_{t-1})_{N_{t-1}}, \sigma_p^2 I) \nn \\
&& p( (x_t)_{N_{t-1}^c} | x_{t-1} ) = \left(\frac{1}{2b_p} \right)^{|N_{t-1}^c|} \exp\left({-\frac{\|(x_t)_{N_{t-1}^c}\|_1}{b_p}}  \right) \nn
%iidLap(0,b_p)
\eea
%and $\n(\cdot; \cdot,\cdot)$ is defined in (\ref{priormod}).
i.e. given $x_{t-1}$ (and hence given $N_{t-1}$), $(x_{t})_{N_{t-1}}$ and $(x_{t})_{N_{t-1}^c}$ are conditionally independent; $(x_{t})_{N_{t-1}}$ is Gaussian with mean $(x_{t-1})_{N_{t-1}}$  while $(x_{t})_{N_{t-1}^c}$ is zero mean Laplace.%
%(conditional independence given $x_{t-1}$)
%Note that $p(y_t|x_t)$ is a re-statement of the observation model $y_t = Ax_t$ while the expression for $p(x_t|x_{t-1})$ indicates that
%The observation likelihood follows from (\ref{obsmod}).
% (follows from the observation model given in (\ref{obsmod}))
% (given $x_{t-1}$, $(x_{t})_{N_{t-1}}$ and $(x_{t})_{N_{t-1}^c}$ are conditionally independent)

\item $x_{t-1}$ is perfectly estimated from $y_0,y_1, \dots y_{t-1}$, and
\bea
p(x_{t-1}|y_{0}, \dots y_{t-1}) = \delta\left(x_{t-1} - \vect{(\xhat_{t-1})_{\Nhat_{t-1}}}{0_{\Nhat_{t-1}^c}}\right) \nn %\ T:=\Nhat_{t-1} \ \ \nn
\eea

\item $\xhat_{t}$ is the solution of (\ref{regmodcs2}) with $\gamma = \frac{b_p}{2\sigma_p^2}$.%  $T = \Nhat_{t-1}$

\een

If the first two assumptions above hold, it is easy to see that the ``causal posterior" at time $t$, $p(x_t|y_{1},\dots y_t)$, satisfies%
\bea
p(x_t|y_{1},\dots y_t) = C \delta(y_t - Ax_t) e^{-\frac{\|(x_t)_T - (\xhat_{t-1})_T\|_2^2}{2\sigma_p^2}} \nn e^{-\frac{\|(x_t)_{T^c}\|_1}{b_p}} \ \ \
\eea
where $T:=\Nhat_{t-1}$ and $C$ is the normalizing constant. Clearly, the second assumption is only an approximation since it assumes that the posterior estimate of $x_{t-1}$ is exactly sparse.

If the last assumption also holds, then  the solution of (\ref{regmodcs2}) is a maximizer of $p(x_t|y_{1},\dots y_t)$, i.e. it is a causal MAP solution.
%, $(C)^{-1} = \int_{\beta:y_t=A\beta} e^{-\frac{\|(\beta)_T - (\xhat)_T\|_2^2}{2\sigma_p^2}} e^{-\frac{\|(\beta)_{T^c}\|_1}{b_p}} d\beta$.%An open question is under what conditions will $\xhat_t$ will be {\em the} unique posterior mode at all times?   at time $t$

The MLE of  $b_p, \sigma_p^2$ can be computed from a training time sequence of signals, $\tx_0, \tx_1,\tx_2, \dots \tx_{t_{\max}}$ as follows. Denote their supports ($b\%$-energy supports in case of compressible signal sequences) by $\tN_0, \tN_1, \dots \tN_{t_{\max}}$. Then the MLE is
\bea
\hat{b}_p \se  \frac{\sum_{t=1}^{t_{\max}} \|(\tx_t)_{\tN_{t-1}^c} \|_1} {\sum_{t=1}^{t_{\max}}|\tN_{t-1}^c|}, \nn \\
\hat{\sigma_p^2} \se  \frac{\sum_{t=1}^{t_{\max}} \| (\tx_t-\tx_{t-1})_{\tN_{t-1}} \|_2^2} {\sum_{t=1}^{t_{\max}} |\tN_{t-1}|} %\ \ \ \ \ \
\label{mle}
\eea

\bibliographystyle{IEEEtran}
%\bibliography{tipnewpfmt_kfcsfullpap,tipnewpfmt_seqcs,tipnewpfmt,visual-tracking-bib,levelsets-bib,occlusion-bib,tip,tipoldcareer,books}
\bibliography{tipnewpfmt_kfcsfullpap,tipnewpfmt_seqcs,tipnewpfmt,visual-tracking-bib,levelsets-bib,occlusion-bib,tip,tipoldcareer,books}

\begin{thebibliography}{10}
\providecommand{\url}[1]{#1}
\csname url@rmstyle\endcsname
\providecommand{\newblock}{\relax}
\providecommand{\bibinfo}[2]{#2}
\providecommand\BIBentrySTDinterwordspacing{\spaceskip=0pt\relax}
\providecommand\BIBentryALTinterwordstretchfactor{4}
\providecommand\BIBentryALTinterwordspacing{\spaceskip=\fontdimen2\font plus
\BIBentryALTinterwordstretchfactor\fontdimen3\font minus
  \fontdimen4\font\relax}
\providecommand\BIBforeignlanguage[2]{{%
\expandafter\ifx\csname l@#1\endcsname\relax
\typeout{** WARNING: IEEEtran.bst: No hyphenation pattern has been}%
\typeout{** loaded for the language `#1'. Using the pattern for}%
\typeout{** the default language instead.}%
\else
\language=\csname l@#1\endcsname
\fi
#2}}

\bibitem{isit09}
N.~Vaswani and W.~Lu, ``Modified-cs: Modifying compressive sensing for problems
  with partially known support,'' in \emph{IEEE Intl. Symp. Info. Theory
  (ISIT)}, June 2009.

\bibitem{icip09}
W.~Lu and N.~Vaswani, ``Modified compressive sensing for real-time dynamic mr
  imaging,'' in \emph{IEEE Intl. Conf. Image Proc. (ICIP)}, 2009.

\bibitem{igorcarron}
I.~Carron, ``Nuit blanche,'' in \emph{\url{http://nuit-blanche.blogspot.com/}}.

\bibitem{rice}
``Rice compressive sensing resources,'' in
  \emph{\url{http://www-dsp.rice.edu/cs}}.

\bibitem{bpdn}
S.~Chen, D.~Donoho, and M.~Saunders, ``Atomic decomposition by basis pursuit,''
  \emph{SIAM Journal of Scientific Computation}, vol.~20, pp. 33–--61, 1998.

\bibitem{sbl}
D.~Wipf and B.~Rao, ``Sparse bayesian learning for basis selection,''
  \emph{IEEE Trans. Sig. Proc.}, vol.~52, pp. 2153--2164, Aug 2004.

\bibitem{candes}
E.~Candes, J.~Romberg, and T.~Tao, ``Robust uncertainty principles: Exact
  signal reconstruction from highly incomplete frequency information,''
  \emph{IEEE Trans. Info. Th.}, vol. 52(2), pp. 489--509, February 2006.

\bibitem{donoho}
D.~Donoho, ``Compressed sensing,'' \emph{IEEE Trans. on Information Theory},
  vol. 52(4), pp. 1289--1306, April 2006.

\bibitem{decodinglp}
E.~Candes and T.~Tao, ``Decoding by linear programming,'' \emph{IEEE Trans.
  Info. Th.}, vol. 51(12), pp. 4203 -- 4215, Dec. 2005.

\bibitem{tropp}
J.~A. Tropp, ``Just relax: Convex programming methods for identifying sparse
  signals,'' \emph{IEEE Trans. Info. Th.}, pp. 1030--1051, March 2006.

\bibitem{dantzig}
E.~Candes and T.~Tao, ``The dantzig selector: statistical estimation when p is
  much larger than n,'' \emph{Annals of Statistics}, 2006.

\bibitem{kfcsicip}
N.~Vaswani, ``Kalman filtered compressed sensing,'' in \emph{IEEE Intl. Conf.
  Image Proc. (ICIP)}, 2008.

\bibitem{kfcspap}
------, ``Ls-cs-residual (ls-cs): Compressive sensing on the least squares
  residual,'' \emph{IEEE Trans. Sig. Proc.}, vol. 58 (8), pp. 4108--4120,
  August 2010.

\bibitem{hassibi}
A.~Khajehnejad, W.~Xu, A.~Avestimehr, and B.~Hassibi, ``Weighted l1
  minimization for sparse recovery with prior information,'' in \emph{IEEE
  Intl. Symp. Info. Theory (ISIT)}, June 2009.

\bibitem{camsap07}
C.~J. Miosso, R.~von Borries, M.~Argez, L.~Valazquez, C.~Quintero, and
  C.~Potes, ``Compressive sensing reconstruction with prior information by
  iteratively reweighted least-squares,'' \emph{IEEE Trans. Sig. Proc.}, vol.
  57 (6), pp. 2424--2431, June 2009.

\bibitem{reclasso}
D.~Angelosante and G.~Giannakis, ``Rls-weighted lasso for adaptive estimation
  of sparse signals,'' in \emph{IEEE Intl. Conf. Acoustics, Speech, Sig. Proc.
  (ICASSP)}, 2009.

\bibitem{dynamicl1}
M.~Asif and J.~Romberg, ``Dynamic updating for sparse time varying signals,''
  in \emph{CISS}, 2009.

\bibitem{reddy}
V.~Cevher, A.~Sankaranarayanan, M.~Duarte, D.~Reddy, R.~Baraniuk, and
  R.~Chellappa, ``Compressive sensing for background subtraction,'' in
  \emph{Eur. Conf. on Comp. Vis. (ECCV)}, 2008.

\bibitem{multiscaleCS}
J.~Park and M.~Wakin, ``A multiscale framework for compressive sensing of
  video,'' in \emph{Picture Coding Symposium (PCS)}, May 2009.

\bibitem{natarajan}
B.~K. Natarajan, ``Sparse approximate solutions to linear systems,'' \emph{SIAM
  J. Comput.}, vol.~24, pp. 227--234, 1995.

\bibitem{boyd}
S.~Boyd and L.~Vandenberghe, \emph{Convex Optimization}.\hskip 1em plus 0.5em
  minus 0.4em\relax Cambridge University Press, 2004.

\bibitem{candes_rip}
E.~Candes, ``The restricted isometry property and its implications for
  compressed sensing,'' \emph{Compte Rendus de l'Academie des Sciences, Paris,
  Serie I}, pp. 589--592, 2008.

\bibitem{foucart_lai}
S.~Foucart and M.~J. Lai, ``Sparsest solutions of underdetermined linear
  systems via ell-q-minimization for 0 <= q <= 1,'' \emph{Applied and
  Computational Harmonic Analysis}, vol.~26, pp. 395--407, 2009.

\bibitem{cosamp}
D.~Needell and J.~Tropp., ``Cosamp: Iterative signal recovery from incomplete
  and inaccurate samples,'' \emph{Appl. Comp. Harmonic Anal.}, To Appear.

\bibitem{RIPsimpleproof}
R.~Baraniuk, M.~Davenport, R.~DeVore, and M.~Wakin, ``A simple proof of the
  restricted isometry property for random matrices,'' \emph{Constructive
  Approximation}, vol. 28(3), pp. 253--263, Dec 2008.

\bibitem{dossal_peyre}
C.~Dossal, G.~Peyre, and J.~Fadili, ``A numerical exploration of compressed
  sampling recovery,'' in \emph{Signal Processing with Adaptive Sparse
  Structured Representations (SPARS)}, 2009.

\bibitem{dossal}
C.~Dossal, ``A necessary and sufficient condition for exact recovery by l1
  minimization,'' in \emph{Preprint}, 2007.

\bibitem{modcsicassp10}
W.~Lu and N.~Vaswani, ``Modified bpdn for noisy compressive sensing with
  partially known support,'' in \emph{IEEE Intl. Conf. Acoustics, Speech, Sig.
  Proc. (ICASSP)}, 2010.

\bibitem{arxiv}
L.~Jacques, ``A short note on compressed sensing with partially known signal
  support,'' \emph{ArXiv preprint 0908.0660}, 2009.

\bibitem{l1magic}
E.~Candes and J.~Romberg, ``{L1 Magic Users Guide},'' October 2005.

\bibitem{weilu_isit}
W.~Lu and N.~Vaswani, ``Regularized modified bpdn for compressive sensing with
  partially known support,'' in \emph{ArXiv preprint}, 2010.

\bibitem{modcs_stab}
N.~Vaswani, ``Stability (over time) of modified-cs and ls-cs for recursive
  causal sparse reconstruction,'' in \emph{ArXiv preprint arXiv:1006.4818},
  2010.

\end{thebibliography}

\section*{Biography}

%{\bf Namrata Vaswani} received a B.Tech. degree from the Indian Institute of Technology (IIT), Delhi, in 1999 and a Ph.D. degree from the University of Maryland, College Park, in 2004, both in electrical engineering. She spent a year as a postdoc and research scientist at Georgia Tech.  Since Fall 2005, she has been an Assistant Professor in the ECE Department at Iowa State University. She is currently serving as an Associate Editor for the IEEE Transactions on Signal Processing (2009-present). Her research interests are in estimation problems in statistical and sequential signal processing and biomedical imaging. Her current research focus is on recursive sparse reconstruction, sequential compressive sensing and large dimensional tracking problems.

{\bf Namrata Vaswani} received a B.Tech. from the Indian Institute of Technology (IIT), Delhi, in August 1999 and a Ph.D. from the University of Maryland, College Park, in August 2004, both in electrical engineering. From 2004 to 2005, she was a research scientist at Georgia Tech. Since Fall 2005, she has been an Assistant Professor in the ECE department at Iowa State University. She is currently serving as an Associate Editor for the IEEE Transactions on Signal Processing (2009-present).

Her research interests are in estimation and detection problems in sequential signal processing and in biomedical imaging. Her current focus is on recursive sparse reconstruction problems, sequential compressive sensing and large dimensional tracking problems.
\\

{\bf Wei Lu} received the B.E. degree from the Department of Electrical Engineering from Nanjing University of Posts and Telecommunications, China,in 2003. He received M.E. degree in Department of Electronic Engineering and Information Science from University of Science and Technology of China in 2006. He is currently a Ph.D student in the Department of Electrical and Computer Engineering in Iowa State University. His current research is focused on Modified Compressive Sensing for reconstruction of sparse signals. His research interests includes signal processing and image processing.

%To Do: (a) Fig 1, Results' figures,   (e) setting \alpha : Section II, V-A.
%Remove LS-CS, mod-LS-CS from 5b(right). %5b: remove LS-CS/mod-LS-CS from right. remove grid minor (just keep grid). ylabel('N-RMSE'), title(''), bigger font for axes, thicker lines, markers.
%Also say LS-CS had to be tweeked to get the result of 5b, left.

%(c) Remark 1 needed or not NOT NEEDED
%(b) introduction: DONE, Monte Carlo-pathological cases: KEEP IT
%, (d) iidLap, other notation (Reviewer 1): DONE, [1:n] was bothering another reviewer so will not change it.

\end{document}